\documentclass[10pt]{article}
\usepackage{graphicx}
\usepackage{amsmath, amsfonts, amssymb, amsthm}
\usepackage[english]{babel}
\usepackage{enumerate}
\usepackage{booktabs, caption}
\usepackage{array}
\usepackage{subcaption}
\usepackage{epstopdf}
\usepackage{natbib}
\usepackage{multicol}
\usepackage{etoolbox}
\usepackage{relsize}
\usepackage{enumitem}
\usepackage{pdflscape}
\usepackage{pifont}
\usepackage[flushleft]{threeparttable}
\newcommand{\blind}{0}

\numberwithin{equation}{section}
\theoremstyle{plain}
\newtheorem{theorem}{Theorem}[section]
\newtheorem{lemma}[theorem]{Lemma}
\newtheorem{prop}[theorem]{Proposition}

\theoremstyle{definition}
\newtheorem{example}{Example}[section]
\newtheorem{defn}{Definition}[section]

\newenvironment{remark}[1][Remark.]{\begin{trivlist}
\item[\hskip \labelsep {\bfseries #1}]}{\end{trivlist}}

\newcommand{\bs}[1]{\boldsymbol{#1}}
\newcommand{\var}{\textnormal{Var}}
\newcommand{\cov}{\textnormal{Cov}}

\newcommand{\te}[1]{\textnormal{#1}}
\newcommand{\nn}{\nonumber \\}
\newcommand{\tr}{\textnormal{Tr}}

\newcommand{\Log}{\textnormal{Log}}
\newcommand{\Exp}{\textnormal{Exp}}

\newcommand{\cmark}{\ding{51}}
\newcommand{\xmark}{\ding{55}}
\DeclareMathOperator*{\argmin}{\arg\!\min}
\makeatletter
\newcommand*\rel@kern[1]{\kern#1\dimexpr\macc@kerna}
\newcommand*\widebar[1]{%
  \begingroup
  \def\mathaccent##1##2{%
    \rel@kern{0.8}%
    \overline{\rel@kern{-0.8}\macc@nucleus\rel@kern{0.2}}%
    \rel@kern{-0.2}%
  }%
  \macc@depth\@ne
  \let\math@bgroup\@empty \let\math@egroup\macc@set@skewchar
  \mathsurround\z@ \frozen@everymath{\mathgroup\macc@group\relax}%
  \macc@set@skewchar\relax
  \let\mathaccentV\macc@nested@a
  \macc@nested@a\relax111{#1}%
  \endgroup
}
\makeatother
\begin{document}

\bibliographystyle{chicago}

\def\spacingset#1{\renewcommand{\baselinestretch}%
{#1}\small\normalsize} \spacingset{1.1}


\if0\blind
{
  \title{\bf Intrinsic wavelet regression for surfaces of Hermitian positive definite matrices}
  \author{Joris Chau\footnote{
	Corresponding author, j.chau@uclouvain.be, Institute of Statistics, Biostatistics, and Actuarial Sciences (ISBA), Universit\'e catholique de Louvain, Voie du Roman Pays 20, B-1348, Louvain-la-Neuve, Belgium. 
  }\quad and Rainer von Sachs\footnote{
  Institute of Statistics, Biostatistics, and Actuarial Sciences (ISBA), Universit\'e catholique de Louvain, Voie du Roman Pays 20, B-1348, Louvain-la-Neuve, Belgium.  
    }
  }
  \date{}
  \maketitle
} \fi

\if1\blind
{
  \bigskip
  \bigskip
  \bigskip
  \begin{center}
    {\LARGE\bf Intrinsic wavelet regression for surfaces of Hermitian positive definite matrices}
\end{center}
  \medskip
} \fi

\begin{abstract}
\noindent
This paper extends the intrinsic wavelet methods for curves of Hermitian positive definite matrices of \cite{CvS17} to \emph{surfaces} of Hermitian positive definite matrices, with in mind the application to nonparametric estimation of the time-varying spectral matrix of a locally stationary time series. First, intrinsic average-interpolating wavelet transforms acting directly on surfaces of Hermitian positive definite matrices are constructed in a curved Riemannian manifold with respect to an affine-invariant metric. Second, we derive the wavelet coefficient decay and linear wavelet thresholding convergence rates of intrinsically smooth surfaces of Hermitian positive definite matrices, and investigate practical nonlinear thresholding of wavelet coefficients based on their trace in the context of intrinsic signal plus noise models in the Riemannian manifold. The finite-sample performance of nonlinear tree-structured trace thresholding is assessed by means of simulated data, and the proposed intrinsic wavelet methods are used to estimate the time-varying spectral matrix of a nonstationary multivariate electroencephalography (EEG) time series recorded during an epileptic brain seizure.
\end{abstract}

\noindent%

\spacingset{1.45} 

\section{Introduction}\label{sec:1}
The Fourier spectral matrix of a second-order stationary multivariate time series can be interpreted as a curve of complex covariance matrices across frequencies in the Fourier domain. More precisely, the Fourier spectrum characterizes the variance-covariance structure of the multivariate time series expanded in terms of sines and cosines (i.e., the Fourier basis functions) oscillating at a particular frequency, and a non-degenerate Fourier spectral matrix therefore always constitutes a curve of Hermitian positive definite (HPD) matrices. In \cite{CvS17}, the authors investigated intrinsic wavelet transforms and wavelet regression for curves of Hermitian positive definite (HPD) matrices, with in mind the application to nonparametric spectral matrix estimation of a stationary multivariate time series. In many real-world applications, however, the assumption of stationarity of the time series may be too strict and one might relax the stationarity assumption to allow for more flexible modeling of spectral charactaristics that vary with time. For instance, in neuroscientific experiments involving electroencephalogram (EEG) or local field potential (LFP) time series recorded during a brain seizure, our aim is to analyze the Fourier spectra locally in time to analyze the evolving spectral behavior during the experiment. There is not a unique way to relax the assumption of stationarity to define a nonstationary time series process with a time-dependent Fourier spectrum. In this paper, we focus on nonparametric spectral estimation for a class of \emph{locally stationary time series} as first defined in \cite{Da97}. The following definition generalizes the Cram\'er representation of a stationary time series (see e.g., \cite[Section 11.8]{BD06}) and is similar in definition to \cite{GDOvS03}, \cite{GD06} and \cite{LK18} among others. This is a modified version of the locally time series model in \cite{Da97} or \cite{D12}, where the original sequences of functions $A^0_{t,T}(\omega)$  in \cite{Da97} and \cite{D12} are replaced by a 2-dimensional surface $A(\omega, t/T)$, typically assumed to be smooth across frequency and time.
\begin{defn} \emph{(Locally stationary vector-valued time series)} \label{def:1.1}
Let $\{ \vec{Y}_t, t = 1,\ldots,T\}$ be a zero-mean vector-valued time series observed at time points $t = 1,\ldots,T$. The time series $\vec{Y}_t$ is said to be \emph{locally stationary} if it admits the following representation with probability 1,
\begin{eqnarray*}
\vec{Y}_t &=& \int_{-\pi}^\pi A(\omega, t/T) \exp(i t \omega)\,d\vec{Z}(\omega).
\end{eqnarray*}
Here, $\{\vec{Z}(\omega), -\pi \leq \omega \leq \pi \}$ is a vector-valued zero-mean orthogonal increment process defined as in \cite[Section 11.8]{BD06}, with for $-\pi \leq \lambda \leq \pi$ and $0 \leq u \leq 1$,
\begin{eqnarray*}
\bs{E}\left[ \left( \int_{-\pi}^\lambda A(\omega, u)\, d\vec{Z}(\omega) \right) \left( \int_{-\pi}^\lambda A(\omega, u)\, d\vec{Z}(\omega) \right)^* \right] &=& \int_{-\pi}^\lambda f(\omega, u)\, d\omega,
\end{eqnarray*} 
such that the cumulants of $d\vec{Z}$ exist and are bounded for all orders. Moreover, $f(\omega, u)$ is the \emph{time-varying} or \emph{evolutionary} spectral density matrix at frequency $\omega \in [-\pi, \pi]$ and at time $u = t/T \in [0,1]$ rescaled to live in the unit interval. In addition, the time-varying spectral density matrix can be expressed in terms of the time-varying transfer function as $f(\omega, u) = A(\omega, u)A(\omega, u)^*$.
\end{defn}
\begin{remark}
In the original locally stationary time series model in \cite{Da97}, the true transfer function $A(\omega, t/T)$ is defined as the limit of a double-indexed sequence $A_{t,T}^0(\omega)$ in order to admit some parametric time series models, such as multivariate time-varying autoregressive (AR) models, not included in the class of locally stationary time series in Definition \ref{def:1.1}. However, this is not a problem in this paper, as we study nonparametric estimation of the time-varying spectral density matrix $f(\omega, u)$ and not estimation of parametric time series models. More precisely, assuming that $A^0_{t,T}(\omega) = A(\omega, t/T)$, with a sufficiently regular limiting surface $A(\omega, u)$, we address time-varying spectral density matrix estimation as a nonparametric 2-dimensional matrix-valued surface estimation problem. 
\end{remark}
In the nondegenerate case, the time-varying spectral density matrix $f(\omega, u)$ constitutes a surface of HPD matrices across time and frequency and any spectral matrix estimator $\hat{f}(\omega,u)$ should preserve these geometric constraints. This is important for several reasons: (i) interpretation of the time-varying spectral matrix estimator as a surface of complex covariance matrices in the time-frequency domain; (ii) well-defined transfer functions in the Cram\'er representation of a locally stationary time series above for the purpose of e.g., simulation of time series and bootstrapping; and (iii) sufficient regularity to avoid computational problems in subsequent inference procedures, such as e.g., computation of the time-varying partial coherences, which require the inverse of the estimated spectrum. This paper studies generalizations of the 1-dimensional (1D) intrinsic wavelet transforms for curves of HPD matrices in \cite{CvS17} extended to 2-dimensional (2D) intrinsic wavelet transforms for surfaces of HPD matrices, such as time-varying spectral density matrices. \\[3mm]
Nonparametric estimation of the time-varying spectral density matrix of a multivariate time series can be grouped in several different categories. In a standard Euclidean framework, straightforward nonparametric estimation of a sufficiently smooth time-varying target spectral matrix can be performed by smoothing segmented (short-time) or localized periodograms across both time and frequency via 2D kernel regression on surfaces of HPD matrices--extending 1D kernel regression as in \cite[Chapter 5]{B81}--  or via localized multitaper spectral estimation as in \cite{BB96} and \cite{XF07} for univariate nonstationary time series. Another approach is to segment the nonstationary time series into approximately stationary blocks and to apply traditional stationary spectral estimation methods (e.g., kernel- or projection-based periodogram smoothing or multitapering) on the segmented blocks, see e.g., \cite{A98} for univariate nonstationary time series or \cite{FO16} for replicated multivariate nonstationary time series. The work by \cite{OvSG05} is based on similar ideas, but with the Fourier basis functions and Fourier spectrum replaced by smooth localized exponential (SLEX) basis functions and its associated SLEX spectrum. We also mention the work by \cite{PEO14}, in which the authors instead consider nonstationary time series data from a vector-valued locally stationary wavelet processes, where estimation of the local wavelet spectral matrix is achieved by kernel smoothing of the wavelet periodogram matrices across wavelet locations. The disadvantage of estimation approaches that equip the space of HPD matrices with the Euclidean metric, is that flexible nonparametric (e.g., wavelet- or spline-) periodogram smoothing across time and frequency generally does not a guarantee positive definite spectral estimate, as such more flexible estimates easily surpass the boundary of the space of HPD matrices, which lies at a finite Euclidean distance. Instead, \cite{GD06}, \cite{Z16b} and \cite{LK18} equip the space of HPD matrices with the Cholesky metric and propose both frequentist and Bayesian procedures to construct time-varying HPD spectral matrix estimates as the square of an estimated surface of Cholesky square root matrices. This allows for more flexible estimation of the time-varying spectrum, such as individual smoothing of Cholesky matrix components, however, the Cholesky metric and Cholesky-based smoothing are not permutation-equivariant with respect to the components of the underlying time series. Essentially, this means that if one observes a reordering of the time series components, the estimated spectrum is not necessarily a rotated version of the estimate under the original ordering of the time series components. The latter implies that the ordering of the time series matters as it has a nontrivial impact on the estimated time-varying spectral matrix. \\[3mm]
The intrinsic surface wavelet transforms in this paper are generalizations of the intrinsic wavelet transforms for curves in \cite{CvS17} and are therefore defined independently of the metric on the space of HPD matrices. However, our primary focus in the remainder of this paper is on estimation in the space of HPD matrices equipped with the \emph{affine-invariant Riemannian metric} for the following reasons: (i) intrinsic 2D wavelet shrinkage of input HPD periodogram matrices guarantees an output HPD spectral estimate as the space of HPD matrices equipped with the affine-invariant metric is a complete metric space, (ii) intrinsic 2D wavelet shrinkage is equivariant to matrix congruence transformation by any invertible matrix, which implies that the time-varying spectral estimator does not nontrivially depend on the coordinate system of the multivariate time series, and (iii) there is no swelling effect as with the Euclidean metric as detailed in \cite{P10}, potentially leading to computational instability. Although we have in mind the application of intrinsic 2D wavelet shrinkage to time-varying HPD spectral matrix estimation, we emphasize that the estimation methods equally apply to other matrix-valued surface estimation or denoising problems, where the target is a surface of symmetric or Hermitian PD matrices. For instance, surface denoising of non-smoothly varying SPD diffusion covariance matrices in diffusion tensor imaging (DTI) as discussed in e.g., \cite{PFA05} or \cite{Y12}.\\[3mm]
The structure of the paper is as follows. In Section \ref{sec:2}, we introduce the necessary geometric tools and notions and consider natural generalizations of the intrinsic average-interpolation subdivision scheme and forward and backward wavelet transforms in \cite{CvS17} for surfaces of HPD matrices in a Riemannian manifold. In Section \ref{sec:3}, we derive wavelet coefficient decay rates of intrinsically smooth surfaces of HPD matrices and convergence rates of linear wavelet thresholding in the manifold of HPD matrices equipped with the affine-invariant Riemannian metric. In Section \ref{sec:4}, we consider intrinsic 2D nonlinear wavelet thresholding in the context of intrinsic i.i.d.\@ signal plus noise models, where the signal is a surface of HPD matrices, such as a time-varying spectral matrix. In particular, we consider nonlinear tree-structured thresholding of the trace of the matrix-valued wavelet coefficients. In Section \ref{sec:5}, we first compare the finite-sample performance of intrinsic wavelet regression to several nonparametric benchmark estimation procedures in the context of simulated surfaces of HPD matrices corrupted by noise, and second we estimate the time-varying Fourier spectrum of multivariate electroencephalography (EEG) time series data recorded during an epileptic brain seizure based on automatic intrinsic 2D wavelet thresholding of localized periodograms. The technical proofs and additional derivations can be found in the supplementary material. The accompanying \texttt{R}-code containing the tools to perform intrinsic 2D wavelet denoising in the space of HPD matrices --and to reproduce all of the illustrations and simulations in this paper-- is publicly available in the R-package pdSpecEst on CRAN, \cite{C17}.

\section{Intrinsic 2D AI wavelet transforms} \label{sec:2}

\subsection{Geometric notions and tools} \label{sec:2.1}
The space of $(d \times d)$-dimensional Hermitian positive definite matrices $\mathbb{P}_{d \times d}$ is an open subset of the real vector space of $(d \times d)$-dimensional Hermitian matrices $\mathbb{H}_{d \times d}$ and as such it is an $d^2$-dimensional smooth manifold, see e.g., \cite{B86}, \cite{L03} or \cite{D92a}. In this work, our primary focus is on the Riemannian manifold of HPD matrices equipped with the so-called \emph{affine-invariant Riemannian metric} according to \cite{PFA05}, \cite{BA11} or \cite{BA11b}. The space of HPD matrices equipped with the affine-invariant metric is a well-studied Riemannian manifold and the affine-invariant metric also appears in the literature as the \emph{natural invariant} metric (\cite{S00}), the \emph{canonical} metric (\cite{H16}), the \emph{trace} metric (\cite{Y12}), the \emph{Rao-Fisher} metric (\cite{S15}) or simply the \emph{Riemannian} metric in \cite[Chapter 6]{B09} or \cite{D09} among others. For ease of notation, in the remainder of this paper we will denote $\mathcal{M} := \mathbb{P}_{d \times d}$. For every $p \in \mathcal{M}$, the tangent space $T_p(\mathcal{M})$ can be identified by $\mathcal{H} := \mathbb{H}_{d \times d}$, and the affine-invariant Riemannian metric $g_R$ on the manifold $\mathcal{M}$ is given by the smooth family of inner products:
\begin{eqnarray} \label{eq:2.1}
\langle h_1, h_2 \rangle_p &=& \tr((p^{-1/2} \ast h_1)(p^{-1/2} \ast h_2)), \quad \quad \forall\: p \in \mathcal{M},
\end{eqnarray}
with $h_1,h_2 \in T_p(\mathcal{M})$. Here and throughout this paper, $y^{1/2}$ always denotes the Hermitian square root matrix of $y \in \mathcal{M}$, and $y \ast x$ is a short notation for the matrix congruence transformation $y^* x y$. The Riemannian distance on $\mathcal{M}$ derived from the Riemannian metric is given by:
\begin{eqnarray}
\delta_R(p_1,p_2) &=& \Vert \Log(p_1^{-1/2} \ast p_2) \Vert_F,  \label{eq:2.2}
\end{eqnarray}
where $\Vert \cdot \Vert_F$ denotes the matrix Frobenius norm and $\Log(\cdot)$ is the matrix logarithm. The mapping $x \mapsto a \ast x$ is an isometry for each invertible matrix $a \in \te{GL}(d,\mathbb{C})$, i.e., it is distance-preserving:
\begin{eqnarray*}
\delta_R(p_1, p_2) &=& \delta_R(a \ast p_1, a \ast p_2),\quad \forall\: a \in \te{GL}(d, \mathbb{C}).
\end{eqnarray*}
By \cite[Prop. 6.2.2]{B09}, the Riemannian manifold $(\mathcal{M}, g_R)$ is a \emph{geodesically complete} manifold, i.e., for each $p \in \mathcal{M}$, every geodesic through $p$ can be extended indefinitely. Moreover, by \cite[Theorem 6.1.6]{B09}, the \emph{geodesic} segment joining any two points $p_1, p_2 \in \mathcal{M}$ is unique and can be parametrized as,
\begin{eqnarray} \label{eq:2.3}
\eta(p_1, p_2, t) &=& p_1^{1/2} \ast \big(p_1^{-1/2} \ast p_2 \big)^t, \quad 0 \leq t \leq 1.
\end{eqnarray}
In the remainder of this paper we make extensive use of the so-called \emph{exponential} and \emph{logarithmic} maps, i.e., diffeomorphic maps between the manifold and its tangent spaces, which can heuristically be viewed as generalized notions of \emph{addition} and \emph{subtraction} on the Riemannian manifold. Since $(\mathcal{M}, g_R)$ is geodesically complete, for every $p \in \mathcal{M}$ the exponential map $\Exp_p$ and the logarithmic map $\Log_p$ are global diffeomorphisms with as domains $T_p(\mathcal{M})$ and $\mathcal{M}$ respectively by the Hopf-Rinow theorem. By (\cite{PFA05}), the exponential map $\Exp_p: T_p(\mathcal{M}) \to \mathcal{M}$ is given by,
\begin{eqnarray} \label{eq:2.4}
\Exp_p(h) &=& p^{1/2} \ast \Exp\left(p^{-1/2} \ast h \right),\quad \forall\: h \in T_p(\mathcal{M}),
\end{eqnarray}
where $\Exp(\cdot)$ denotes the matrix exponential. The logarithmic map $\Log_p: \mathcal{M} \to T_p(\mathcal{M})$ is given by the inverse exponential map:
\begin{eqnarray} \label{eq:2.5}
\Log_p(q) &=& p^{1/2} \ast \Log\left(p^{-1/2} \ast q \right).
\end{eqnarray}
The Riemannian distance may now also be expressed in terms of the logarithmic map as:
\begin{eqnarray} \label{eq:2.6}
\delta_R(p_1, p_2) \ \ = \ \ \Vert \Log_{p_1}(p_2) \Vert_{p_1} \ \ = \ \ \Vert \Log_{p_2}(p_1) \Vert_{p_2}, \quad \forall\: p_1, p_2 \in \mathcal{M},
\end{eqnarray}
where $\Vert h \Vert_p := \langle h, h \rangle_p$ denotes the norm of $h \in T_p(\mathcal{M})$ induced by the Riemannian metric.
\paragraph{Intrinsic means and averages}
Finally, an important tool that is used throughout this paper is the notion of the mean or average of a sample or distribution of HPD matrix observations \emph{intrinsic} to the affine-invariant Riemannian metric. A manifold-valued random variable $X: \Omega \to \mathcal{M}$ is defined to be a measurable function from a probability space $(\Omega, \mathcal{A}, \nu)$ to the measurable space $(\mathcal{M}, \mathcal{B}(\mathcal{M}))$, where $\mathcal{B}(\mathcal{M})$ is the Borel algebra in the complete separable metric space $(\mathcal{M}, \delta_R)$. By $P(\mathcal{M})$, we denote the set of all probability measures on $(\mathcal{M}, \mathcal{B}(\mathcal{M}))$ and $P_p(\mathcal{M})$ denotes the subset of probability measures in $P(\mathcal{M})$ that have finite moments of order $p$ with respect to the Riemannian distance,
\begin{eqnarray*}
P_p(\mathcal{M}) &:=& \left\{ \nu \in P(\mathcal{M}) : \exists\: y_0 \in \mathcal{M},\ \te{s.t.} \int_{\mathcal{M}} \delta(y_0, x)^p\: \nu(dx) < \infty \right\}
\end{eqnarray*}
Note that if $\int_{\mathcal{M}} \delta(y_0, x)^p\: \nu(dx) < \infty$ for some $y_0 \in \mathcal{M}$ and $1 \leq p < \infty$, this is true for any $y \in \mathcal{M}$, which follows by the triangle inequality and the fact that $\delta(p_1, p_2) < \infty$ for any $p_1,p_2 \in \mathcal{M}$. \\[3mm]
In the subsequent wavelet refinement scheme, the center of a random variable $X \sim \nu$ is charactarized by its intrinsic (also Karcher or Fr\'echet) mean, see e.g.\@ \cite{P06}. The set of intrinsic means is given by the points that minimize the second moment with respect to the Riemannian distance, 
\begin{eqnarray*}
\mu\ =\ \mathbb{E}_{\nu}[X] \ := \ \arg\min_{y \in \te{supp}(\nu)} \int_\mathcal{M} \delta(y,x)^2\ \nu(dx).
\end{eqnarray*}
If $\nu \in P_2(\mathcal{M})$, then at least one intrinsic mean exists and since the Riemannian manifold $(\mathcal{M}, g_R)$ is a space of non-positive curvature with no cut-locus (see e.g., \cite{PFA05} or \cite{S84}), by \cite[Proposition 1]{Le95} the intrinsic mean $\mu$ is also unique.
Recall that the cut-locus at a point $p \in \mathcal{M}$ is the complement of the image of the exponential map $\Exp_p$, which is the empty set for each $p \in \mathcal{M}$ as the image of $\Exp_p$ is the entire manifold $\mathcal{M}$. By \cite[Corollary 1]{P06}, the intrinsic mean can also be represented by the point $\mu \in \mathcal{M}$ that satisfies,
\begin{eqnarray} \label{eq:2.7}
\mu &=& \Exp_{\mu}\left( \int_{\mathcal{M}} \Log_{\mu}(x)\, \nu(dx) \right).
\end{eqnarray}
In general, the sample intrinsic mean of a set of observations $\{X_1,\ldots,X_n\} \in \mathcal{M}$ has no closed-form solution, but it can be computed efficiently through a gradient descent algorithm as described in e.g., \cite{P06}. Throughout this paper, we use the shorthand notation $\te{Ave}(\{X_i\}_{i=1}^n)$ for the (unweighted) intrinsic sample mean of $\{X_1,\ldots, X_n\} \in \mathcal{M}$. A weighted intrinsic sample mean of $\{X_1,\ldots,X_n\} \in \mathcal{M}$, with weights $\{ w_1,\ldots,w_n \} \in [0,1]$ summing up to one, will be denoted as $\te{Ave}(\{X_i\}_{i=1}^n; \{w_i\}_{i=1}^n)$ and is given by:
\begin{eqnarray*}
\te{Ave}(\{X_i\}_{i=1}^n, \{w_i\}_{i=1}^n) &=& \left\{ \bar{X}_n \in \mathcal{M} \, :\, \bar{X}_n = \Exp_{\bar{X}_n}\left( \sum_{i=1}^n w_i \Log_{\bar{X}_n}(X_i) \right) \right\}
\end{eqnarray*}
\begin{remark}
The representation of the intrinsic mean in eq.(\ref{eq:2.7}) above has an intuitive interpretation if we view the exponential and logarithmic maps as generalized notions of addition and subtraction on the Riemannian manifold. In particular, if we equip the Riemannian manifold of HPD matrices with the Euclidean metric, the exponential and logarithmic map correspond to ordinary matrix addition and subtraction and the above representation reduces to the usual expectation of a random variable.
\end{remark}
Table \ref{tab:1} above provides a quick overview of the different geometric tools used throughout this paper in the Riemannian manifold of HPD matrices equipped with the affine-invariant metric.
\begin{table}[t]
\begin{tabular}{lll}
  \toprule
  Manifold: & \hspace{.15cm} & $\mathbb{P}_{n \times n} := \{ p \in \mathbb{C}^{n \times n}\, :\, p = p^* \te{ and } \vec{z}^* p \vec{z} > 0, \te{ for } \vec{z} \in \mathbb{C}^n, \vec{z} \neq \vec{0} \}$\\
  Tangent spaces: & \hspace{.15cm} & $T_p(\mathbb{P}_{n \times n}) \cong \mathbb{H}_{n \times n} := \{ h \in \mathbb{C}^{n \times n}\, :\, h = h^* \}$ \\
  Riemannian metric: & \hspace{.15cm} & $\langle h_1, h_2 \rangle_p = \tr((p^{-1/2} \ast h_1)(p^{-1/2} \ast h_2))$\\
  Distance: & \hspace{.15cm} & $\delta_R(p_1,p_2) = \Vert \Log(p_1^{-1/2} \ast p_2)\Vert_F$\\
  Exponential map: & \hspace{.15cm} & $\Exp_p(h) = p^{1/2} \ast \Exp(p^{-1/2} \ast h)$ \\
  Logarithmic map: & \hspace{.15cm} & $\Log_p(q) = p^{1/2} \ast \Log(p^{-1/2} \ast q)$ \\
  Weighted average: & \hspace{.15cm} & $\te{Ave}(\{X_i\}_i; \{w_i\}_i) = \Big\{ \mu \in \mathbb{P}_{d \times d}\, :\, \mu = \Exp_{\mu}\big( \sum_i w_i \Log_{\mu}(X_i) \big) \Big\}$ \\
  \bottomrule
\end{tabular}
\caption{Geometric tools for the Riemannian manifold of HPD matrices equipped with the affine-invariant Riemannian metric. \label{tab:1}}
\end{table}

\subsection{Intrinsic 2D AI subdivision scheme} \label{sec:2.2}
\begin{figure}[ht]
\centering
\begin{subfigure}{1\linewidth}
\includegraphics[scale=0.53]{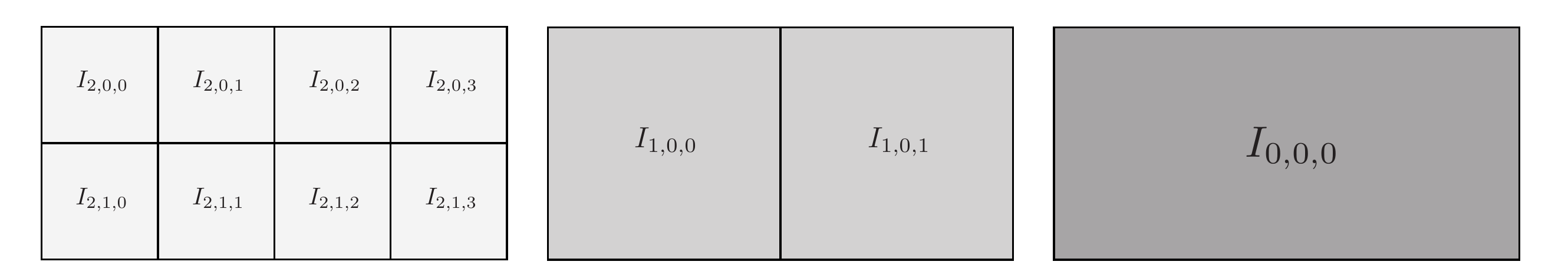}
\subcaption{Natural dyadic refinement pyramid (\cmark)} \label{fig:1a}
\end{subfigure}\\
\begin{subfigure}{1\linewidth}
\includegraphics[scale=0.53]{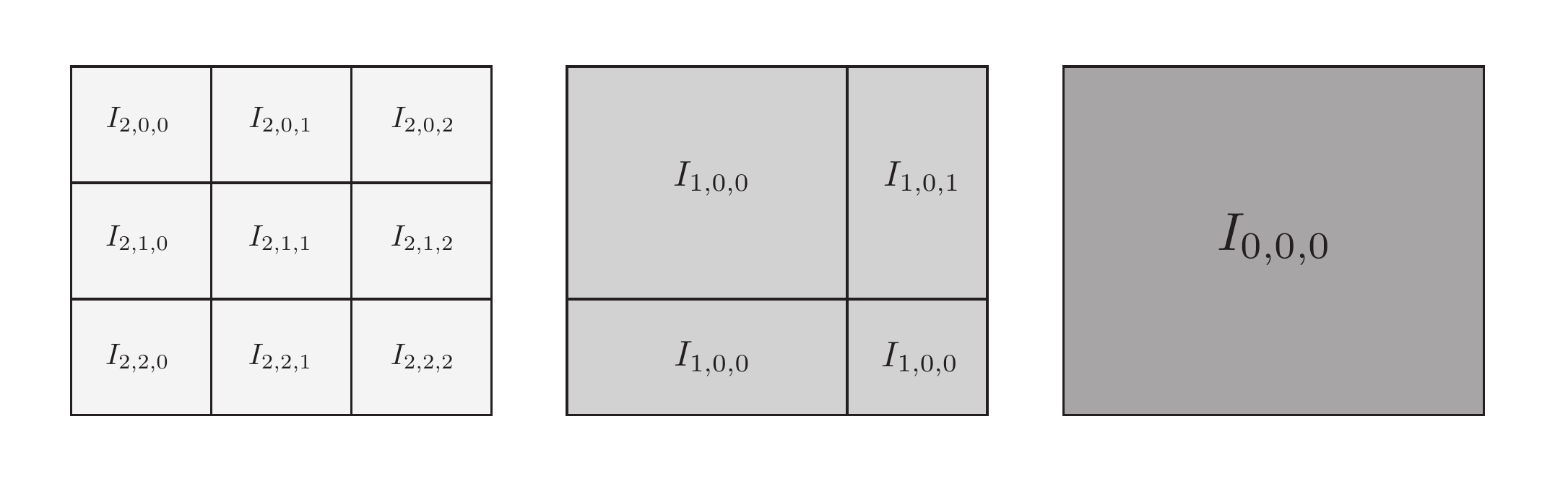}
\subcaption{Non-dyadic refinement pyramid (\cmark)} \label{fig:1b}
\end{subfigure}\\[3mm]
\begin{subfigure}{0.49\linewidth}
\includegraphics[scale=0.5]{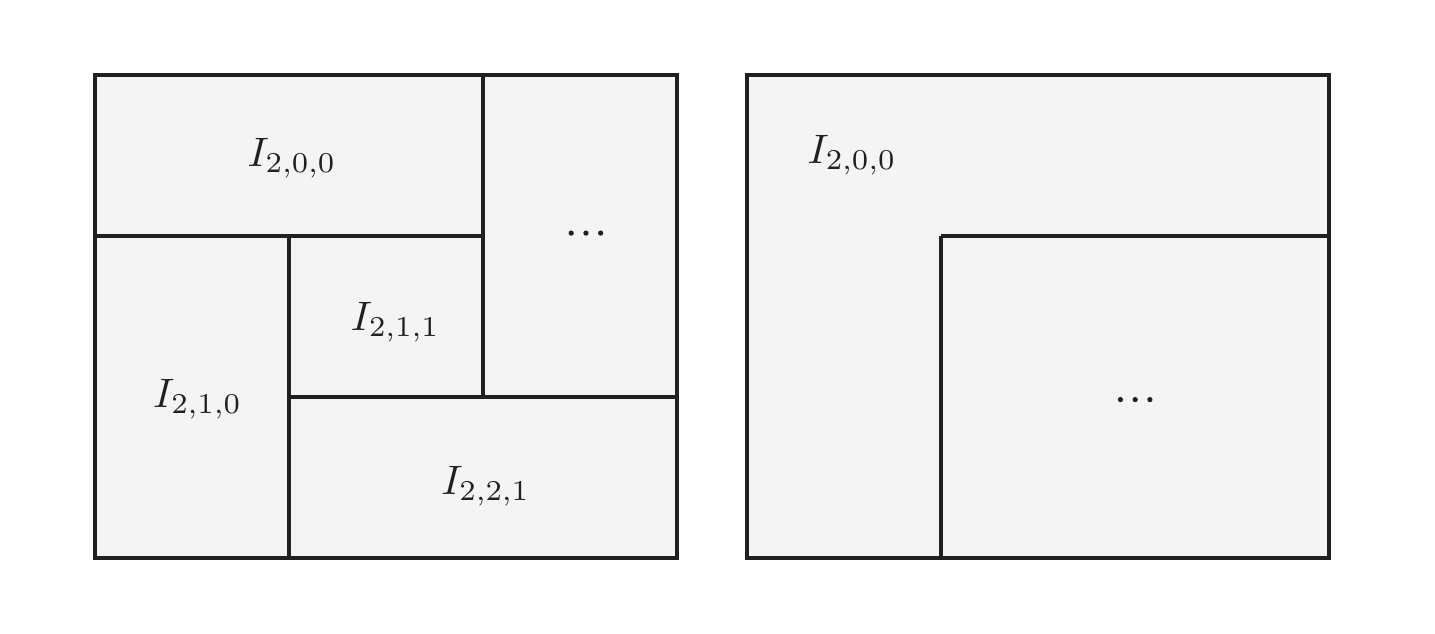}
\subcaption{Invalid partitions (Shape condition \xmark)} \label{fig:1c}
\end{subfigure}
\begin{subfigure}{0.49\linewidth}
\includegraphics[scale=0.5]{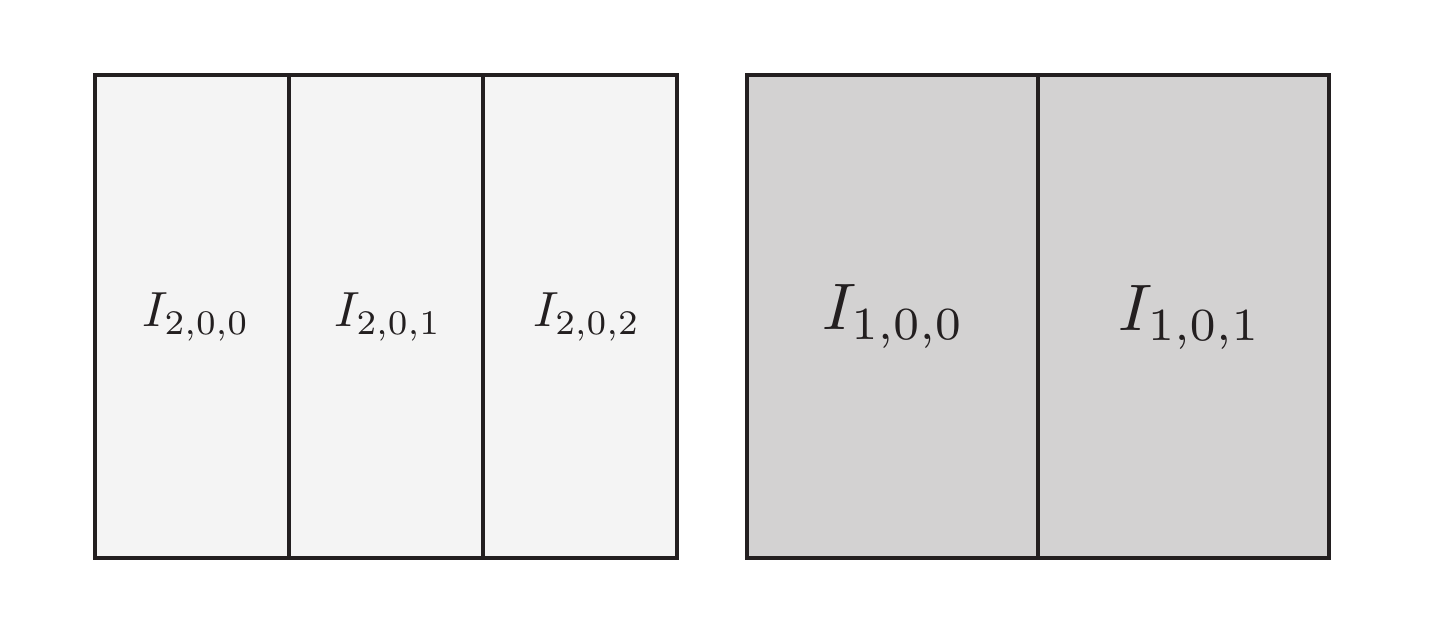}
\subcaption{Invalid refinement (Refinement condition \xmark)} \label{fig:1d}
\end{subfigure}
\end{figure}

The aim of this section is to construct intrinsic 2D average-interpolation (AI) wavelet transforms for surfaces in the space of HPD matrices as generalizations of the intrinsic 1D AI wavelet transforms in \cite{CvS17}. Let $\gamma: \mathcal{I} \to \mathcal{M}$ be a surface of HPD matrices on a bounded domain $\mathcal{I} \subset \mathbb{R}^2$, and suppose that we observe an $(n_1 \times n_2)$-dimensional discretized grid of local averages $M_{J,k_1,k_2} = \te{Ave}_{I_{J,k_1,k_2}}(\gamma)$ of the surface $\gamma$ across equally-sized, non-overlapping, closed rectangles $(I_{J,k_1,k_2})_{k_1,k_2}$ with $0 \leq k_1 \leq n_1-1$ and $0 \leq k_2 \leq n_2-1$, such that $\bigcup_{k_1,k_2} I_{J,k_1,k_2} = \mathcal{I}$. Here, $\te{Ave}_{I_{J,k_1,k_2}}(\gamma)$ denotes the intrinsic mean of $\gamma$ over the rectangle $I_{J,k_1,k_2}$ as described above, i.e.,
\begin{eqnarray*} 
\te{Ave}_{I_{J,k_1,k_2}}(\gamma) &=& \left\{ \mu \in \mathcal{M} \, :\, \mu = \Exp_{\mu}\bigg( \int_I \Log_{\mu}(\gamma(x_1, x_2))\ d(x_1,x_2) \bigg) \right\}.
\end{eqnarray*} 
For instance, if $\mathcal{I} = [0,1] \times [0,1]$, we may observe a $(2^J \times 2^J)$-dimensional grid of local averages across squares $I_{J,k_1,k_2} = [k_1 2^{-J}, (k_1 + 1) 2^{-J}] \times [k_2 2^{-J}, (k_2 + 1) 2^{-J}]$ for $0 \leq k_1, k_2 \leq 2^J - 1$. In general, wavelet transforms are most straightforward to build in a dyadic framework, i.e., the sizes $n_1$ and $n_2$ are both dyadic numbers. In the construction of the wavelet transforms below, we do not impose dyadic constraints on $n_1$ and $n_2$. Instead, we assume that we are given (or have chosen) a redundant refinement pyramid of closed rectangles $(I_{j,k_1,k_2})_{j, k_1,k_2}$ at resolution scales $j=0,\ldots,J$, satisfying the following set of constraints: 
\begin{enumerate}
\item \textbf{Shape condition}: at each resolution scale $j=0,\ldots,J$, the union of a rectangle $I_{j, \cdot, \cdot}$ and its top-, bottom-, left- or right-connected neighboring rectangle --if existing-- form again a rectangle. This property ensures that we can assign unambiguous location indices $(k_1,k_2)$ to each rectangle in the refinement pyramid, as columns and rows of rectangles are well-defined.
\item \textbf{Partitioning condition}: at each scale $j=0,\ldots,J$, the collection of rectangles $(I_{j,k_1,k_2})_{k_1,k_2}$ covers the entire domain, i.e., $\bigcup_{k_1,k_2} I_{j,k_1,k_2} = \mathcal{I}$, and their overlap has measure zero, i.e., $\lambda_2\left( \bigcap_{k_1,k_2} I_{j,k_1,k_2} \right) = 0$, where throughout this paper $\lambda_2$ denotes the Lebesgue measure on $\mathbb{R}^2$. 
\item \textbf{Refinement condition}: at each scale $j=0,\ldots,J-1$, every coarser-scale rectangle $I_{j,k_1,k_2}$ can be expressed as the union of a non-empty set of finer-scale rectangles, i.e., $I_{j,k_1,k_2} = \bigcup_{\ell_1,\ell_2} I_{j+1,\ell_1,\ell_2}$ for some set of locations $(\ell_1,\ell_2)$, such that $I_{j+1, \ell_1, \ell_2} \subset I_{j,k_1,k_2}$. In addition, the number of rectangles is strictly increasing from coarser to finer scales, i.e., $\# (I_{j,k_1,k_2})_{k_1,k_2} < \# (I_{j+1,\ell_1,\ell_2})_{\ell_1,\ell_2}$.
\end{enumerate}
In the remainder, for convenience, it is assumed that the coarsest scale $j = 0$ contains a single covering rectangle $I_{0,0,0} = \mathcal{I}$, this is not a constraint as we can always add an additional resolution scale in the refinement pyramid combining the rectangles at the coarsest scale into a single rectangle covering the complete domain $\mathcal{I}$. Figures \ref{fig:1a} and \ref{fig:1b} give two examples of valid rectangle pyramids using both dyadic and non-dyadic refinement steps and Figures \ref{fig:1c} and \ref{fig:1d} give several examples of invalid refinement partitions.
\paragraph{Midpoint pyramid}
Given the grid of local averages $(M_{J,k_1,k_2})_{k_1,k_2}$ at resolution scale $J$ and the refinement pyramid $(I_{j,k_1,k_2})_{j,k_1,k_2}$ with $j = 0,\ldots, J$, we can build a redundant midpoint or scaling coefficient pyramid associated to the rectangles $(I_{j,k_1,k_2})_{k_1,k_2}$ at each resolution scale $j$. Note that the construction of the midpoint pyramid is similar to that in \cite{CvS17} for curves of HPD matrices, except that the size of the grid is not necessarily dyadic. At the finest resolution scale $j = J$, the local averages $(M_{J,k_1,k_2})_{k_1,k_2}$ are given. At the next coarser scale $j=J-1$, with indices $(\ell_1,\ell_2)$ such that $I_{j+1,\ell_1,\ell_2} \subset I_{j,k_1,k_2}$, compute the weighted intrinsic average,
\begin{eqnarray} \label{eq:2.8}
M_{j,k_1,k_2} &=& \te{Ave}\left( \left\{ M_{j+1, \ell_1, \ell_2} \right\}_{\ell_1,\ell_2}\, ;\, \left\{ 
\frac{\lambda_2(I_{j+1,\ell_1,\ell_2})}{\lambda_2(I_{j,k_1,k_2})}\right\}_{\ell_1,\ell_2} \right),
\end{eqnarray}
where the ratio $\lambda_2(I_{j+1,\ell_1,\ell_2})/\lambda_2(I_{j,k_1,k_2})$ corresponds to the relative size of the finer-scale rectangle $I_{j+1,\ell_1,\ell_2}$ in the coarser-scale rectangle $I_{j,k_1,k_2}$. In particular, by the \emph{refinement condition} above, the weights in the intrinsic averages always sum up to one. The above weighted averaging operation is iterated up to the coarsest scale $j=0$, which contains a single grand average $M_{0,0,0}$ over the domain $\mathcal{I}$, as $I_{0,0,0} = \mathcal{I}$ by assumption.

\paragraph{Intrinsic polynomial surfaces} In \cite{CvS17}, intrinsic polynomial curves of degree $k$ in the Riemannian manifold $(\mathcal{M}, g_R)$ are defined as  manifold-valued curves with vanishing $k$-th and higher-order \emph{covariant derivatives}. For the construction of the intrinsic 2D AI subdivision scheme, we need the notion of intrinsic polynomial surfaces, i.e., bivariate polynomials. Let $\gamma: \mathcal{I} \to \mathcal{M}$ with $\mathcal{I} \subseteq \mathbb{R}^2$ be a surface of HPD matrices. Throughout the remainder, $\gamma$ is always implicitly assumed to be a square integrable surface in the sense that $\int_{\mathcal{I}} \delta_R(\gamma(t,s), y_0)^2\, d(t,s) < \infty$ for some $y_0 \in \mathcal{M}$. We say that $\gamma$ is a polynomial surface or bivariate polynomial of bi-degree $(k_1, k_2)$ if for all $(t,s) \in \mathcal{I}$, 
\begin{eqnarray*}
\left\{ \begin{array}{ccc}
\nabla^{\ell_1}_{\gamma'_t} \gamma'_t(t, s) &=& \bs{0}, \quad \quad \forall\, \ell_1 \geq k_1, \\
\nabla^{\ell_2}_{\gamma'_s} \gamma'_s(t, s) &=& \bs{0}, \quad \quad \forall\, \ell_2 \geq k_2.
\end{array} \right.
\end{eqnarray*}
Here, $\gamma'_t(t,s)$ and $\gamma'_s(t,s)$ are partial derivatives of $\gamma(t,s)$ in the marginal directions $t$ and $s$ respectively, $\bs{0}$ is the zero matrix, and $\nabla^{\ell}$ denotes the $\ell$-th order (partial) covariant derivative. $\gamma'_t(t,s)$ and $\gamma'_s(t,s)$ are tangent vectors in $T_{\gamma(t,s)}(\mathcal{M})$ and can be represented as,
\begin{eqnarray*}
\gamma'_t(t,s)\ :=\ \frac{\partial}{\partial t} \gamma(t,s) &=& \lim_{\Delta t \to 0} \frac{\Log_{\gamma(t,s)}(\gamma(t + \Delta t, s))}{\Delta t} \nn
 \gamma'_s(t,s)\ :=\ \frac{\partial}{\partial s} \gamma(t,s) &=& \lim_{\Delta s \to 0} \frac{\Log_{\gamma(t,s)}(\gamma(t, s + \Delta s))}{\Delta s}.
\end{eqnarray*}
The partial covariant derivative $\nabla_{\gamma'_t} \gamma'_t(t,s)$ of $\gamma(t,s)$ in the first variable $t$ can then be written in terms of the parallel transport as,
\begin{eqnarray*}
\nabla_{\gamma'_t} \gamma'_t(t,s) &=& \lim_{\Delta t \to 0} \frac{\Gamma(\gamma(\cdot, s))_{t + \Delta t}^t(\gamma'_t(t + \Delta t, s)) - \gamma'_t(t, s)}{\Delta t},
\end{eqnarray*}
where $\Gamma(\gamma(\cdot, s))_{t + \Delta t}^t(\gamma'(t+\Delta t, s))$ denotes the \emph{parallel transport} on the Riemannian manifold $(\mathcal{M}, g_R)$ of the tangent vector $\gamma'(t + \Delta t, s) \in T_{\gamma(t + \Delta t, s)}(\mathcal{M})$ to the tangent space $T_{\gamma(t, s)}(\mathcal{M})$ transported along the surface $\gamma(\cdot, s)$ in its first coordinate, (i.e., along a curve), keeping the second coordinate fixed. For a more detailed description of the parallel transport in the Riemannian manifold $(\mathcal{M}, g_R)$, we also refer to \cite[Chapter 2]{C18}. The definition is analogous for $\nabla_{\gamma'_s} \gamma'_s(t,s)$, the partial covariant derivative of $\gamma(t,s)$ in the second variable $s$. Essentially, the above definition of an intrinsic polynomial surface says that for fixed $s$, $\gamma(\cdot,s)$ is a polynomial curve of degree $k_1$ in its first argument, and conversely, for fixed $t$, $\gamma(t,\cdot)$ is a polynomial curve of degree $k_2$ in its second argument. For instance, a $(0,0)$-degree polynomial is a surface with constant values for each $(t,s)$; a $(0,1)$- or $(1,0)$-degree polynomial is a translated geodesic curve, either along the $t$- or the $s$-axis; and a $(1,1)$-degree polynomial is a geodesic surface. Note that discretized polynomial surfaces are straightforward to generate by extending the numerical integration algorithm in \cite{HFJ14} for discretized polynomial curves.

\paragraph{Intrinsic polynomial surface interpolation}
In the intrinsic 2D AI subdivision scheme described below, in order to predict finer scale midpoints from a collection of coarse scale midpoints, we need the notion of polynomial surface interpolation in the Riemannian manifold of HPD matrices. To be precise, given control points $P_{i,j} \in \mathcal{M}$ on a rectangular grid at the nodes $(t_{i}, s_j)$ for $i=0,\ldots,n$ and $j=0,\ldots,m$, we wish to evaluate the bivariate polynomial $P(t,s)$ of bi-degree $(n,m)$ going through the $(n+1)(m+1)$ control points at some location $(\tilde{t}, \tilde{s})$, with $(t_0, s_0) \leq (\tilde{t},\tilde{s}) \leq (t_n, s_m)$. To solve this bivariate problem we proceed as follows: 
\begin{enumerate}
\item Let $P_j(t)$ for $j=0,\ldots,m$ be the interpolating polynomial curve according to \cite{CvS17} through the control points $P_{0,j},\ldots,P_{n,j}$ at the nodes $t_0,\ldots,t_n$, i.e., $P_j(t_i) = P_{i,j}$. These polynomials evaluated at $\tilde{t}$, i.e., $P_j(\tilde{t})$, can be constructed directly by $m+1$ applications of the univariate intrinsic version of Neville's algorithm outlined in \cite{CvS17} or \cite[Chapter 3]{C18}.
\item Let $P(\tilde{t},s)$ be the interpolating polynomial curve through the points $P_0(\tilde{t}),\ldots,P_m(\tilde{t})$ at the nodes $s_0,\ldots,s_m$, i.e., $P(\tilde{t}, s_j) = P_j(\tilde{t})$. Then, the bivariate polynomial of bi-degree $(n,m)$ evaluated at $(\tilde{t}, \tilde{s})$, i.e., $P(\tilde{t}, \tilde{s})$, can be constructed by a single application of the univariate intrinsic version of Neville's algorithm. By this construction, $P(t,s)$ is the unique bivariate polynomial of degree $(n,m)$ that satisfies $P(t_i, s_j) = P_{i,j}$ for each $i=0,\ldots,n$ and $j=0,\ldots,m$.
\end{enumerate}
Intrinsic polynomial interpolation with respect to the affine-invariant metric for a surface of HPD matrices by means of Neville's algorithm is implemented in \texttt{R} through the function \texttt{pdNeville()} in the \texttt{pdSpecEst}-package, and we refer to the package documentation for additional details and information about this function. 
\begin{figure}[t]
\centering
\includegraphics[scale=0.22]{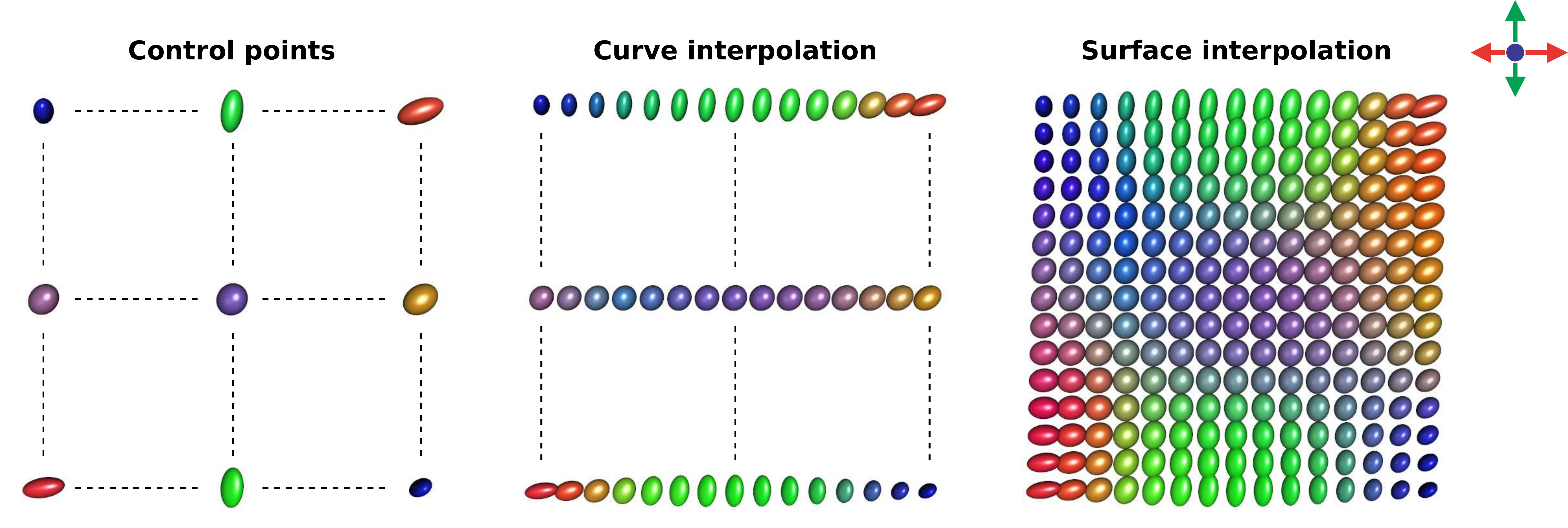}
\caption{Illustration of intrinsic polynomial surface interpolation for a set of random $(3 \times 3)$-dimensional SPD matrix-valued control points represented as 3D-ellipsoids. The colors indicate the direction of the eigenvector associated to the largest eigenvalue of each matrix-valued object. \label{fig:1e}}
\end{figure}

\subsubsection{Midpoint prediction via intrinsic average-interpolation}
Equipped with the notion of intrinsic polynomial surfaces and a practical procedure for polynomial surface interpolation, we outline the intrinsic 2D AI subdivision scheme. The aim of the intrinsic 2D AI subdivision scheme is to reconstruct an intrinsic polynomial surface $\tilde{\gamma}(t, s)$ with $j$-scale midpoints $(M_{j,k_1',k_2'})_{k_1',k_2'}$ as given by the midpoint pyramid. This is an extended version of the intrinsic 1D AI subdivision scheme in \cite{CvS17}, and it is instructive to compare the steps listed in this paper to their analogous counterparts in \cite{CvS17}. The average-interpolating surface $\tilde{\gamma}(t,s)$ with $j$-scale midpoints $(M_{j,k_1',k_2'})_{k_1',k_2'}$ does not follow directly from reconstructing the intrinsic polynomial surface passing through the $j$-scale midpoints. Instead, consider the cumulative intrinsic mean of $\tilde{\gamma}(t, s)$, \@ $M_{t_0, s_0} : \mathcal{I} \cap ([t_0, \infty) \times [\nu_0,\infty)) \to \mathcal{M}$, given by:
\begin{eqnarray} \label{eq:2.9}
M_{t_0, s_0}(t,s) &=& \te{Ave}_{[t_0, t] \times [s_0, s]}(\tilde{\gamma}).
\end{eqnarray}
That is, $M_{t_0, s_0}(t,s)$ solves:
\begin{eqnarray*}
M_{t_0, s_0}(t,s) &=& \Exp_{M_{t_0,s_0}(t,s)}\left(\int_{s_0}^s \int_{t_0}^t \Log_{M_{t_0,s_0}(t,s)}(\tilde{\gamma}(t,s))\, dt\, ds \right). 
\end{eqnarray*}
To illustrate, suppose that $\mathcal{I} = [0,1] \times [0,1]$ and the refinement rectangles at scale $j$ are given by $I_{j, k_1,k_2} = [k_12^{-j}, (k_1+1)2^{-j}] \times [k_2 2^{-j}, (k_2+1)2^{-j}]$. If $\tilde{\gamma}(t, \nu)$ is the intrinsic polynomial surface with $j$-scale midpoints $(M_{j,k_1,k_2})_{k_1,k_2}$, then $M_{0,0}((k_1+1)2^{-j}, (k_2+1)2^{-j})$ equals the cumulative intrinsic average of the midpoints $\{M_{j, \ell_1,\ell_2} \}_{\ell_1,\ell_2}$, with $(\ell_1,\ell_2) \in \{ 0,\ldots,k_1 \} \times\{0,\ldots,k_2\}$. The key observation is that the cumulative intrinsic mean of an intrinsic polynomial of bi-degree $(r_1, r_2)$ is again an intrinsic polynomial of bi-degree smaller than or equal to $(r_1, r_2)$, analogous to the standard Euclidean case, where the integral of a polynomial also remains a polynomial. The intrinsic 2D AI subdivision scheme proceeds as follows: (i) collect a set of coarse-scale midpoints close to the refinement location of interest; (ii) using Neville's algorithm, reconstruct an intrinsic polynomial surface passing through the cumulative intrinsic averages based on the set of coarse-scale midpoints; and (iii) predict the finer-scale midpoints at the refinement location of interest based on the fitted intrinsic polynomial surface. These steps are described in detail below:
\begin{enumerate}
\item \textbf{Collect coarse-scale midpoints}.
Fix a midpoint location $(k_1,k_2)$ at scale $j-1$ and an average-interpolation order $(N_1,N_2) = (2L_1 + 1, 2L_2 + 1)$, with $L_1,L_2 \geq 0$. Collect the $N_1N_2$ closest neighboring $(j-1)$-scale midpoints $(M_{j-1,k_1 + \ell_1, k_2+\ell_2})_{\ell_1,\ell_2}$ to $M_{j-1,k_1,k_2}$, with $(\ell_1, \ell_2) \in \{-L_1,\ldots, L_1 \} \times \{-L_2, \ldots, L_2\}$. If the symmetric neighboring midpoints at the locations $(\ell_1, \ell_2) \in \{-L_1,\ldots, L_1 \} \times \{-L_2, \ldots, L_2\}$ do not all exist, instead, we either collect the non-symmetric $N_1N_2$ closest neighboring midpoints, or we reduce the average-interpolation order, such that all symmetric neighbors exist. For instance, at a boundary location $(0,0)$ corresponding to the corner rectangle $I_{j-1,0,0}$, we may collect the set of closest existing neighboring midpoints to $M_{j-1,0,0}$ given by $(M_{j-1,\ell_1, \ell_2})_{\ell_1,\ell_2}$, with $(\ell_1, \ell_2) \in \{0,\ldots, 2L_1 \} \times \{0, \ldots, 2L_2\}$.
\item \textbf{Reconstruct intrinsic polynomial surface}.
For convenience, suppose that the $N_1N_2$ coarse-scale midpoints $(M_{j-1,k_1 + \ell_1, k_2+\ell_2})_{\ell_1,\ell_2}$, with symmetric neighboring locations $(\ell_1, \ell_2) \in \{-L_1,\ldots, L_1 \} \times \{-L_2, \ldots, L_2\}$ exist. For non-symmetric neighboring midpoints, the procedure is completely analogous. Set $(t_0, s_0) \in \mathbb{R}^2$ equal to the bottom left-corner of the refinement rectangle $I_{j-1,k_1-L_1,k_2-L_2}$, such that $(t_0, s_0) = \min\big(\bigcup_{\ell_1,\ell_2} I_{j-1,k_1+\ell_1, k_2+\ell_2}\big)$. Define the cumulative intrinsic averages $(\widebar{M}_{j-1, r_1, r_2})_{r_1,r_2}$ with $(r_1,r_2) \in \{0,\ldots, N_1-1\} \times \{0, \ldots,N_2-1\}$ as: 
\begin{eqnarray} \label{eq:2.10}
\widebar{M}_{j-1, r_1, r_2} &=& \te{Ave}\left( \{ M_{j-1,i_1,i_2} \}_{(i_1, i_2) \in \mathcal{R}} \ ; \ \dfrac{\lambda_2(I_{j-1,i_1,i_2})}{\lambda_2\Big( \bigcup\limits_{\substack{(s_1,s_2) \in \mathcal{R}}} I_{j-1,s_1,s_2} \Big)} \right) \nn
& = & M_{t_0, s_0}(\max(I_{j-1,k_1-L_1+r_1, k_2-L_2+r_2})). 
\end{eqnarray}
Here, $\mathcal{R} = \{k_1-L_1,\ldots,k_1-L_1+r_1\} \times \{k_1-L_2,\ldots,k_2-L_2+r_2\}$, $M_{t_0,s_0}(t,s)$ is as in eq.(\ref{eq:2.9}) based on the average-interpolating surface $\tilde{\gamma}(t,s)$ with $(j-1)$-scale midpoints $(M_{j-1,\vec{\ell}})_{\vec{\ell}}$ at $(I_{j-1,\vec{\ell}})_{\ell}$, and $\max(I_{j-1,k_1-L_1+r_1, k_2-L_2+r_2}) \in \mathbb{R}^2$ is the upper right-corner of the rectangle $I_{j-1,k_1-L_1+r_1, k_2-L_2+r_2}$. To illustrate, $\widebar{M}_{j-1,0,0} = M_{j-1,k_1-L_1,k_2-L_2}$ and $\widebar{M}_{j-1,N_1-1,N_2-1}$ is the intrinsic average of the total set of midpoints $(M_{j-1,k_1 + \ell_1, k_2+\ell_2})_{\ell_1,\ell_2}$.  Note that the expression after the second equality in the above equation is valid due to the \emph{shape condition} of the refinement pyramid. Given the set of cumulative intrinsic averages, we fit an interpolating polynomial surface $\widehat{M}_{t_0,s_0}(t,s)$ of order $(N_1-1,N_2-1)$ through the $N_1N_2$ cumulative midpoints $(\widebar{M}_{j-1, r_1, r_2})_{r_1,r_2}$, this serves as a polynomial estimate of the surface $M_{t_0,s_0}(t,s)$. In practice, it is not necessary to reconstruct the entire interpolating polynomial surface, it suffices to evaluate the interpolating surface at a finite number of locations, which is done efficiently by Neville's algorithm as described above.
\item \textbf{Predict finer-scale midpoints}
The predicted finer-scale midpoints $(\widetilde{M}_{j,i_1,i_2})_{i_1,i_2}$ at locations $(i_1,i_2)$ corresponding to the finer-scale rectangles $I_{j,i_1,i_2} \subset I_{j-1,k_1,k_2}$ are now uniquely determined by the intrinsic local averages over the regions $I_{j,i_1,i_2}$ obtained from the fitted polynomial surface $\widehat{M}_{t_0,s_0}(t,s)$ and the midpoint relation in eq.(\ref{eq:2.8}). In general, the exact expressions for the predicted finer-scale midpoints $(\widetilde{M}_{j,i_1,i_2})_{i_1,i_2}$ inherently depend on the shapes and sizes of the refinement rectangles $(I_{j,i_1,i_2})_{i_1,i_2}$. Below, in a dyadic framework, we discuss finer-scale midpoint prediction in more detail based on a natural choice of equally-sized square refinement rectangles at each resolution scale. We also refer to the (simplistic) illustration in Figure \ref{fig:4} for a visual description of the prediction procedure.
\end{enumerate}
\begin{remark}
An important observation is that if the coarse midpoints $(M_{j-1,k_1 + \ell_1, k_2+\ell_2})_{\ell_1,\ell_2}$ are generated from an intrinsic polynomial surface $\gamma(t, s)$ of degree $\leq (N_1-1,N_2-1)$, then the finer-scale midpoints $(M_{j,i_1,i_2})_{i_1,i_2}$ are perfectly reconstructed. This is analogous to the intrinsic 1D AI subdivision scheme in \cite{CvS17} and is referred to as the \emph{intrinsic polynomial reproduction} property. 
\end{remark}

\subsubsection{Dyadic midpoint prediction} 
\begin{figure}[t]
\begin{subfigure}{0.3\linewidth}
\includegraphics[scale=0.36]{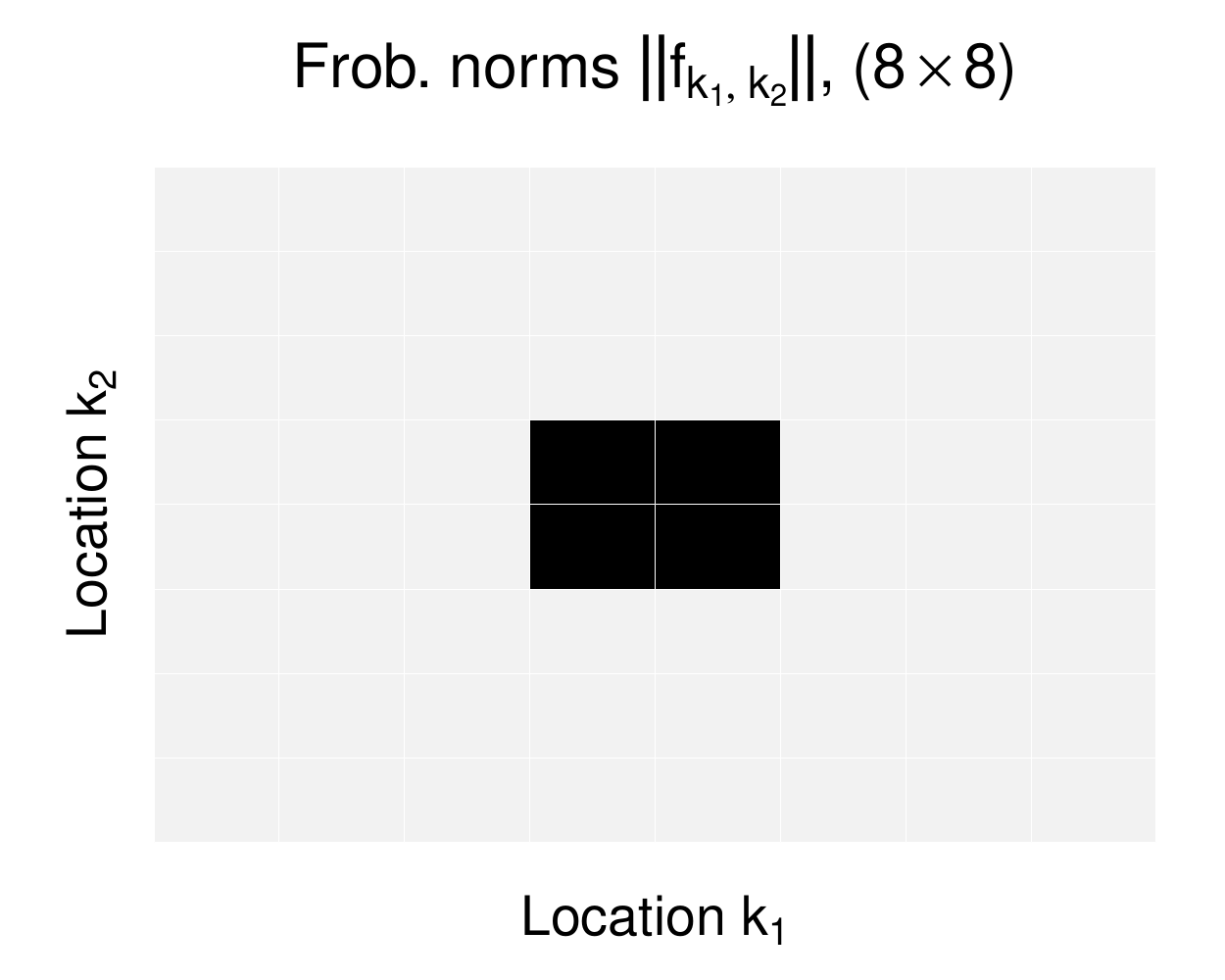}
\end{subfigure}
\begin{subfigure}{0.3\linewidth}
\includegraphics[scale=0.36]{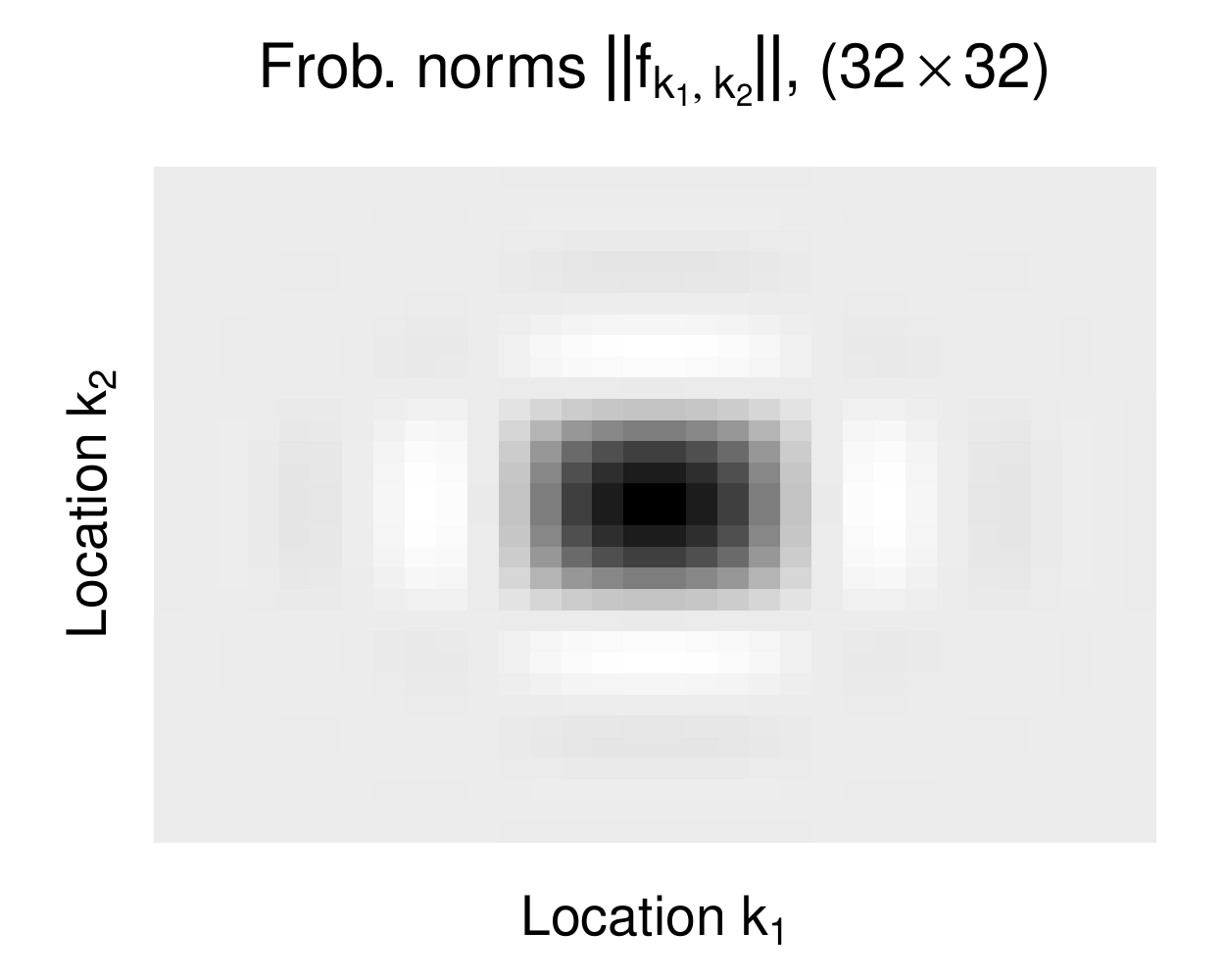}
\end{subfigure}
\begin{subfigure}{0.3\linewidth}
\includegraphics[scale=0.36]{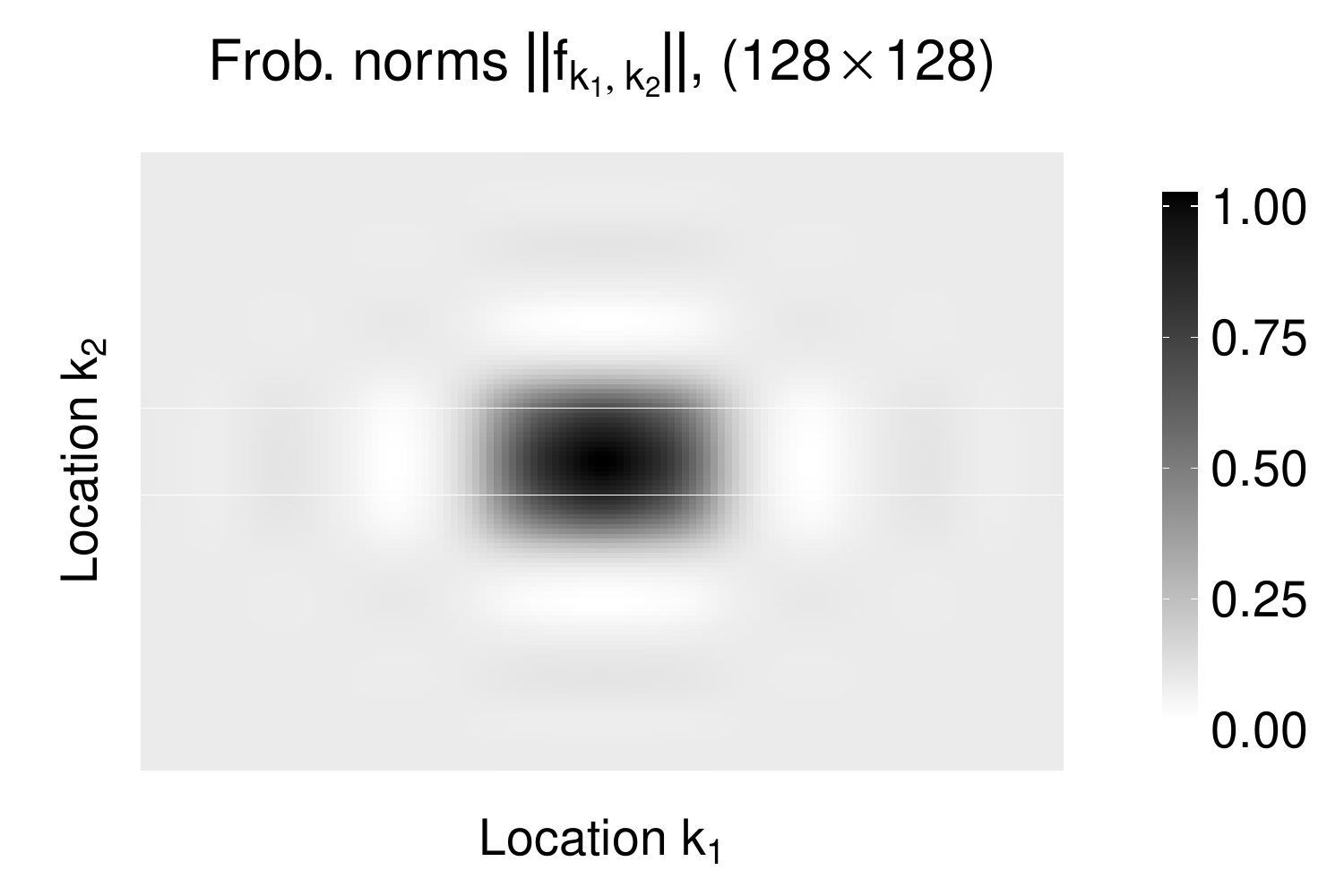}
\end{subfigure}
\caption{(Away from boundary) Frobenius norm of $(f_{k_1,k_2})_{k_1,k_2}$ after successive applications of the subdivision scheme of order $(N_1,N_2) = (5,5)$ starting from a square grid of size $(n_1,n_2) = (8,8)$.} \label{fig:2}
\end{figure}
As the midpoint prediction is most straightforward in a dyadic framework, we discuss the dyadic intrinsic 2D AI subdivision scheme in more detail. 
Suppose that we observe an $(n_1 \times n_2)$-dimensional grid of local averages $(M_{J,k_1,k_2})_{k_1,k_2}$ at the finest resolution scale $J$ across equally-sized rectangles $(I_{J,k_1,k_2})_{k_1,k_2}$, such that $n_1 = 2^{J_1}$ and $n_2 = 2^{J_2}$ are both dyadic numbers. A natural choice for the refinement pyramid $(I_{j,k_1,k_2})_{j, k_1,k_2}$, with equally-sized square refinement rectangles at each resolution scale, is already shown in Figure \ref{fig:1a}. Without loss of generality, let us assume that $\mathcal{I} = [0,1] \times [0,1]$, then the maximum resolution scale is $J = \max(J_1,J_2)$ and the rectangles in the natural refinement pyramid at scale $j = 0,\ldots,J$ are given by:
\begin{eqnarray*}
I_{j,k_1,k_2} &=& ([k_1 2^{-j+J-J_1}, (k_1 + 1) 2^{-j + J - J_1}] \cap [0,1]) \times ([k_2 2^{-j + J - J_2}, (k_2 + 1) 2^{-j + J - J_2}] \cap [0,1]), 
\end{eqnarray*}
with locations $(k_1,k_2) \in \{0,\ldots,(2^{j-J+J_1}-1) \vee 0 \} \times \{0,\ldots,(2^{j-J+J_2}-1) \vee 0 \}$ at resolution scale $j$. In particular, if $J_1 \neq J_2$; for $j > |J_1 - J_2|$, $I_{j-1,k_1,k_2}$ is the union of four finer-scale rectangles $I_{j,2k_1,2k_2}$, $I_{j,2k_1+1,2k_2}$, $I_{j,2k_1,2k_2+1}$ and $I_{j,2k_1+1,2k_2+1}$; for $j \leq |J_1 - J_2|$, $I_{j-1, k_1,k_2}$ is the union of two finer-scale rectangles $I_{j,2k_1,k_2}$ and $I_{j,2k_1+1,k_2}$ if $J_1 > J_2$, or $I_{j,k_1,2k_2}$ and $I_{j,k_1,2k_2+1}$ if $J_1 < J_2$. We observe that for $j \leq |J_1 - J_2|$, the dyadic 2D subdivision scheme  essentially reduces to the dyadic 1D subdivision scheme for curves of HPD matrices in \cite{CvS17}.\\[3mm]
In the supplementary material, we give the exact expressions of the predicted midpoints in a dyadic framework based on intrinsic average-interpolation via Neville's algorithm as described above. In Figures \ref{fig:2} and \ref{fig:3}, we demonstrate successive applications of dyadic average-interpolation refinement for an interior and a boundary midpoint starting from 64 dummy HPD matrix-valued observations $(f_{k_1,k_2})_{k_1,k_2}$ on a square grid of size $(n_1,n_2) = (8,8)$. Analogous to \cite{CvS17}, the intrinsic version of Neville's algorithm essentially interpolates a polynomial surface through weighted geodesic combinations of the input set of coarse-scale midpoints, with weights depending on the average-interpolation order $(N_1,N_2)$. For this reason, the predicted midpoints $(\widetilde{M}_{j,i_1,i_2})_{i_1,i_2}$ can effectively be expressed as weighted intrinsic averages of the inputs $(M_{j-1,k_1 + \ell_1, k_2+\ell_2})_{\ell_1,\ell_2}$. For $j \leq |J_1 - J_2|$, the expressions for the predicted midpoints and their corresponding weights are exactly equivalent to the expressions in \cite{CvS17}, as the 2D subdivision scheme reduces to a 1D subdivision scheme. For $j > |J_1 - J_2|$, with $(i_1,i_2) \in \{(0,0),(0,1),(1,0),(1,1)\}$, the predicted midpoints can be represented as the following intrinsic weighted averages:
\begin{eqnarray} \label{eq:2.11}
\widetilde{M}_{j, 2k_1+i_1, 2k_2 + i_2} &=& \te{Ave}\left( \{ M_{j-1, k_1 + \ell_1, k_2 + \ell_2} \}_{\ell_1,\ell_2} \ ; \ \left\{ C^{N_1,N_2,i_1,i_2}_{L_1 + \ell_1, L_2 + \ell_2} \right\}_{\ell_1,\ell_2} \right), \quad 
\end{eqnarray}
with $(\ell_1,\ell_2) \in \{-L_1,\ldots,L_1\} \times \{ -L_2,\ldots,L_2 \}$. The weights $\bs{C}^{N_1,N_2,i_1,i_2} = (C_{m_1,m_2}^{N_1,N_2,i_1,i_2})_{m_1,m_2}$ with $(m_1,m_2) \in \{ 0,\ldots, N_1-1\} \times \{0,\ldots,N_2-1\}$ depend on the location indices $(i_1,i_2)$ and the average-interpolation order $(N_1,N_2) \geq 1$ and sum up to 1. For instance, away from the boundary, for $(i_1,i_2) = (0,0)$, the weights are as follows:
\begin{figure}[t]
\begin{subfigure}{0.3\linewidth}
\includegraphics[scale=0.36]{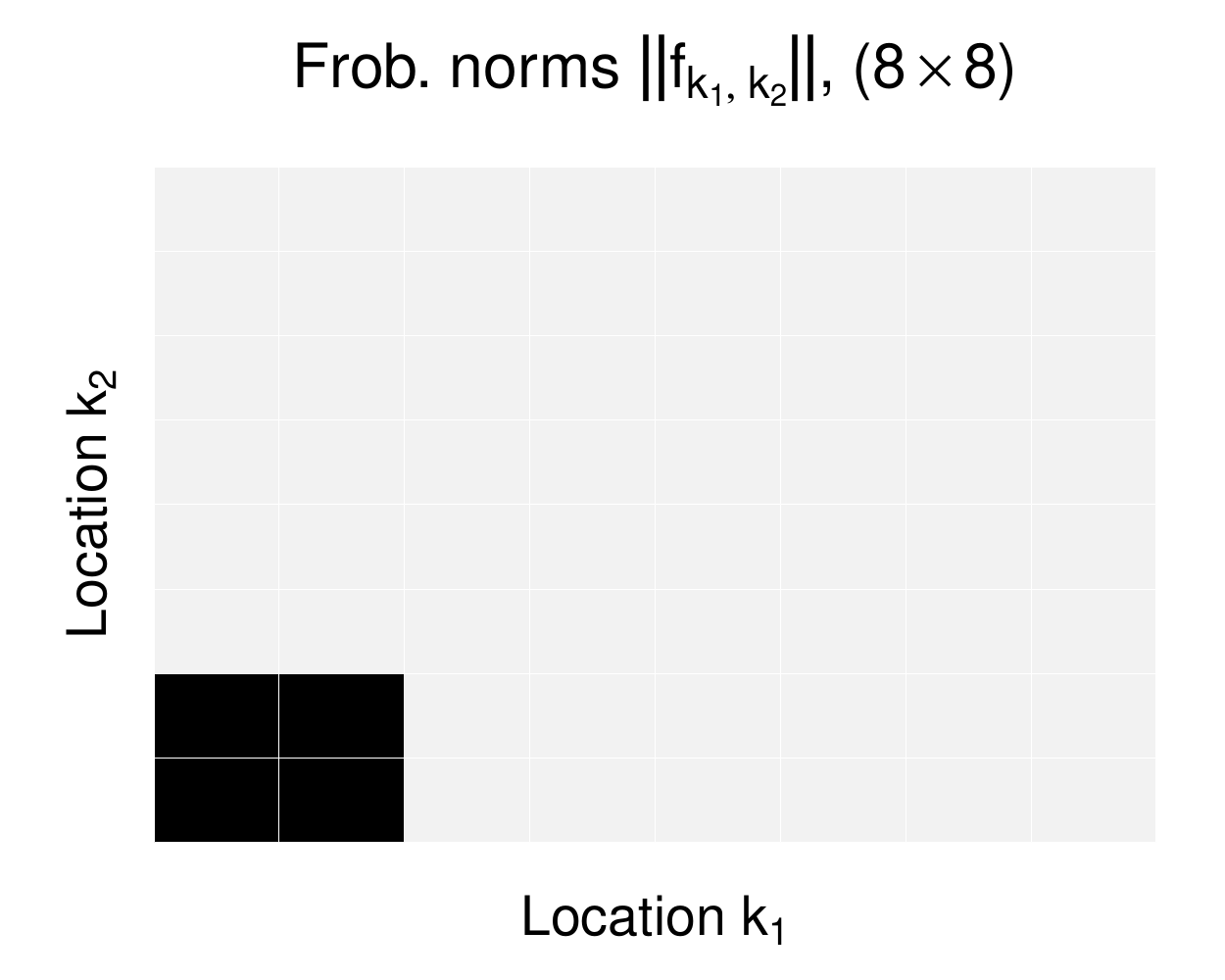}
\end{subfigure}
\begin{subfigure}{0.3\linewidth}
\includegraphics[scale=0.36]{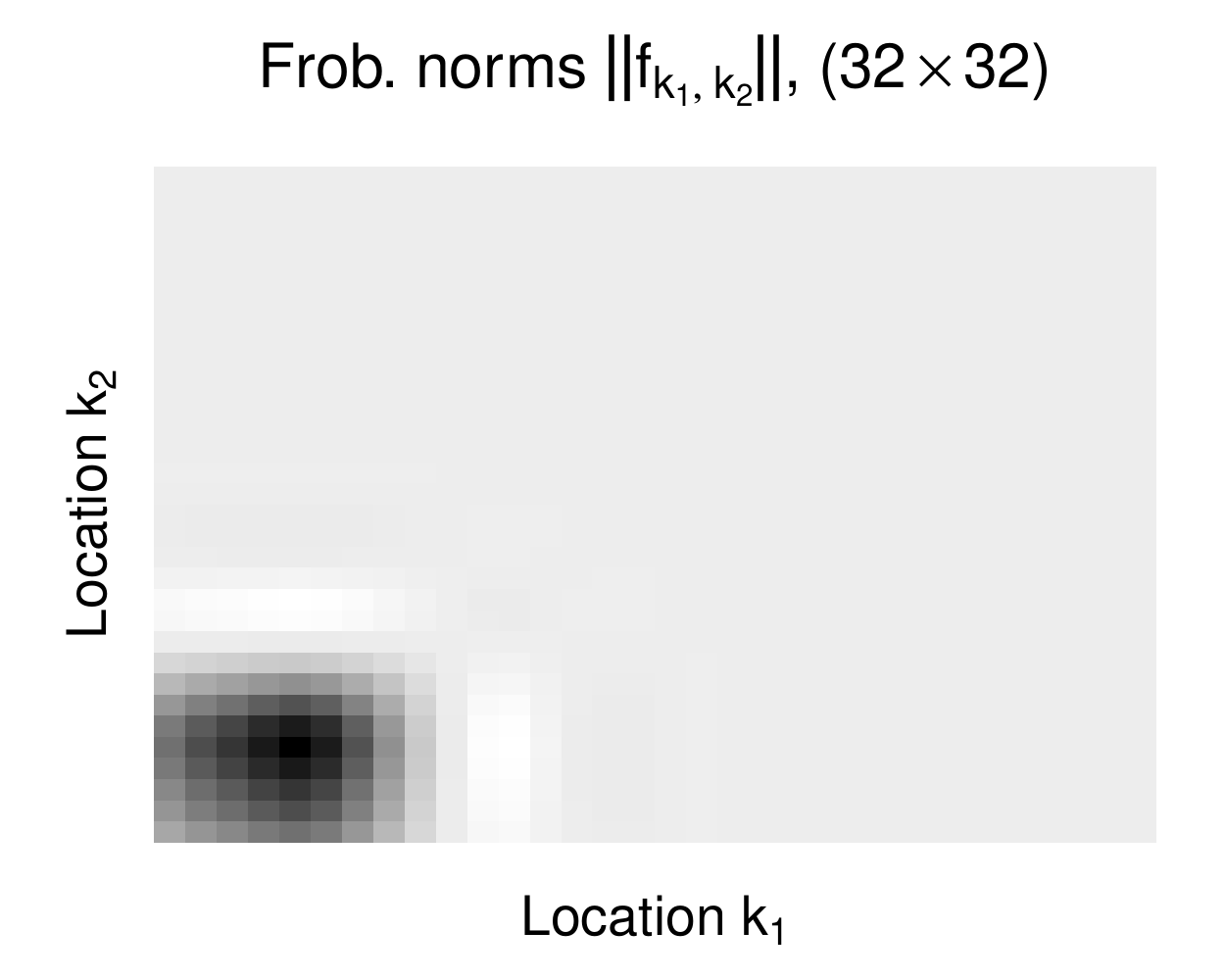}
\end{subfigure}
\begin{subfigure}{0.3\linewidth}
\includegraphics[scale=0.36]{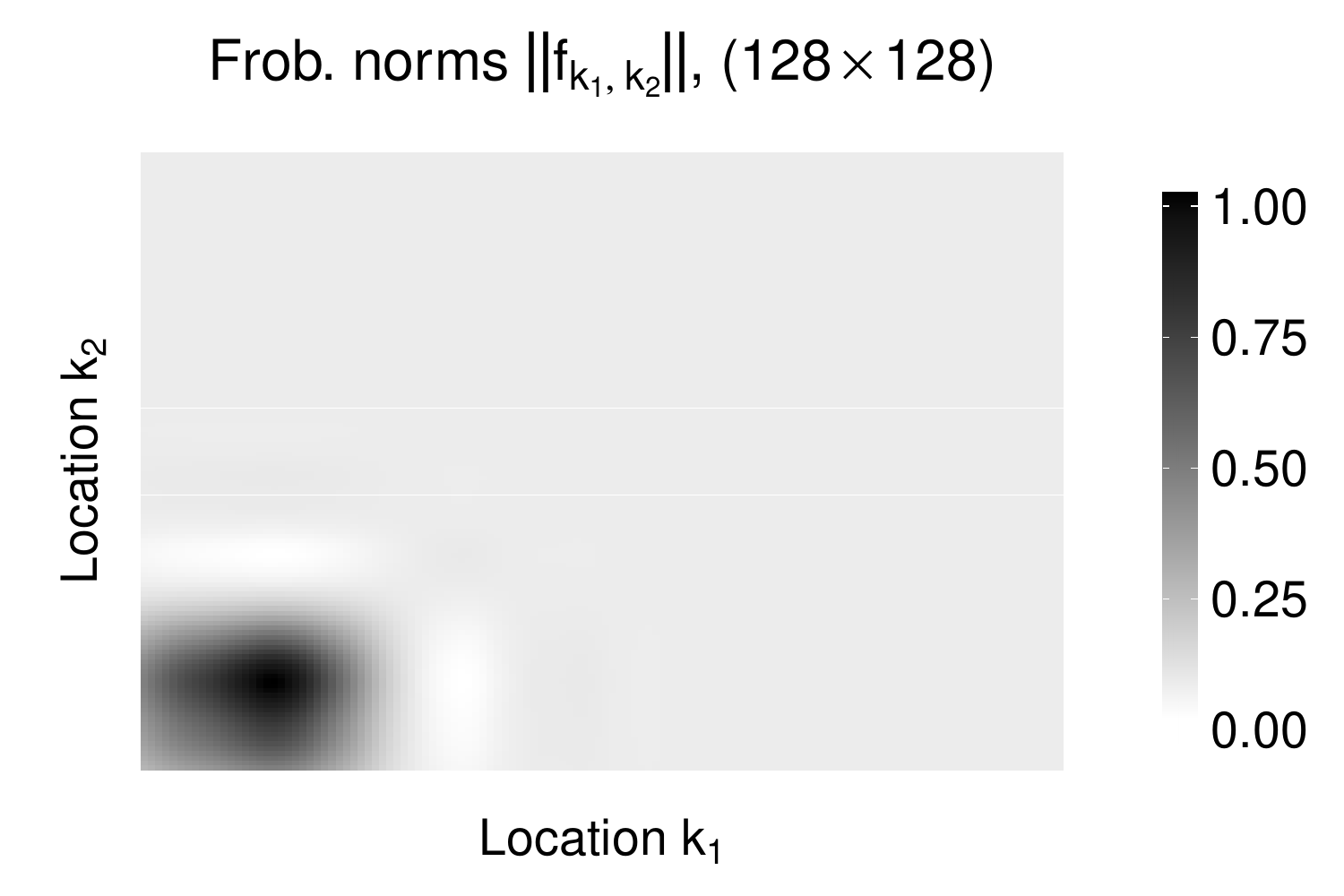}
\end{subfigure}
\caption{(At the boundary) Frobenius norm of $(f_{k_1,k_2})_{k_1,k_2}$ after successive applications of the subdivision scheme of order $(N_1,N_2) = (5,5)$ starting from a square grid size $(n_1,n_2) = (8,8)$.} \label{fig:3}
\end{figure}
\begin{itemize}
\item If $(N_1,N_2) = (1,1)$, then $C_{0,0}^{N_1,N_2,0,0} = 1$. 
\item If $(N_1,N_2) = (3,3)$, with rows $m_1 = \{0,1,2\}$ and columns $m_2 = \{0,1,2\}$,
\begin{eqnarray*}
\bs{C}^{N_1,N_1,0,0} &=& \left[ \begin{matrix}
\frac{1}{64} &  \frac{1}{8} & -\frac{1}{64} \\
\frac{1}{8} & 1 & -\frac{1}{8} \\
-\frac{1}{64} & -\frac{1}{8} & \frac{1}{64}
\end{matrix} \right]
\end{eqnarray*}
\item If $(N_1,N_2) = (5,5)$, with rows $m_1 = \{0,\ldots,4\}$ and columns $m_2 = \{0,\ldots,4\}$,
\begin{eqnarray*}
\bs{C}^{N_1,N_1,0,0} &=& \left[ \begin{matrix}
\frac{9}{16384} & -\frac{33}{8192} & -\frac{3}{128} & \frac{33}{8192} & -\frac{9}{16384} \\
-\frac{33}{8192} & \frac{121}{4096} & \frac{11}{64} & -\frac{121}{4096} & \frac{33}{8192} \\
-\frac{3}{128} & \frac{11}{64} & 1 & -\frac{11}{64} & \frac{3}{128} \\
\frac{33}{8192} & -\frac{121}{4096} & -\frac{11}{64} & \frac{121}{4096} & -\frac{33}{8192} \\
-\frac{9}{16384} & \frac{33}{8192} & \frac{3}{128} & -\frac{33}{8192} & \frac{9}{16384} \\
\end{matrix} \right]
\end{eqnarray*} 
\end{itemize}
In the R-package \texttt{pdSpecEst}, these prediction weights have been pre-determined for all combinations $(N_1,N_2) \leq (9,9)$ at locations $(k_1,k_2)$ away from the boundary, such that the symmetric $N_1N_2$ neighboring coarse-scale midpoints exist. This allows for faster computation of the predicted midpoints in practice using a weighted version of the gradient descent algorithm in \cite{P06}, with the function \texttt{pdMean()}. For higher average-interpolation orders (i.e., $N_1 \vee N_2 > 9$), or for predicted midpoints close to the boundary, (such that the symmetric neighboring midpoints are no longer available), the midpoints are predicted via Neville's algorithm as explained above. We point out that if $N_1 \wedge N_2 = 1$, the refinement weights reduce exactly to the 1D refinement weights discussed in \cite{CvS17}, as the dyadic 2D subdivision scheme reduces to a dyadic 1D subdivision scheme.

\subsection{Intrinsic forward and backward 2D AI wavelet transform} \label{sec:2.3}
\begin{figure}
\centering
\includegraphics[scale=0.45]{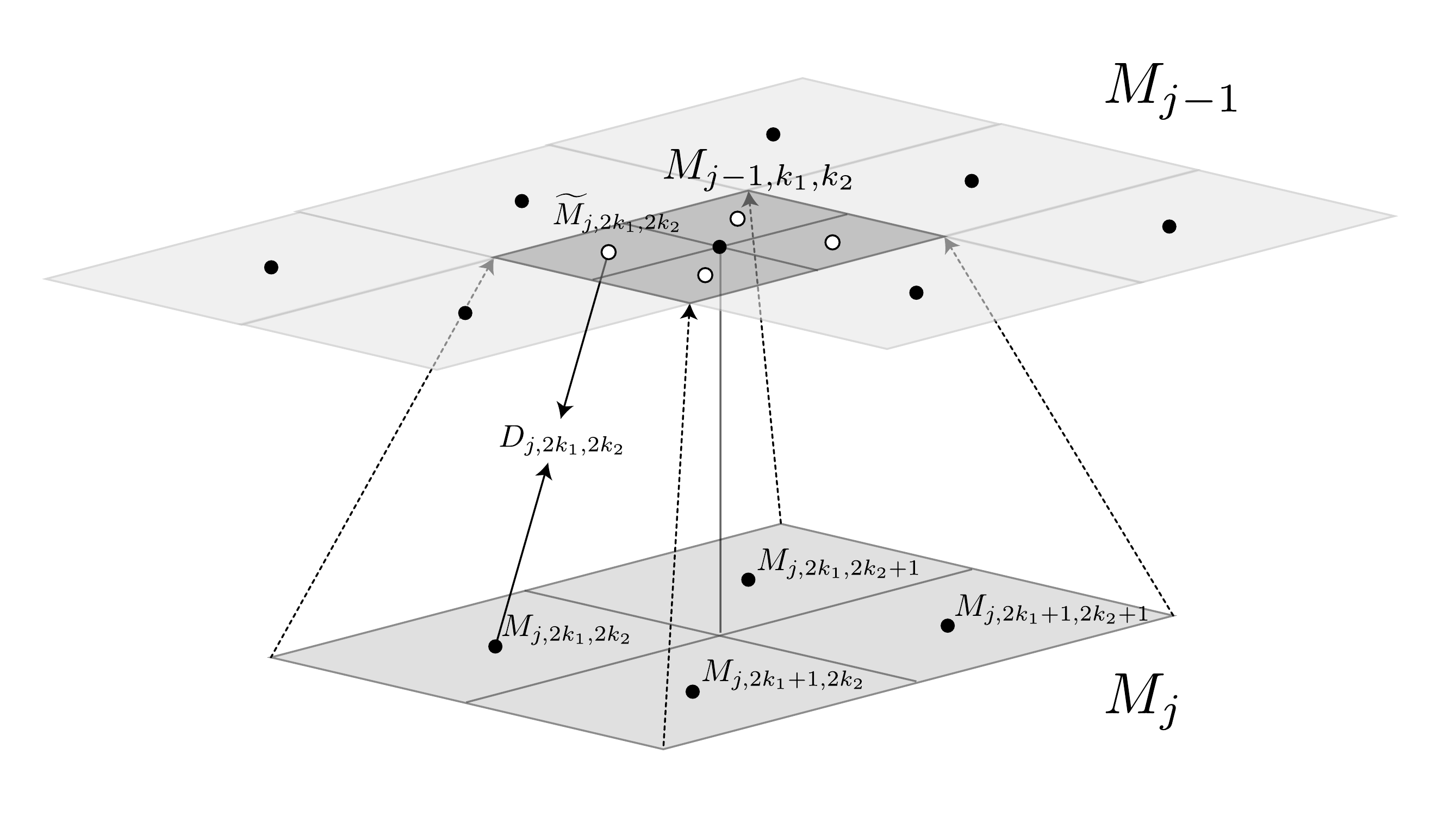}
\caption{(Simplistic) illustration of the forward 2D AI wavelet transform passing from $j$-scale midpoints to $(j-1)$-scale midpoints and $j$-scale wavelet coefficients: (i) compute coarse-scale midpoints $(M_{j-1,k_1',k_2'})_{k_1',k_2'}$, (ii) predict fine-scale midpoints $(\widetilde{M}_{j,k_1,k_2})_{k_1,k_2}$, (iii) compute wavelet coefficients $(D_{j,k_1,k_2})_{k_1,k_2}$.\label{fig:4}}
\end{figure}

\paragraph{Forward wavelet transform} The intrinsic 2D AI subdivision scheme leads to an intrinsic 2D AI wavelet transform passing from $j$-scale midpoints to $(j-1)$-scale midpoints plus $j$-scale wavelet coefficients as follows:
\begin{enumerate}
\item \textbf{Coarsen/Refine:} given the set of $j$-scale midpoints $(M_{j,k_1,k_2})_{k_1,k_2}$ corresponding to the refinement rectangles $(I_{j,k_1,k_2})_{k_1,k_2}$, compute the $(j-1)$-scale midpoints $(M_{j-1,k_1',k_2'})_{k_1',k_2'}$ corresponding to the refinement rectangles $(I_{j,k_1',k_2'})_{k_1',k_2'}$ through the coarsening step in eq.(\ref{eq:2.2}). Select an average-interpolation order $(N_1,N_2) \geq (1,1)$ and generate the predicted midpoints $(\widetilde{M}_{j,k_1,k_2})_{k_1,k_2}$ based on the $(j-1)$-scale midpoints $(M_{j-1,k_1',k_2'})_{k_1',k_2'}$ via the 2D AI subdivision scheme in Section \ref{sec:2.2}.
\item \textbf{Difference:} given the true and predicted $j$-scale midpoints $(M_{j,k_1,k_2})_{k_1,k_2}, (\widetilde{M}_{j,k_1,k_2})_{k_1,k_2}$, define the wavelet coefficients as scaled intrinsic differences according to,
\begin{eqnarray} \label{eq:2.12}
D_{j,k_1,k_2} &=& \sqrt{\frac{\lambda_2(I_{j,k_1,k_2})}{\lambda_2(\mathcal{I})}} \cdot \Log_{\widetilde{M}_{j,k_1,k_2}}\big(M_{j,k_1,k_2} \big)  \ \in \  T_{\widetilde{M}_{j,k_1,k_2}}(\mathcal{M}). \quad
\end{eqnarray}
Note that $\Vert D_{j,k_1,k_2} \Vert_{\widetilde{M}_{j,k_1,k_2}}^2 = \frac{\lambda_2(I_{j,k_1,k_2})}{\lambda_2(\mathcal{I})} \delta_R(M_{j,k_1,k_2}, \widetilde{M}_{j,k_1,k_2})^2$ by definition of the Riemannian distance, giving the wavelet coefficients the interpretation of a (scaled) difference between $M_{j,k_1,k_2}$ and $\widetilde{M}_{j,k_1,k_2}$. In the remainder, we also keep track of the \emph{whitened} wavelet coefficients, 
\begin{eqnarray} \label{eq:2.13}
\mathfrak{D}_{j,k_1,k_2} &=& \sqrt{\frac{\lambda_2(I_{j,k_1,k_2})}{\lambda_2(\mathcal{I})}} \cdot \widetilde{M}_{j,k_1,k_2}^{-1/2} \ast \Log_{\widetilde{M}_{j,k_1,k_2}}\big(M_{j,k_1,k_2}\big) \ \in \ T_{\te{Id}}(\mathcal{M}). \quad \quad 
\end{eqnarray}
The whitened coefficients correspond to the coefficients in eq.(\ref{eq:2.12}) transported to the same tangent space (at the identity matrix) via the so-called \emph{whitening transport} $\Gamma_{p}^{\te{Id}}(w) = p^{-1/2} \ast w$ that parallel transports a tangent vector $w \in T_p(\mathcal{M})$ to $T_{\te{Id}}(\mathcal{M})$ along a geodesic curve in the Riemannian manifold $(\mathcal{M}, g_R)$, similar to e.g., \cite{Y12}. This allows for straightforward comparison of coefficients across scales and locations in Section \ref{sec:3} and \ref{sec:4}. Note in particular that for the whitened coefficients $\Vert \mathfrak{D}_{j,k_1,k_2} \Vert_F^2 = \Vert D_{j,k_1,k_2} \Vert_{\widetilde{M}_{j,k_1,k_2}}^2$. Figure \ref{fig:4} gives a visual description of the construction of the $j$-scale wavelet coefficients in a dyadic framework, where the coarse-scale refinement rectangle $I_{j-1, k_1,k_2}$ corresponds to the union of the equally-sized finer-scale rectangles $I_{j,2k_1+\ell_1,2k_2 + \ell_2}$ with $\ell_1, \ell_2 \in \{ 0, 1\}$.
\end{enumerate}
\paragraph{Inverse wavelet transform} The inverse wavelet transform passes from coarse $(j-1)$-scale midpoints plus $j$-scale wavelet coefficients to finer $j$-scale midpoints and follows directly by reverting the above operations:
\begin{enumerate}
\item \textbf{Predict/Refine}: given the $(j-1)$-scale midpoints $(M_{j-1,k_1',k_2'})_{
k_1',k_2'}$ corresponding to the refinement rectangles $(I_{j-1,k_1',k_2'})_{k_1',k_2'}$ and an average-interpolation order $(N_1,N_2) \geq (1,1)$, generate the predicted $j$-scale midpoints $(\widetilde{M}_{j,k_1,k_2})_{k_1',k_2'}$ via the 2D AI subdivision scheme in Section \ref{sec:2.2}. 
\item \textbf{Complete}: recover the $j$-scale midpoint at the location $(k_1,k_2)$ from the predicted midpoint $(\widetilde{M}_{j,k_1,k_2})_{k_1,k_2}$ and the wavelet coefficients $(D_{j,k_1,k_2})_{k_1,k_2}$ through:
\begin{eqnarray*}
M_{j,k_1,k_2} &=& \Exp_{\widetilde{M}_{j,k_1,k_2}}\left(\sqrt{\frac{\lambda_2(\mathcal{I})}{\lambda_2(I_{j,k_1,k_2})}} \cdot D_{j,k_1,k_2}\right).
\end{eqnarray*} 
\end{enumerate}
Given the coarsest midpoint $M_{0,0,0}$ at scale $j=0$, the pyramid of refinement rectangles $(I_{j,k_1,k_2})_{j,k_1,k_2}$ and the pyramid of wavelet coefficients $(D_{j,k_1,k_2})_{j,k_1,k_2}$ at scales $j=1,\ldots,J$, repeating the reconstruction procedure above up to scale $J$, we recover the $(n_1 \times n_2)$-dimensional discretized grid of local averages $(M_{J,k_1,k_2})_{k_1,k_2} \in \mathcal{M}$ given as input to the forward wavelet transform.  
\section{Wavelet thresholding for smooth HPD surfaces} \label{sec:3}
In this section, we derive the wavelet coefficient decay and linear wavelet thresholding convergence rates of intrinsically smooth surfaces of HPD matrices $\gamma: \mathcal{I} \to \mathcal{M}$, with existing partial covariant derivatives of degree $(N_1,N_2) \geq (1,1)$ or higher. 
Without loss of generality, assume that $\mathcal{I} = [0,1] \times [0,1]$. Suppose that we observe an $(n_1 \times n_2)$-dimensional discretized grid of random independent local averages  $X_{k_1,k_2} \in \mathcal{M}$ across equally-sized, non-overlapping, closed rectangles $(I_{J,k_1,k_2})_{k_1,k_2}$ with $0 \leq k_1 \leq n_1-1$ and $0 \leq k_2 \leq n_2-1$, such that $\bigcup_{k_1,k_2} I_{J,k_1,k_2} = \mathcal{I}$. For the random variables $X_{k_1,k_2}$, we assume that $X_{k_1,k_2} \sim \nu_{k_1,k_2}$, with intrinsic mean $\mathbb{E}_{\nu_{k_1,k_2}}[X] = \te{Ave}_{I_{J,k_1,k_2}}(\gamma)$ and $\nu_{k_1,k_2} \in P_2(\mathcal{M})$ for each $(k_1,k_2)$, using the same notation as in Section \ref{sec:2.1}. The derivations of the coefficient decay and convergence rates below inherently depend on the sizes of the refinement rectangles $(I_{j,k_1,k_2})_{k_1,k_2}$. For this reason, in this section, we assume that $n_1 = 2^{J_1}$ and $n_2 = 2^{J_2}$ are dyadic and the refinement pyramid is given by the natural refinement rectangles $(I_{j,k_1,k_2})_{k_1,k_2}$ as in Section \ref{sec:2.2} for $j = 0,\ldots,J$, with $J = \max(J_1,J_2)$. In particular, within each scale $j = 0,\ldots,J$, the sizes of the refinement rectangles are constant across locations and given by:
\begin{eqnarray} \label{eq:3.1}
\lambda_2(I_{j,k_1,k_2}) &=& (2^{-j+J-J_1} \wedge 1)(2^{-j+J-J_2}\wedge 1), \quad \te{for all } k_1,k_2
\end{eqnarray}
This assumption can be relaxed to rectangles with sizes of the same order $\lambda_2(I_{j,k_1,k_2}) = O((2^{-j+J-J_1} \wedge 1)(2^{-j+J-J_2}\wedge 1))$ at each scale $j$, (and $n_1, n_2$ not necessarily dyadic), but for the sake of precise arguments, we focus on the exact natural dyadic refinement pyramid given in Section \ref{sec:2.2}.
\paragraph{Empirical wavelet coefficient error}
The first proposition below gives the estimation error of the wavelet coefficients based on the empirical finest-scale midpoints $M_{J,k_1,k_2,n} = X_{k_1,k_2}$ with respect to the wavelet coefficients obtained from the finest-scale midpoints $M_{J,k_1,k_2} = \te{Ave}_{I_{J,k_1,k_2}}(\gamma)$ of the target surface $\gamma$.
\begin{prop} \label{prop:3.1} (Estimation error) Let $(M_{J,k_1,k_2})_{k_1,k_2}$ and $(X_{k_1,k_2})_{k_1,k_2}$ be as defined above, such that $(k_1,k_2) \in \{0,\ldots,n_1-1\} \times \{0,\ldots,n_2-1\}$, with $n_1 = 2^{J_1}$ and $n_2 = 2^{J_2}$ for $J_1,J_2$ sufficiently large. Consider $(I_{j,k_1,k_2})_{j,k_1,k_2}$ to be the natural dyadic refinement pyramid on $\mathcal{I} = [0,1] \times [0,1]$ for $j=0,\ldots,J$, with $J = \max(J_1,J_2)$. Then, for each scale $j > 0$ sufficiently small and location $(k_1,k_2)$, 
\begin{eqnarray*}
\bs{E}\Vert \widehat{\mathfrak{D}}_{j,k_1,k_2,n} - \mathfrak{D}_{j,k_1,k_2} \Vert_F^2 & \lesssim & (n_1n_2)^{-1},
\end{eqnarray*}
where, 
\begin{eqnarray*}
\widehat{\mathfrak{D}}_{j,k_1,k_2,n} &=&  \sqrt{\frac{\lambda_2(I_{j,k_1,k_2})}{\lambda_2(\mathcal{I})}} \, \Log\big(\widetilde{M}_{j,k_1,k_2,n}^{-1/2} \ast M_{j,k_1,k_2,n}\big)
\end{eqnarray*}
is the empirical whitened wavelet coefficient at scale-location $(j,k_1,k_2)$ based on some subdivision order $(N_1,N_2) \geq (1,1)$. Here, $M_{j,k_1,k_2,n}$ is the empirical midpoint at scale-location $(j, k_1,k_2)$ constructed from the observations $(X_{k_1,k_2})_{k_1,k_2}$, and $\widetilde{M}_{j,k_1,k_2,n}$ is the predicted midpoint based on the estimated midpoints $(M_{j-1,k_1',k_2',n})_{k_1',k_2'}$. Similarly, $\mathfrak{D}_{j,k_1,k_2}$ is the whitened wavelet coefficient at scale-location $(j,k_1,k_2)$ based on the finest-scale midpoints $M_{J,k_1,k_2} = \te{Ave}_{I_{J,k_1,k_2}}(\gamma)$ of the target surface $\gamma$ subject to the same subdivision order $(N_1,N_2)$.
\end{prop}
\paragraph{Wavelet coefficient decay} 
In order to derive the wavelet coefficient decay of intrinsically smooth surfaces, we rely on the \emph{intrinsic polynomial reproduction} property, which ensures that the subdivision scheme reproduces intrinsic polynomial surfaces without error. This is a generalization of \cite[Proposition 4.2]{CvS17} from intrinsically smooth (1D) curves to intrinsically smooth (2D) surfaces in the Riemannian manifold. 
\begin{prop} \label{prop:3.2} (Coefficient decay) Given a subdivision order $(N_1,N_2) \geq (1,1)$, suppose that $\gamma: \mathcal{I} \to \mathcal{M}$ is a smooth surface with existing partial covariant derivatives of order $(N_1,N_2)$ or higher. Let $(I_{j,k_1,k_2})_{j,k_1,k_2}$ be the natural dyadic refinement pyramid on the domain $\mathcal{I} = [0,1] \times [0,1]$ for $j = 0,\ldots,J$, with $J = \max(J_1,J_2)$ and $J_1,J_2$ sufficiently large. Then, for each scale $j > |J_1-J_2|$ sufficiently large and location $(k_1,k_2)$,
\begin{eqnarray*}
\Vert \mathfrak{D}_{j,k_1,k_2} \Vert_F & \lesssim & 2^{-j + |J_2 - J_1|/2}\big( 2^{(-j + J - J_1)N_1} \vee 2^{(-j + J-J_2)N_2}\big),
\end{eqnarray*}
where $\mathfrak{D}_{j,k_1,k_2}$ denotes the whitened wavelet coefficient at scale-location $(j,k_1,k_2)$ based on the finest-scale midpoints $M_{J,k_1,k_2} = \te{Ave}_{I_{J,k_1,k_2}}(\gamma)$, with subdivision order $(N_1,N_2)$, similar to Proposition \ref{prop:3.1} above.
\end{prop}
\begin{remark}
The parameters $J_1$ and $J_2$ determine the sizes of the refinement rectangles in eq.(\ref{eq:3.1}), i.e., as $J_1,J_2 \to \infty$, $\lambda_2(I_{J,k_1,k_2}) \to 0$. Note that if $J = J_1 = J_2$, which implies that the refinement pyramid $(I_{j,k_1,k_2})_{j,k_1,k_2}$ consists of equally-sized squares at each scale $j$, then the coefficient decay rate simplifies to $\Vert \mathfrak{D}_{j,k_1,k_2} \Vert_F \lesssim 2^{-j} 2^{-j(N_1 \wedge N_2)}$.
\end{remark}
\paragraph{Linear wavelet thresholding}
Combining Propositions \ref{prop:3.1} and \ref{prop:3.2}, the main theorem below gives the integrated mean squared error in terms of the Riemannian distance $\delta_R$ of a linear wavelet thresholded estimator of a smooth surface $\gamma(t,\nu)$. The wavelet estimator is based on the input sample of local average observations $(X_{k_1,k_2})_{k_1,k_2}$ associated to the refinement rectangles $(I_{J,k_1,k_2})_{k_1,k_2}$. In the theorem below, as before, it is assumed that the  random observations $X_{k_1,k_2}$ are sampled on a dyadic grid of dimension $(n_1 \times n_2)$, with $n_1 = 2^{J_1}$ and $n_2 = 2^{J_2}$, with $n = 2^{J_1+J_2}$ the total number of observations, and the refinement rectangles $(I_{j,k_1,k_2})_{j,k_1,k_2}$ are the rectangles associated to the natural dyadic refinement pyramid.
\begin{theorem} \label{thm:3.3} (Convergence rates linear thresholding) Given a subdivision order $(N_1,N_2) \geq (1,1)$, suppose that $\gamma : \mathcal{I} \to \mathcal{M}$ is a smooth surface with existing partial covariant derivatives of order $(N_1,N_2)$ or higher. Let $(X_{k_1,k_2})_{k_1,k_2}$ be as defined above, with $(k_1,k_2) \in \{0,\ldots, n_1-1\} \times \{0,\ldots,n_2-1\}$, such that $(n_1,n_2) = (2^{J_1}, 2^{J_2})$ and $n = n_1n_2$, and let $(I_{j,k_1,k_2})_{j,k_1,k_2}$ be the natural dyadic refinement pyramid for $j=0,\ldots,J$, with $J = \max(J_1,J_2)$. Consider a linear wavelet estimator based on the observations $(X_{k_1,k_2})_{k_1,k_2}$ that thresholds all wavelet coefficients at scales $j \geq J_0$, such that $J_0 = (\log_2(n)+|J_1-J_2|(1+2(N_1\vee N_2))/(2 + 2(N_1 \wedge N_2))$, with $(N_1,N_2)$ the average-interpolation order of the wavelet transform. For $n$ sufficiently large, 
\begin{eqnarray} \label{eq:3.2}
\sum_{j,k_1,k_2} \bs{E} \Vert \widehat{\mathfrak{D}}_{j,k_1,k_2} - \mathfrak{D}_{j,k_1,k_2} \Vert_F^2 & \lesssim & 2^{\frac{2(N_1\vee N_2)}{1+(N_1\wedge N_2)}|J_1-J_2|}\,(n_1 \vee n_2)^{-\frac{2(N_1 \wedge N_2)}{1 + (N_1 \wedge N_2)}},
\end{eqnarray}
with the sum ranging over all locations at scales $0 < j \leq J$. Here, $\widehat{\mathfrak{D}}_{j,k_1,k_2}$ is the empirical whitened wavelet coefficient based on the observations $(X_{k_1,k_2})_{k_1,k_2}$ after linear thresholding of wavelet scales $j \geq J_0$, and $\mathfrak{D}_{j,k_1,k_2}$ is the whitened wavelet coefficient based on the grid of finest-scale midpoints $M_{J,k_1,k_2} = \te{Ave}_{I_{J,k_1,k_2}}(\gamma)$ of the smooth surface $\gamma$. Moreover, denote by $(\widehat{M}_{J,k_1,k_2,n})_{k_1,k_2}$ the grid of finest-scale midpoints based on the linear wavelet thresholded estimator. Then, for $n$ sufficiently large, also,
\begin{eqnarray} \label{eq:3.3}
\frac{1}{n}\sum_{k_1,k_2} \bs{E}\left[\delta_R(\widehat{M}_{J,k_1,k_2}, M_{J,k_1,k_2,n})^2\right] & \lesssim & 2^{\frac{2(N_1\vee N_2)}{1+(N_1\wedge N_2)}|J_1-J_2|}\,(n_1 \vee n_2)^{-\frac{2(N_1 \wedge N_2)}{1 + (N_1 \wedge N_2)}}.
\end{eqnarray}
\end{theorem} 
\noindent Examining the convergence rates in the theorem above, we may obtain simplified rates under several specific scenarios:
\begin{enumerate}
\item[(i)] If $N := N_1 = N_2$, the product of the two powers can be combined into a single term as,
\begin{eqnarray*}
2^{\frac{2(N_1\vee N_2)}{1+(N_1\wedge N_2)}|J_1-J_2|}\,(n_1 \vee n_2)^{-\frac{2(N_1 \wedge N_2)}{1 + (N_1 \wedge N_2)}} \ = \ \left( 2^{(J_1 \vee J_2) - |J_1 - J_2|} \right)^{-\frac{2N}{1 + N}} \ = \ 
(n_1 \wedge n_2)^{-\frac{2N}{1 + N}}.
\end{eqnarray*}
\item[(ii)] If $n_1 = O(n_2)$, or in other words $|J_1 - J_2| = O(1)$, the first power reduces to a constant, and we can bound, 
\begin{eqnarray*}
2^{\frac{2(N_1\vee N_2)}{1+(N_1\wedge N_2)}|J_1-J_2|}\,(n_1 \vee n_2)^{-\frac{2(N_1 \wedge N_2)}{1 + (N_1 \wedge N_2)}} & \lesssim & n_1^{-\frac{2(N_1 \wedge N_2)}{1 + (N_1 \wedge N_2)}}.
\end{eqnarray*}
Such a situation arises when the shape of the rectangular observation grid remains constant as $n$ increases, i.e., the ratio $n_1/n_2$ is fixed.
\end{enumerate}

\section{Nonlinear wavelet thresholding for HPD surfaces} \label{sec:4}
In this section, we study nonlinear wavelet denoising for surfaces of HPD matrices corrupted by noise, where the target signal is not necessarily an intrinsically smooth surface $\gamma: \mathcal{I} \to \mathcal{M}$, and may be subject to e.g., varying degrees of local smoothness, or local spikes or jump discontinuities. In such cases, more flexible nonlinear thresholding of wavelet coefficients outperforms linear thresholding of wavelet scales, as the nonlinear wavelet thresholded estimator is able to adapt to different degrees of local smoothness in the signal. Our main focus is on nonlinear wavelet thresholding in the context of generalized intrinsic signal \emph{plus} i.i.d.\@ noise models in the Riemannian manifold $(\mathcal{M}, g_R)$ as described in \cite[Chapter 2]{C18}. As in Section \ref{sec:3}, without loss of generality assume that $\mathcal{I} = [0,1] \times [0,1]$ and suppose that we observe an $(n_1 \times n_2)$-dimensional discretized grid of random independent local averages  $X_{k_1,k_2}$ taking values in the space of HPD matrices across equally-sized, non-overlapping, closed rectangles $(I_{J,k_1,k_2})_{k_1,k_2}$ with $0 \leq k_1 \leq n_1-1$ and $0 \leq k_2 \leq n_2-1$, such that $\bigcup_{k_1,k_2} I_{J,k_1,k_2} = \mathcal{I}$. In this section, the random variates $\{X_{k_1,k_2} \}_{k_1,k_2}$ are assumed to follow an intrinsic discretized signal plus i.i.d.\@ noise model with respect to the affine-invariant metric according to:
\begin{eqnarray} \label{eq:4.1}
X_{k_1,k_2} &=& \gamma_{k_1,k_2}^{1/2} \ast \epsilon_{k_1,k_2}, \quad 0 \leq k_1 \leq n_1-1, \quad 0 \leq k_2 \leq n_2 - 1.
\end{eqnarray}
Here, $(\epsilon_{k_1,k_2})_{k_1,k_2} \overset{\te{iid}}{\sim} \nu$, with $\nu \in P_2(\mathcal{M})$ having intrinsic mean equal to the identity, i.e., $\mathbb{E}_{\nu}[\epsilon] = \te{Id}$, and $\gamma_{k_1,k_2} := \te{Ave}_{I_{J,k_1,k_2}}(\gamma)$ corresponds to the intrinsic mean of the square integrable surface $\gamma$ over the refinement rectangle $I_{J,k_1,k_2}$. In order to derive the results in Proposition \ref{prop:4.1}, we assume again that $n_1 = 2^{J_1}$ and $n_2 = 2^{J_2}$ are dyadic and the refinement pyramid is given by the natural dyadic refinement rectangles $(I_{j,k_1,k_2})_{k_1,k_2}$ for $j = 0,\ldots,J$, with $J = \max(J_1,J_2)$, as in Section \ref{sec:2.2}.
\begin{example} \emph{(Time-varying spectral matrix estimation).} \label{ex:4.1}
As an illustration, consider estimating the time-varying spectral density matrix $f(\omega, u) \in \mathcal{M}$ of a locally stationary Gaussian time series of length $T$. Suppose that we have computed time-varying (e.g., localized or segmented) periodograms $I_T(\omega_{k_1}, u_{k_2})$ on an $(n_{1,T} \times n_{2,T})$-dimensional rectangular grid at equidistant time-frequency points $(\omega_{k_1}, u_{k_2})_{k_1,k_2} \in [0,\pi] \times [0,1]$. If we let the size of the time-frequency grid grow at a sufficiently slower rate than the length of the time series $T$, the time-varying periodograms $I_T(\omega_{k_1}, u_{k_2})$ are asymptotically independent between time-frequency points as $T \to \infty$ and asymptotically complex Wishart distributed according to e.g., \cite{B81}. Let $(X_{k_1,k_2})_{k_1,k_2} \in \mathcal{M}$ be noisy multitapered HPD periodograms at time-frequencies $(\omega_{k_1}, u_{k_2})_{k_1,k_2}$ with $L \geq d$ tapers (see e.g., \cite{W00}), where $d$ is the dimension of the time series. The asymptotic distribution of the \emph{bias-corrected} HPD periodograms is of the form $e^{-c(d,L)}W_d^C(L, L^{-1} f(\omega_{k_1},u_{k_2}))$, where the factor $e^{-c(d,L)}$ corresponds to the Wishart bias-correction in \cite[Theorem 5.1]{CvS17}, so that its intrinsic mean with respect to the affine-invariant metric is equal to $f(\omega_{k_1}, u_{k_2})$. We observe that the noisy HPD periodogram observations $(X_{k_1,k_2})_{k_1,k_2}$ asymptotically follow an intrinsic signal plus i.i.d.\@ noise model, since as $T \to \infty$, for each $(k_1,k_2)$,
\begin{eqnarray*}
X_{k_1,k_2} & \overset{d}{\to} & f_{k_1,k_2}^{1/2} \ast \epsilon_{k_1,k_2}, \quad \quad \epsilon_{k_1,k_2} \overset{\te{iid}}{\sim} \nu, 
\end{eqnarray*}
with target signal $f_{k_1,k_2} = f(\omega_{k_1}, u_{k_2})$ and noise distribution $\nu = e^{-c(d,L)}W_d^C(L, L^{-1}\te{Id}) \in P_2(\mathcal{M})$, such that $\mathbb{E}_\nu[\epsilon] = \te{Id}$. In order to make the precise correspondence with the model in eq.(\ref{eq:4.1}), we can take equally-sized, non-overlapping, closed rectangles $(I_{J,k_1,k_2})_{k_1,k_2}$, such that each rectangle contains a single time-frequency point $(\omega_{k_1},u_{k_2})$ and $\bigcup_{k_1,k_2} I_{J,k_1,k_2} = [0,\pi] \times [0,1]$. As we are only interested in estimating the spectrum $f(\omega, u)$ at the discretized time-frequency points, we may set $f_{k_1,k_2} = \te{Ave}_{I_{J,k_1,k_2}}(f) = f(\omega_{k_1}, u_{k_2})$.
\end{example}
\subsection{Trace thresholding of coefficients}\label{sec:4.1}
Nonlinear wavelet-denoised surfaces in the space of HPD matrices can be constructed by shrinkage or thresholding of individual components of the matrix-valued wavelet coefficients, or by simultaneous shrinkage or thresholding of entire wavelet coefficients. Analogous to \cite{CvS17}, in this section we focus on hard thresholding of entire wavelet coefficients based on the trace of the whitened wavelet coefficients because of its simplicity and appealing properties, but we emphasize that more flexible componentwise shrinkage or thresholding of the wavelet coefficients may also be appropriate. For sampled observations from an intrinsic discretized signal plus i.i.d.\@ noise model as in eq.(\ref{eq:4.1}), the trace of the noisy whitened coefficients decomposes into an additive signal plus mean-zero noise sequence model, and we can derive analytic expressions for the variance of the trace of the noisy whitened coefficients across wavelet scales. Moreover, since the trace operator outputs a scalar, we can directly apply ordinary scalar thresholding or shrinkage to the matrix-valued coefficients. Thresholding or shrinkage of the trace of the whitened coefficients is \emph{general linear} congruence equivariant. In the context of spectral estimation of multivariate time series, this means that the estimator does not nontrivially depend on the chosen basis or coordinate system of the time series, as the spectral estimator is equivariant under a change of basis of the time series. All these properties are generalized versions of the properties outlined in \cite{CvS17}, extended from the setting of 1D curves of HPD matrices to 2D surfaces of HPD matrices.
\begin{lemma} (General linear congruence equivariance) \label{lem:4.1}
Let $(X_{k_1,k_2})_{k_1,k_2}$ be a surface of HPD matrices and $(\hat{f}_{k_1,k_2})_{k_1,k_2}$ its wavelet-denoised estimate based on linear or nonlinear shrinkage of the trace of the whitened wavelet coefficients. The estimator is equivariant under general linear congruence transformation of the observations in the sense that the wavelet-denoised estimate $(\hat{f}_{a, k_1,k_2})_{k_1,k_2}$ of $(a \ast X_{k_1,k_2})_{k_1,k_2}$ equals $(a \ast \hat{f}_{k_1,k_2})_{k_1,k_2}$ for each $a \in \te{GL}(d, \mathbb{C})$.
\end{lemma}
\noindent The proof is a straightforward extension of the proofs of Proposition 5.2 and Lemma 5.3 in \cite{CvS17} and is therefore omitted here. In general, using again the notation $\mathcal{H}$ for the space of Hermitian matrices, if we write $\lambda: \mathcal{H} \to \mathcal{H}$ for the thresholding or shrinkage operator, with $\lambda(D) \in \mathcal{H}$ the thresholded or shrunken version of the wavelet coefficient $D \in \mathcal{H}$. Then, if $\lambda(\cdot)$ is general linear congruence equivariant, in the sense that, $\lambda(a \ast D) = a \ast \lambda(D)$ for each $a \in \te{GL}(d, \mathbb{C})$, the nonlinear wavelet estimator is also general linear congruence equivariant. If $\lambda(\cdot)$ is only unitary congruence equivariant, i.e., $\lambda(a \ast D) = a \ast \lambda(D)$ only for $a \in \mathcal{U}$, where $\mathcal{U}$ denotes the set of unitary matrices, then the nonlinear wavelet estimator is also only unitary congruence equivariant.\\[3mm]
In the proposition below, we derive several useful trace properties of the whitened wavelet coefficients analogous to \cite[Proposition 5.3]{CvS17}. In the context of an intrinsic signal plus noise model, at scale-location $(j,k_1',k_2')$: $\mathfrak{D}_{j,k_1',k_2'}^X$ denotes the random whitened coefficient based on the noisy sample observations with finest-scale midpoints $(X_{k_1,k_2})_{k_1,k_2}$ associated to the refinement rectangle $(I_{J,k_1,k_2})_{k_1,k_2}$; $\mathfrak{D}_{j,k_1',k_2'}^{\gamma}$ denotes the deterministic whitened coefficient based on the target surface $\gamma$ with finest-scale midpoints $(\gamma_{k_1,k_2})_{k_1,k_2}$; and $\mathfrak{D}_{j,k_1',k_2'}^\epsilon$ denotes the random whitened coefficient based on the i.i.d.\@ noise terms with finest-scale midpoints $(\epsilon_{k_1,k_2})_{k_1,k_2}$.
\begin{prop} \label{prop:4.1} (Trace properties)
Let $(X_{k_1,k_2})_{k_1,k_2} \in \mathcal{M}$, with $(k_1,k_2) \in \{0,\ldots,n_1-1\} \times \{0,\ldots,n_2-1\}$ be sampled from an intrinsic discretized signal plus i.i.d.\@ noise model according to:
\begin{eqnarray*}
X_{k_1,k_2} &=& \gamma_{k_1,k_2}^{1/2} \ast \epsilon_{k_1,k_2}, \quad 0 \leq k_1 \leq n_1-1, \quad 0 \leq k_2 \leq n_2-1,
\end{eqnarray*}
with $(\epsilon_{k_1,k_2})_{k_1,k_2} \overset{\te{iid}}{\sim} \nu$, such that $\nu \in P_2(\mathcal{M})$ and $\mathbb{E}_{\nu}[\epsilon] = \te{Id}$. Assume that $n_1 = 2^{J_1}$, $n_2 = 2^{J_2}$ with $n = n_1n_2$, then for each scale-location $(j,k_1,k_2)$, the whitened wavelet coefficients obtained from the intrinsic 2D AI wavelet transform of order $(N_1,N_2) \geq (1,1)$, based on the natural dyadic refinement pyramid $(I_{j,k_1,k_2})_{j,k_1,k_2}$, satisfy:
\begin{eqnarray*}
\tr(\mathfrak{D}_{j,k_1,k_2}^X) &=& \tr(\mathfrak{D}_{j,k_1,k_2}^{\gamma}) + \tr(\mathfrak{D}_{j,k_1,k_2}^\epsilon).
\end{eqnarray*}
Moreover, $\bs{E}[\tr(\mathfrak{D}_{j,k_1,k_2}^X)] \ = \ \tr(\mathfrak{D}_{j,k_1,k_2}^{\gamma})$, and,
\begin{eqnarray} \label{eq:4.2}
\var(\tr(\mathfrak{D}_{j,k_1,k_2}^X)) &=& 
\left\{ 
\begin{array}{ll}
\frac{1}{2n} \Big( \sum_{\vec{\ell} \in \Lambda_{k_1,k_2}} \big(C_{\vec{\ell}}^{N_1,N_2,k_1,k_2} \big)^2 \Big) \var(\tr(\zeta)), & \te{if } j < |J_1-J_2|, \\
\frac{1}{2n} \Big( 1 + \frac{1}{2}\sum_{\vec{\ell} \in \Lambda_{k_1,k_2}} \big(C_{\vec{\ell}}^{N_1,N_2,k_1,k_2} \big)^2 \Big) \var(\tr(\zeta)), & \te{if } j \geq |J_1 - J_2|.
\end{array}
\right.\quad \quad \quad
\end{eqnarray}
Here, $\Lambda_{k_1,k_2}$ denotes the set of locations $\vec{\ell}=(\ell_1,\ell_2)$ of the neighboring $(j-1)$-scale midpoints needed to predict the $j$-scale midpoint at scale-location $(j,k_1,k_2)$, with $(C_{\vec{\ell}}^{N_1,N_2,k_1,k_2})_{\vec{\ell}}$ the corresponding prediction weights as explained in Section \ref{sec:2.2}. Also, $\zeta \overset{d}{=} \Log(\epsilon)$, where $\epsilon \sim \nu$. 
\end{prop}
\noindent An important observation is that, for locations away from the boundary, i.e., with constant prediction weights across scales, the variance of the trace of the whitened coefficients $\var(\tr(\mathfrak{D}_{j,k_1,k_2}^X))$ is constant across wavelet scales $j < |J_1-J_2|$ and across wavelet scales $j \geq |J_1 - J_2|$. In particular; at scales $j < |J_1 - J_2|$, the prediction weights $(C_{\vec{\ell}}^{N_1,N_2,k_1,k_2})_{\vec{\ell}}$ simplify to the filter coefficients in the 1D AI subdivision scheme as in \cite{CvS17}; and at scales $j \geq |J_1-J_2|$, the prediction weights are as outlined in Section \ref{sec:2.2}. If the sample grid is square, i.e., $J_1 = J_2$, then for locations away from the boundary, the variance $\var(\tr(\mathfrak{D}_{j,k_1,k_2}^X))$ is constant across all wavelet scales, and we can directly apply any preferred classical nonlinear thresholding or shrinkage procedure well-suited to additive signal plus noise models for scalar surfaces, with homogeneous variances across wavelet scales. If the sample grid is rectangular, i.e., $J_1 \neq J_2$, there are several ways to homogenize the variances $\var(\tr(\mathfrak{D}_{j,k_1,k_2}^X))$ across wavelet scales:
\begin{enumerate}
\item[(i)] \emph{(Parametric)}. If the variance $\var(\tr(\zeta))$ is known or given (e.g., asymptotically), and we have access to the filter weights (1D and 2D) of the subdivision scheme with given order $(N_1,N_2)$, we can normalize the variances $\var(\tr(\mathfrak{D}_{j,k_1,k_2}^X))$ across wavelet scales to unit variances directly via the analytic expressions in eq.(\ref{eq:4.2}). 
\item[(ii)] \emph{(Nonparametric)}. If the variance $\var(\tr(\zeta))$ is unknown, but we have access to the filter weights (1D and 2D) of the subdivision scheme with given order $(N_1,N_2)$, we can robustly estimate $\sigma_{\epsilon,\te{2D}}^2 = \var(\tr(\mathfrak{D}_{j,k_1,k_2}^X))$ for $j \geq |J_1 - J_2|$ from the finest wavelet scale through $\hat{\sigma}_{\epsilon,\te{2D}}^2 = \linebreak[1] \te{MAD}\{ (\tr(\mathfrak{D}_{J,k_1,k_2}^X))_{k_1,k_2}\}^2$, as the finest wavelet scale contains primarily noise for sufficiently large samples. The variance  $\sigma_{\epsilon,\te{1D}}^2 = \var(\tr(\mathfrak{D}_{j,k_1,k_2}^X))$ for $j < |J_1 - J_2|$ is then estimated from $\hat{\sigma}_{\epsilon,2D}^2$ using the analytic expression in eq.(\ref{eq:4.2}) and the 1D- and 2D-filter weights.
\item[(iii)] \emph{(Semiparametric)}. If the variance $\var(\tr(\zeta))$ is known or given (e.g., asymptotically), for $j < |J_1-J_2|$, we can compute the variances $\sigma_{\epsilon,\te{1D}}^2 = \var(\tr(\mathfrak{D}_{j,k_1,k_2}^X))$ from the 1D-filter weights and the analytic expression in eq.(\ref{eq:4.2}). For $j \geq |J_1 - J_2|$, we robustly estimate the variances from the finest wavelet scale through $\hat{\sigma}_{\epsilon,2D}^2 = \te{MAD}\{ (\tr(\mathfrak{D}_{j,k_1,k_2}^X))_{k_1,k_2}\}^2$. This procedure does not require the 2D-filter weights.
\end{enumerate}
\begin{example} \emph{(Time-varying spectral matrix estimation continued).} \label{ex:4.2}
Consider again the application of nonlinear wavelet thresholding to time-varying spectral matrix estimation as in Example \ref{ex:4.1} above. For a random variable $\epsilon \sim \nu$, with $\nu = e^{-c(d,B)}W_d^C(B, B^{-1}\te{Id})$, it is shown in the proof of \cite[Proposition 5.3]{CvS17} that $\var(\tr(\zeta)) = \var(\tr(\Log(\epsilon))) = \sum_{i=1}^d \psi'(B - (d - i))$, with $\psi'(\cdot)$ the trigamma function. In the R-package \texttt{pdSpecEst}, the function \texttt{pdSpecEst2D()} performs time-varying spectral matrix estimation via nonlinear trace thresholding of wavelet coefficients of noisy HPD local periodogram observations on a rectangular dyadic time-frequency grid. The variances $\var(\tr(\mathfrak{D}_{j,k_1,k_2}^X))$ are homogenized across wavelet scales via the semiparametric method (iii) above, using the asymptotic variance $\var(\tr(\zeta)) = \var(\tr(\Log(\epsilon))) = \sum_{i=1}^d \psi'(B - (d - i))$ and the 1D-filter weights given in \cite{CvS17} for refinement orders $(N_1,N_2)$, with $N_1 \vee N_2 \leq 9$.
\end{example}
\begin{example} \emph{(Gaussian noise models).}
The space of $(d \times d)$-Hermitian matrices $\mathcal{H}$ is a real vector space. Denote $\bs{1}_{i,j}$ for the $(d \times d)$-matrix of zeros with a one at location $(i,j)$, then an orthonormal basis $(H_{i,j})_{i,j}$ of $(\mathcal{H}, \langle \cdot, \cdot \rangle_F)$, is given by the following collection of $d^2$ matrices: (i) $H_{i,j} = \bs{1}_{ii}$ for $i = j$, with $1 \leq i \leq d$, (ii) $H_{i,j} = 2i/\sqrt{2} \bs{1}_{i,j}$ for $1 \leq i < j \leq d$, (iii) $H_{i,j} = 2/\sqrt{2} \bs{1}_{i,j}$ for $1 \leq j < i \leq d$. Suppose that the observations $(X_{\ell_1,\ell_2})_{\ell_1,\ell_2}$ are obtained from an intrinsic signal plus i.i.d.\@ \emph{Gaussian} noise model according to eq.(\ref{eq:4.1}), such that $\zeta_{\ell_1,\ell_2} = \Log(\epsilon_{\ell_1,\ell_2}) \overset{d}{=} \sum_{i,j} z_{\ell_1,\ell_2}^{i,j} H_{i,j}$, with $\te{vec}((z_{\ell_1,\ell_2}^{i,j})_{i,j}) \sim N(\bs{0}, \Sigma)$. Here, $\te{vec}(Z)$ denotes the vectorization of a matrix $Z$. From the proof of Proposition \ref{prop:4.1}, for each scale-location $(j, k_1,k_2)$, it follows that $\tr(\mathfrak{D}^X_{j,k_1,k_2})$ is a weighted linear combination of Gaussian random variables $(\tr(\zeta_{\ell_1,\ell_2}))_{\ell_1,\ell_2}$, and we therefore obtain a Gaussian sequence model in the wavelet domain:
\begin{eqnarray*}
\tr(\mathfrak{D}^X_{j,k_1,k_2}) &=& \tr(\mathfrak{D}^{\gamma}_{j,k_1,k_2}) + \tr(\mathfrak{D}^\epsilon_{j,k_1,k_2}), \quad \te{with } \tr(\mathfrak{D}^\epsilon_{j,k_1,k_2}) \sim N(0, \sigma_e^2).
\end{eqnarray*}
Here, $\sigma_e^2 = \var(\tr(\mathfrak{D}^X_{j,k_1,k_2}))$ according to eq.(\ref{eq:4.2}), with  $\var(\tr(\zeta)) = \sum_{i,j=1}^d \cov(z_{i,i},z_{j,j})$, where $(z_{i,i})_{i,i}$ are the diagonal components of a random Hermitian matrix distributed as $\te{vec}((z_{i,j})_{i,j}) \sim N(\bs{0}, \Sigma)$.
\end{example}

\section{Illustrative data examples} \label{sec:5}
\subsection{Finite-sample performance} \label{sec:5.1}

\begin{figure}[t]
\includegraphics[scale=0.23]{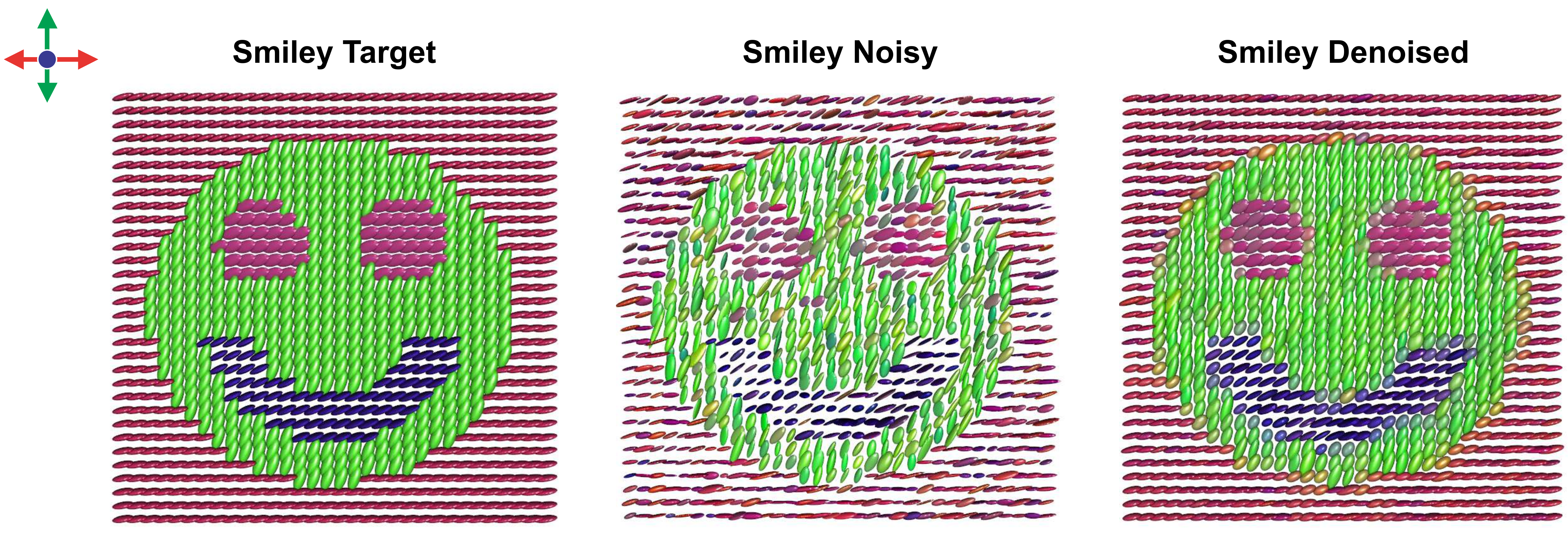}
\caption{Left: target piecewise constant \texttt{smiley} surface on a $(64 \times 64)$ dyadic grid. Middle: noisy HPD observations from \texttt{smiley} surface according to a Riemannian signal plus i.i.d.\@ intrinsic normal noise model. Right: intrinsic tree-structure wavelet denoised HPD surface with \texttt{pdSpecEst2D()}, (based on Riemannian metric $g_R$, with order $(3,3)$ and oracle penalty $\lambda$).\label{fig:6}}
\end{figure}

\begin{figure}[ht]
\includegraphics[scale=0.23]{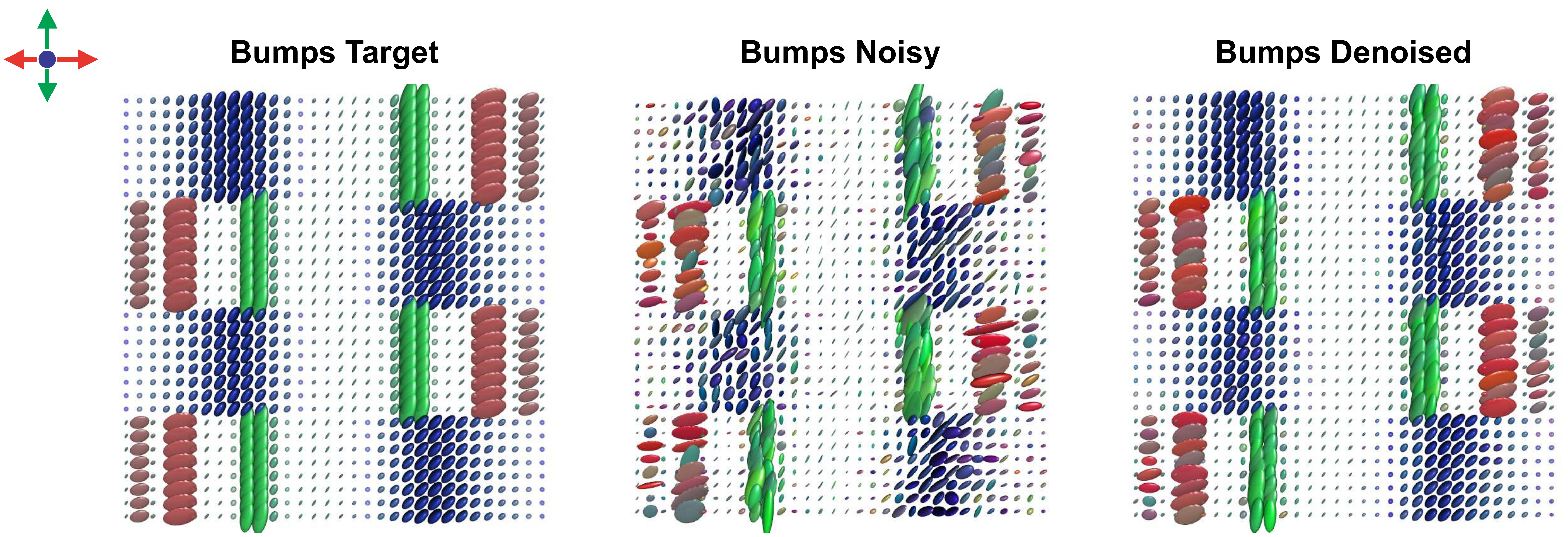}
\caption{Left: target \texttt{bumps} surface with varying local smoothness on a $(64 \times 64)$ dyadic grid. Middle: noisy HPD observations from \texttt{bumps} surface according to a Riemannian signal plus i.i.d.\@ intrinsic normal noise model. Right: intrinsic tree-structure wavelet denoised HPD surface with \texttt{pdSpecEst2D()}, (based on Riemannian metric $g_R$, with order $(3,1)$ and oracle penalty $\lambda$).\label{fig:7}}
\end{figure}

\begin{figure}[ht]
\includegraphics[scale=0.23]{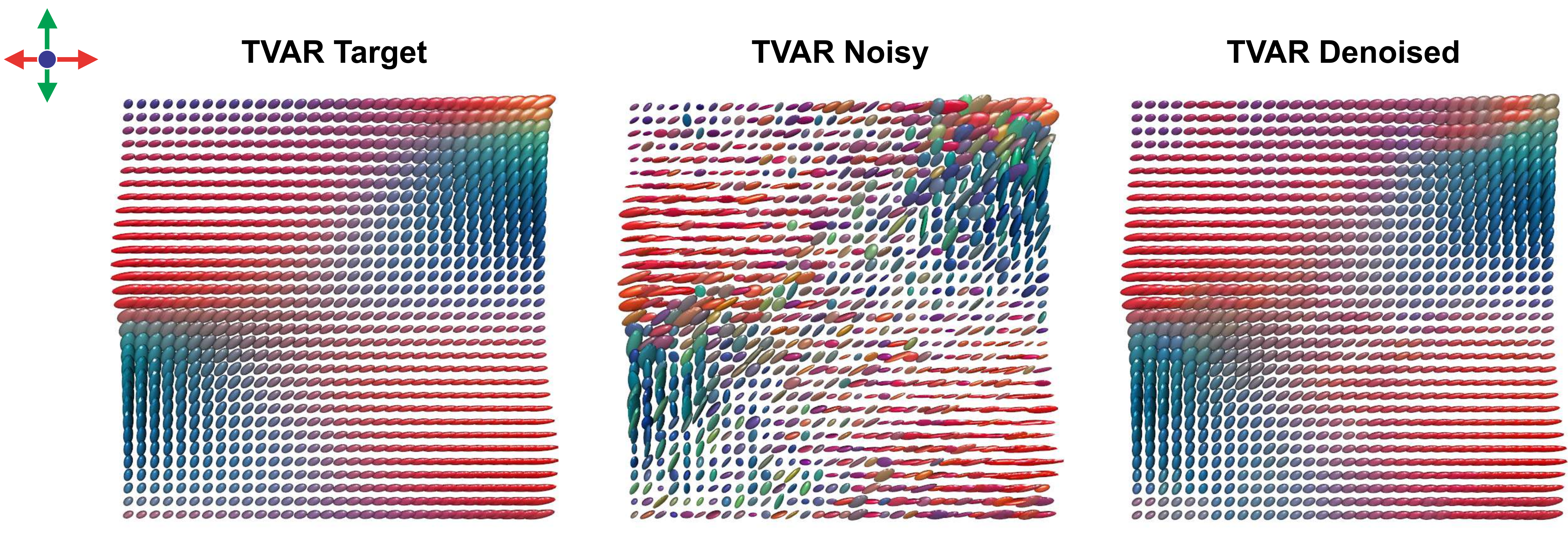}
\caption{Left: target smooth  \texttt{tvar} surface (time-varying vAR(1) spectrum) on a $(64 \times 64)$ dyadic grid. Middle: noisy HPD observations from \texttt{tvar} surface according to a Riemannian signal plus i.i.d.\@ intrinsic normal noise model. Right: intrinsic tree-structure wavelet denoised HPD surface with \texttt{pdSpecEst2D()}, (based on Riemannian metric $g_R$, with order $(3,3)$ and oracle penalty $\lambda$).\label{fig:8}}
\end{figure}

In this section, we assess the finite-sample performance of intrinsic wavelet smoothing of noisy surfaces in the space of HPD matrices and benchmark the performance against several alternative nonparametric surface smoothing procedures in the Riemannian manifold $(\mathcal{M}, g_R)$. In particular, we consider four ($3 \times 3$)-dimensional HPD target test surfaces, which display both globally homogeneous and locally varying smoothness behavior. The HPD test surfaces are available through the function \texttt{rExamples2D()} in the package \texttt{pdSpecEst} by specifying the argument \texttt{example} as \texttt{"blocks"}, \texttt{"smiley"}, \texttt{"bumps"} and \texttt{"tvar"} respectively. The left-hand images in Figures \ref{fig:5} to \ref{fig:8} display 3D-ellipsoids similar to Figure \ref{fig:1e} corresponding to the SPD modulus of the ($3 \times 3$)-dimensional HPD matrix-valued target surfaces in the $x$- and $y$-directions. The colors represent the direction of the eigenvector associated to the largest eigenvalue of the matrix objects, i.e., red, green and blue represent dominant eigenvector directions in the right-left, anterior-posterior and superior-interior orientations. 

\paragraph{Estimation procedures} 
In the simulation studies below, we consider intrinsic wavelet denoising of dyadic surfaces of HPD matrices based on nonlinear hard thresholding of entire wavelet coefficients based on the traces of the whitened wavelet coefficients as described in Section \ref{sec:4.1}. The wavelet coefficients are obtained from a dyadic intrinsic 2D AI wavelet transform with a given average-interpolation order $(N_1,N_2) \geq (1,1)$. For data generated from the four different test surfaces, we fix the average-interpolation orders to: $(1,1)$ for the \texttt{blocks} surface, $(3,3)$ for the \texttt{smiley} surface, $(3,1)$ for the \texttt{bumps} surface and $(3,3)$ for the \texttt{tvar} surface. The impact of the choice of the average-interpolation order is relatively small in terms of the estimation error in comparison to the choice of the threshold tuning parameter $\lambda$ . It is important, however, in terms of visualization of the estimator as demonstrated in e.g., Figure \ref{fig:5}, where the Haar wavelet transform of order $(1,1)$ allows for the reconstruction of piecewise constant surfaces of HPD matrices.\\
As a straightforward nonlinear thresholding method, we consider scalar dyadic tree-structured thresholding of the traces of the wavelet coefficients analogous to \cite{CvS17}, but extended to the setting of 2D dyadic pyramids of coefficients. In particular, for each scale-location $(j,k_1,k_2)$, denote $d_{j,k_1,k_2} = \tr(\mathfrak{D}^X_{j,k_1,k_2})$ for the trace of the noisy whitened wavelet coefficient and let $w_{j,k_1,k_2} \in \{0,1\}$ be a binary label. Given a penalty parameter $\lambda \geq 0$, we optimize the CPRESS criterion:
\begin{eqnarray*}
\argmin_{\bs{w}} \te{CPRESS}(\bs{w}) &=& \argmin_{\bs{w}} \sum_{j,k_1,k_2} \big|d_{j,k_1,k_2} w_{j,k_1,k_2} - d_{j,k_1,k_2}\big|^2 + \lambda^2 \sum_{j,k_1,k_2} w_{j,k_1,k_2}, 
\end{eqnarray*}
under the constraint that the nonzero labels $\{w_{j,k_1,k_2} | w_{j,k_1,k_2} = 1\}$ form a dyadic rooted tree. That is, for each scale $j$ and location $(k_1,k_2)$, if any of the labels $w_{j,2k_1,2k_2}$, $w_{j,2k_1+1,2k_2}$, $w_{j,2k_1,2k_2}$ or $w_{j,2k_1+1,2k+1}$ (if existing) are nonzero, then the label $w_{j-1,k_1,k_2}$ also has to be nonzero. The sums in the above expression range across all scale-locations $(j,k_1,k_2)$ in the pyramid of wavelet coefficients. For 2D dyadic pyramids of coefficients, the constrained optimization problem can still be solved in $O(N)$ computations, with $N$ the total number of coefficients, by a 2D version of the dyadic tree-pruning algorithm outlined in e.g., \cite[Chapter 2]{C18}. The estimated wavelet coefficients are then given by $\widehat{D}_{j,k_1,k_2} = w_{j,k_1,k_2} D_{j,k_1,k_2}^X$ for each scale-location $(j,k_1,k_2)$. \\[3mm]
Intrinsic nonlinear tree-structured wavelet thresholding for dyadic surfaces of HPD matrices is directly available in the package \texttt{pdSpecEst} through \texttt{pdSpecEst2D()}. By default, \texttt{pdSpecEst2D()} performs intrinsic wavelet denoising of the surface of HPD matrices with respect to the affine-invariant metric as described in this paper, but the function can also perform intrinsic wavelet denoising with respect to a number of other metrics, such as (i) the Log-Euclidean metric, the Euclidean inner product between matrix logarithms; (ii) the Cholesky metric, the Euclidean inner product between Cholesky matrices; or (iii) the ordinary Euclidean metric. In the simulation studies in this section, we focus solely on intrinsic wavelet denoising with respect to the affine-invariant metric $g_R$, as none of the other metrics can guarantee general linear congruence invariance of the wavelet estimates combined with the fact that the estimates are constrained to live in the space of HPD matrices. The performance of the intrinsic wavelet-based estimator with respect to the affine-invariant metric is benchmarked against intrinsic NN-regression and intrinsic NW-regression:
\begin{figure}[t]
\includegraphics[scale=0.23]{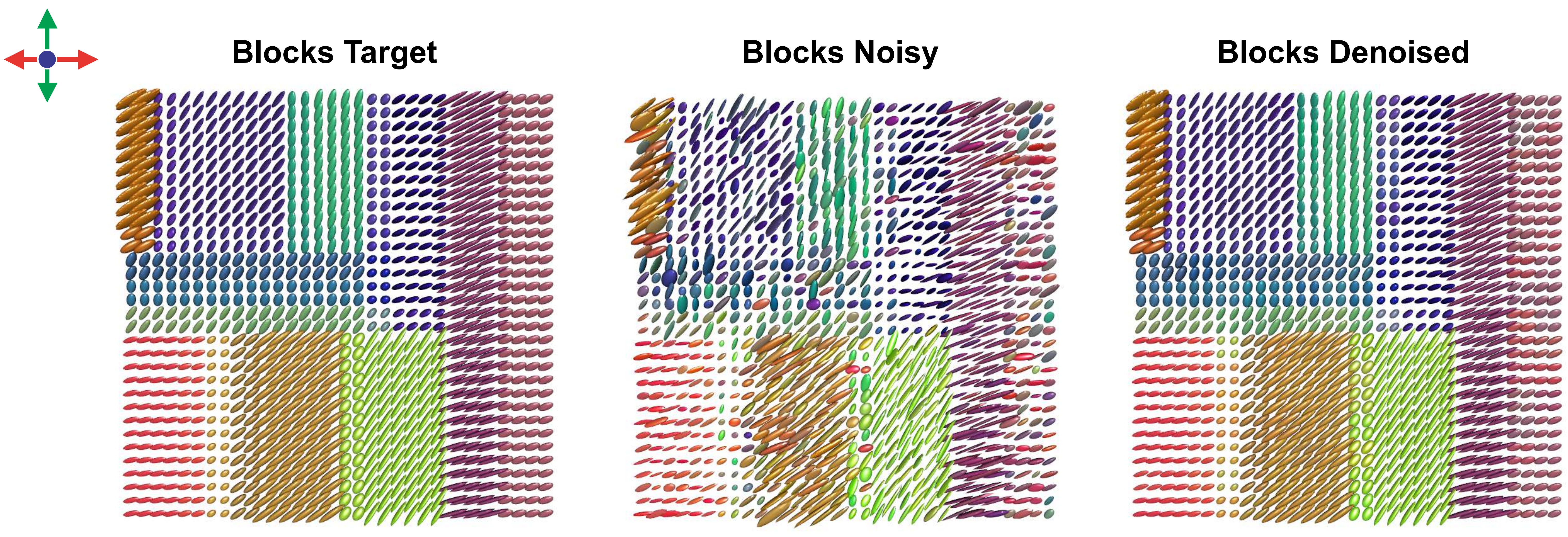}
\caption{Left: target piecewise constant \texttt{blocks} surface on a $(64 \times 64)$ dyadic grid. Middle: noisy HPD observations from \texttt{blocks} surface according to a Riemannian signal plus i.i.d.\@ intrinsic normal noise model. Right: intrinsic tree-structure wavelet denoised HPD surface with \texttt{pdSpecEst2D()}, (based on Riemannian metric $g_R$, with order $(1,1)$ and oracle penalty $\lambda$).\label{fig:5}}
\end{figure}
\begin{enumerate}
\item{\textbf{Intrinsic NN-regression}}: we consider intrinsic nearest-neighbor regression in the Riemannian manifold $(\mathcal{M}, g_R)$ by replacing ordinary local Euclidean averages by their intrinsic counterparts based on the affine-invariant metric. To be precise, the local intrinsic averages are calculated efficiently by the gradient descent algorithm in \cite{P06} available via the function \texttt{pdMean()}. 
\item{\textbf{Intrinsic NW-regression}}: we consider 2D (Nadaraya-Watson) kernel regression in the Riemannian manifold $(\mathcal{M}, g_R)$ based on the affine-invariant metric. To control the involved computational effort, we use a radially symmetric 2D Epanechnikov kernel, which has compact 2D support in contrast to e.g., a radially symmetric 2D Gaussian kernel. The kernel weighted averages are again computed efficiently by gradient descent using the function \texttt{pdMean()}.
\end{enumerate}
\begin{remark}
In \cite{CvS17}, additional nonparametric benchmark estimation procedures under the affine-invariant Riemannian metric for curves of HPD matrices include the intrinsic cubic spline approach in \cite{BA11} and intrinsic local polynomial regression according to \cite{Y12}. Although there exist generalizations of cubic spline or local polynomial regression to estimate (e.g., real-valued) surfaces in a Euclidean context, the methods in \cite{BA11} and \cite{Y12} do not have immediate straightforward generalizations to estimate surfaces of HPD matrices with respect to the affine-invariant Riemannian metric. 
\end{remark}
The intrinsic tree-structured wavelet estimator depends only on a single tuning parameter, i.e., the penalty parameter $\lambda$ in the CPRESS criterion. To make an objective comparison between the wavelet and benchmark estimation procedures, we also consider only a single tuning parameter for the benchmark procedures. For both the intrinsic NN- and NW-estimator, we consider a single isotropic bandwidth parameter $\lambda$ determining the size of the kernel or smoothing window. For the intrinsic NN-estimator, this means that at each location, the NN-estimate is an intrinsic local average over a square grid of closest neighboring observations. For the intrinsic NW-estimator, this means that the bandwidth matrix $H = \lambda\cdot \te{Id}$ of the radially symmetric Epanechnikov kernel is diagonal and is determined by the single bandwidth parameter $\lambda$. 

\paragraph{Simulation setup and results}
Given the test functions as target HPD surfaces, we generate observations $(Y_{ij})_{ij} \in \mathbb{P}_{3 \times 3}$ from the target surface $(f_{ij})_{ij} \in \mathbb{P}_{3 \times 3}$ corrupted by noise $(E_{ij})_{ij} \in \mathbb{P}_{3 \times 3}$ according to a discretized intrinsic signal plus i.i.d.\@ noise model with respect to the affine-invariant Riemannian metric as in eq.(\ref{eq:4.1}), i.e.,
\begin{eqnarray*}
Y_{ij} &=& f_{ij}^{1/2} \ast E_{ij} \ \in \ \mathbb{P}_{3 \times 3}, \quad 1 \leq i,j \leq n,
\end{eqnarray*}
The noise random variables $E_{ij} \sim \nu$ are i.i.d.\@ and generated from two different noise distributions $\nu \in P_2(\mathbb{P}_{3 \times 3})$.
\begin{itemize}
\item \emph{Intrinsic normal distribution}: for each location $(i,j)$, $E_{ij} \overset{d}{=} \Exp(Z_{ij})$, where $Z_{ij} \overset{d}{=} \sum_{k=1}^{9} z_k e^k$, with $z_1,\ldots, z_9 \overset{\te{iid}}{\sim} N(0, 1/2)$ and $\{e^1,\ldots,e^9\} \in \mathbb{H}_{3 \times 3}$ an orthonormal basis of $(\mathbb{H}_{3 \times 3}, \langle \cdot, \cdot \rangle_F)$. In particular, the intrinsic mean of $E_{ij}$ equals the identity matrix, i.e., $\mathbb{E}_{\nu}[E_{ij}] = \te{Id}$, which immediately follows from the observation that $\bs{E}_{\nu}[ \Log_{\te{Id}}(E_{ij}) ] = \bs{E}_{\nu} [ Z_{ij} ] = \bs{0}$ combined with eq.(\ref{eq:2.7}), as the logarithmic map at the identity matrix reduces to the ordinary matrix logarithm $\Log_{\te{Id}}(\cdot) = \Log(\cdot)$.  
\item \emph{Rescaled Wishart distribution}: for each location $(i,j)$, $E_{ij} \overset{d}{=} c(3, 4)\cdot W_{ij}$, where $W_{ij} \sim W_3^c(4, \te{Id} / 4)$ is a $(3 \times 3)$-dimensional complex Wishart matrix with 4 degrees of freedom and Euclidean mean equal to the identity matrix. The rescaling factor $c(3,4) = \exp(\log(4) - \frac{1}{3}\sum_{i=1}^3 \psi(1 + i))$ corresponds to the bias-correction in \cite[Theorem 5.1]{CvS17}, such that the intrinsic mean of $E_{ij}$ equals the identity matrix, i.e., $\mathbb{E}_{\nu}[E_{ij}] = \te{Id}$.
\end{itemize} 
To assess the performance of the nonparametric surface estimation procedures, we approximate, by means of simulation, the intrinsic integrated squared estimation error (IISE) with respect to the target test surface based on the Riemannian distance $\delta_R$. For the tree-structured wavelet estimators, we compute the IISE based on the following choices of the tuning parameter: (i) $\lambda$ equal to the universal threshold, (ii) $\lambda$ a semi-oracular penalty and (iii) $\lambda$ the oracular penalty. The oracular penalty minimizes the IISE with respect to the true target surface at each individual simulation. The semi-oracular penalty is a fixed penalty pre-determined by averaging the oracle penalty over a number of simulation runs. The semi-oracular penalty is included because it is computationally too expensive to compute the oracle penalty at each individual simulation for the intrinsic NW-estimator, in particular for large bandwidth parameters, due to the relatively expensive computation of locally weighted averages under the affine-invariant metric. For this estimator, the results are displayed only for the semi-oracular bandwidth parameter, which are expected to be relatively close to the estimation results for the oracular penalty. For each of the other estimators, the displayed results include both a semi-oracular and oracular choice of the tuning parameter.\\[3mm]
\begin{figure}[!ht]
\begin{subfigure}{1\linewidth}
\centering
\includegraphics[scale=0.6]{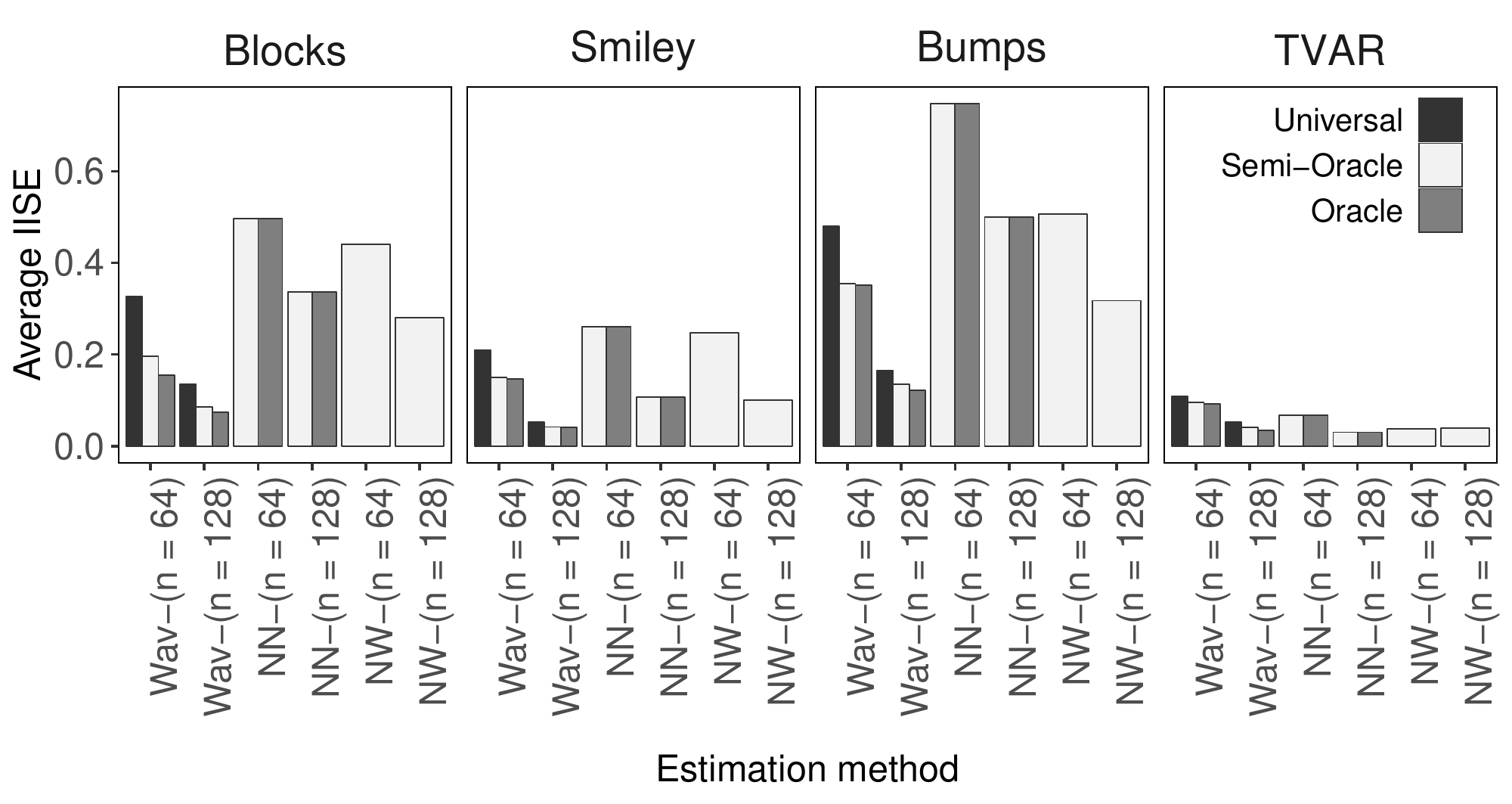}
\subcaption{Intrinsic normal noise distribution}
\end{subfigure}\\
\begin{subfigure}{1\linewidth}
\centering
\includegraphics[scale=0.6]{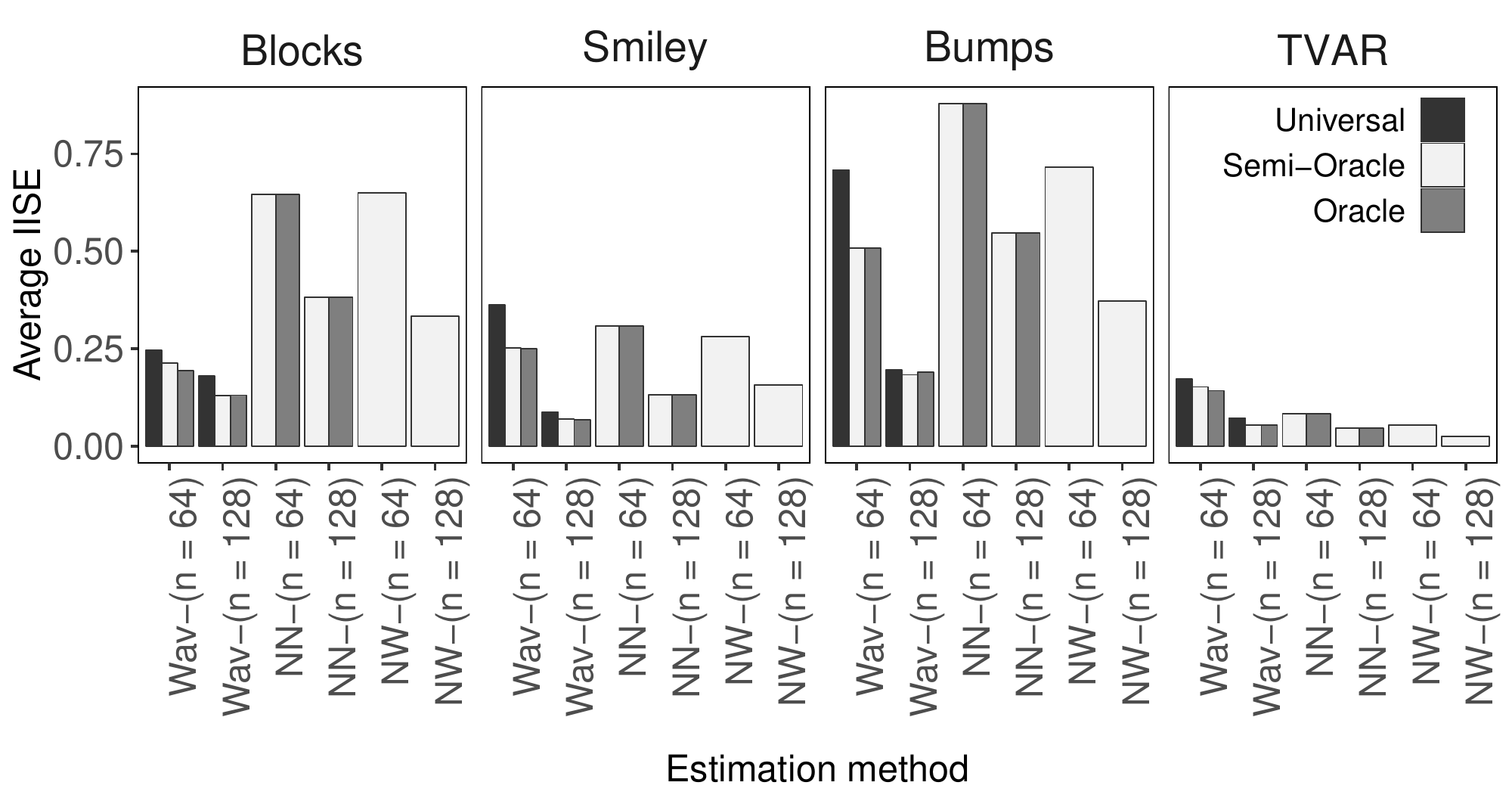}
\subcaption{Rescaled Wishart noise distribution}
\end{subfigure}
\caption{Average intrinsic integrated squared error (IISE) based on the Riemannian distance $\delta_R$, with $\te{Wav}-$, $\te{NN}-$ and $\te{NW}-$ respectively the intrinsic (tree-structure) wavelet thresholded, intrinsic nearest-neighbor, and intrinsic 2D kernel smoothed surface estimates. Data generated from a Riemannian signal-noise model on a $(64 \times 64)-$ and $(128 \times 128)$-grid. \label{fig:9}}
\end{figure} 
Figure \ref{fig:9} displays the average IISEs based on a number of $M$ simulation runs for noisy HPD surface data generated from an intrinsic signal plus i.i.d.\@ noise model on square dyadic observation grids of size $(n \times n)$, with $n = 64$ and $n = 128$. For $n = 64$, we computed the average IISE over $M = 1\,350$ surface estimates, and for $n = 128$ we averaged the IISE over $M = 350$ surface estimates. In the top panel of Figure \ref{fig:9} the noise is generated from an intrinsic normal distribution, and in the bottom panel from a rescaled Wishart distribution. We have performed the same simulated experiments for data generated from several rectangular dyadic observation grids. The simulation results are roughly similar to the results displayed in Figures \ref{fig:9} and for this reason have been omitted here. \\[3mm]
According to Figure \ref{fig:9}, the intrinsic wavelet estimator outperforms the benchmark estimators in terms of the IISE in the majority of the simulated scenarios based on the test surfaces \texttt{blocks}, \texttt{smiley} and \texttt{bumps}, each of which are not globally smooth HPD surfaces, as illustrated in Figures \ref{fig:5} to \ref{fig:7}. This is attributed to the fact that, in contrast to the benchmark procedures, the wavelet-based estimator is able to capture varying degrees of smoothness in the HPD surface, such as local peaks or discontinuities in the surface combined with highly regular behavior. The benchmark procedures do outperform the wavelet-based estimators in terms of the IISE in the highly smooth \texttt{tvar} test surface, (see also Figure \ref{fig:8}), as a single global smoothing parameter in the benchmark procedures is sufficient to capture the smooth behavior in the HPD surface. Although the wavelet-based estimator does not significantly improve upon the estimation error for globally smooth surfaces, from a computational perspective the wavelet-estimator may still remain the preferred option, as it provides a fast heuristic choice of the penalty parameter (e.g., a simple universal threshold). For both benchmark procedures there is no simple heuristic choice for the bandwidth parameter(s) and in real-world applications we need to resort either to computationally expensive cross-validation methods or manual bandwidth tuning, as the (semi-)oracle bandwidths are not available in practice.

\subsection{Epileptic seizure EEG recordings}
\begin{figure}[t]
\centering
\includegraphics[scale=0.6]{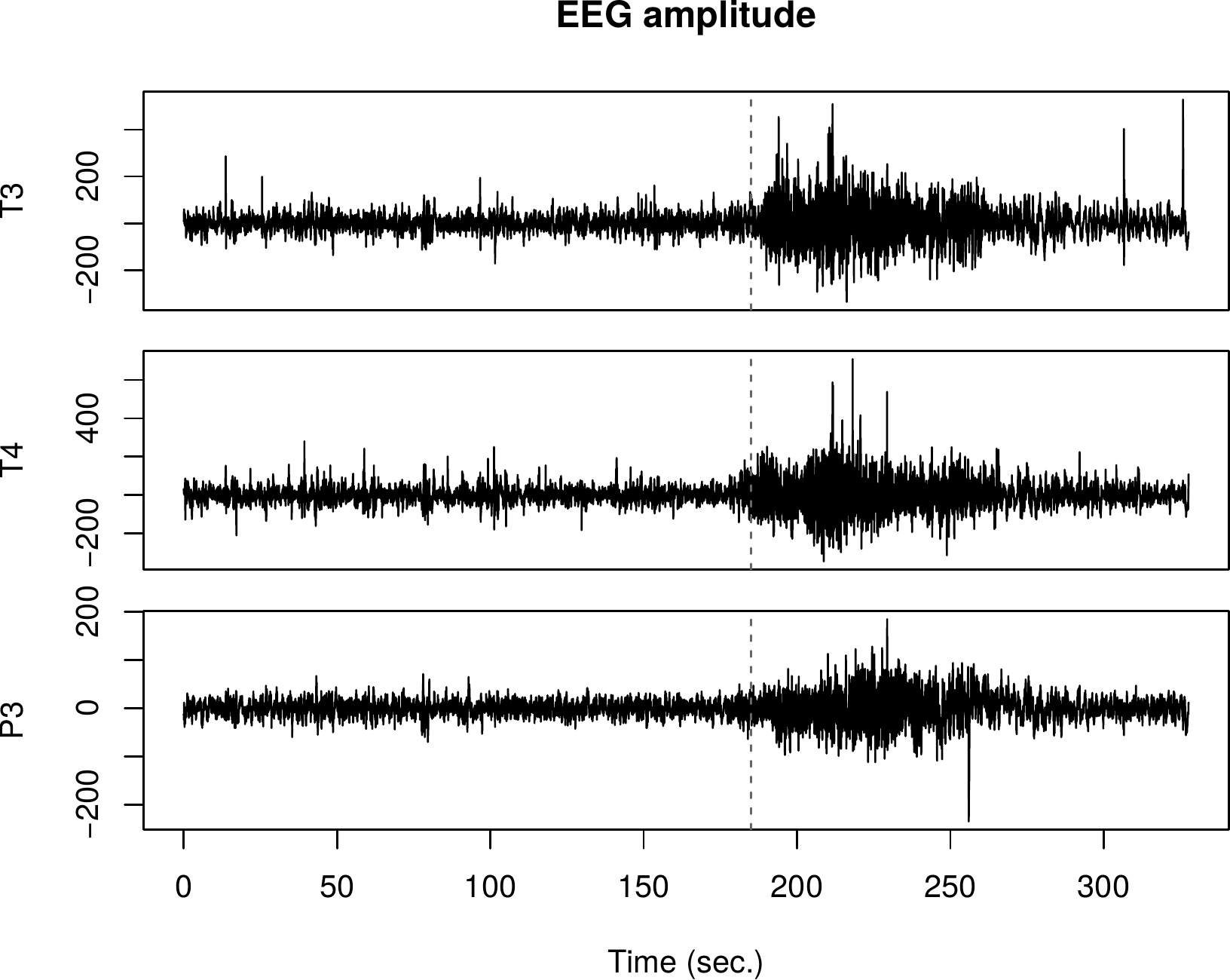}
\caption{Multichannel EEG amplitude time series, displaying the \texttt{T3}, \texttt{T4} and \texttt{P3} channels corresponding respectively to the EEG recordings at the left temporal, right temporal and left parietal lobe. The time onset of the epileptic seizure is indicated by the vertical dashed gray line.\label{fig:11}}
\end{figure}

To demonstrate the intrinsic wavelet denoising methods in the context of time-varying spectral matrix estimation, we analyze a brain signal dataset of multichannel electroencephalogram (EEG) time series recorded during an epileptic seizure in the brain of a patient diagnosed with left temporal lobe epilepsy. The spectral characteristics of this multivariate nonstationary EEG dataset have previously been investigated in \cite{O01}, \cite{OvSG05} and \cite{OH06} taking into account multiple EEG channels and in \cite{GDOvS03} for single EEG channels. Analogous to the cited papers, our direct aim is to study the evolving spectral characteristics in the EEG time series before, during and after the epileptic seizure. The available EEG time series data is recorded at 21 spatial locations, i.e., channels, on the patient's scalp. In this section, we extract a subset of 3 EEG channels of interest located at the left temporal lobe (\texttt{T3}), the right temporal lobe (\texttt{T4}) and the left parietal lobe (\texttt{P3}). Note that these channels are also included in the set of analyzed EEG channels in \cite{OvSG05}. The main reason for restricting our analysis to a subset of 3 EEG channels, instead of considering the complete 21-dimensional EEG dataset, is that we cannot easily display detailed information across time and frequency for the complete ($21 \times 21$)-dimensional time-varying spectral matrix. From a computational perspective, there is no issue with estimating the full-blown ($21 \times 21$)-dimensional time-varying spectral matrix. The available EEG time series consists of $T = 32\, 768$ recordings of the EEG amplitude in the patient's brain recorded at 100 Hz, thus roughly corresponding to five and a half minutes of EEG amplitude data, with the onset of the epileptic seizure occurring around $185$ seconds after the start of the recordings, according to the neurologist. We point out that this EEG time series dataset is similar to the dataset analyzed in \cite{O01}, where the authors in \cite{O01} only investigate the \texttt{T3} and \texttt{P3} channels. Figure \ref{fig:11} displays an equidistant sample of 10\,000 vector-valued EEG time series observations from the start to the end of the experiment. The vertical dashed gray line indicates the approximate onset of the epileptic seizure, which is followed by a nonstationary power burst in the EEG time series.

\begin{figure}[t]
\centering
\hspace{-3mm}
\includegraphics[scale=0.55]{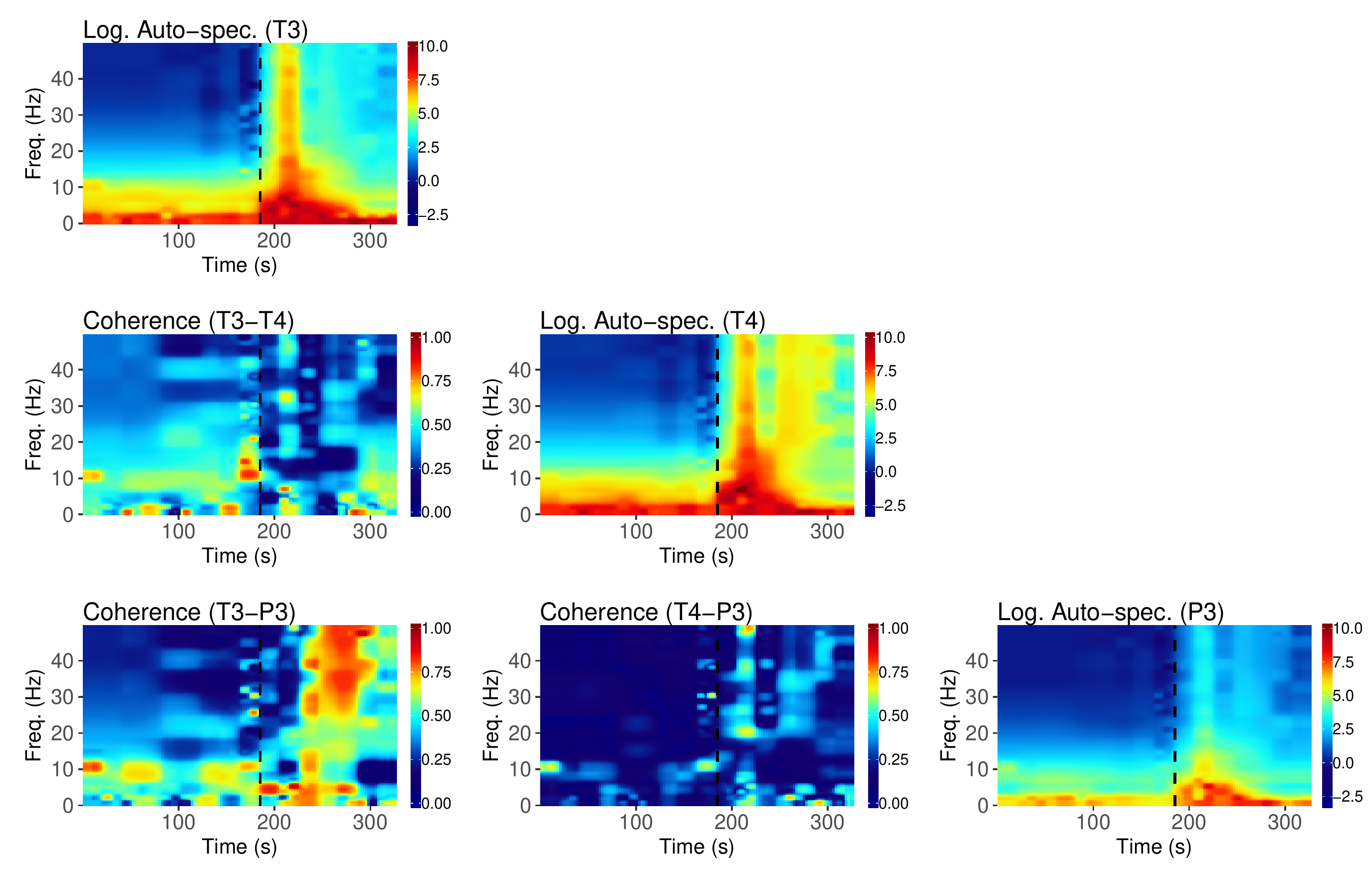}
\caption{Intrinsic wavelet-thresholded HPD spectral estimate of $3$-dimensional EEG time series, with channels \texttt{T3}, \texttt{T4} and \texttt{P3}, at a $(128 \times 128)$ time-frequency grid, obtained via the function \texttt{pdSpecEst2D()}, with affine-invariant metric $g_R$, average-interpolation order $(3,3)$ and universal penalty parameter $\lambda$. The time onset of the epileptic seizure is indicated by the dashed line. \label{fig:13}}
\end{figure}

\paragraph{Spectral estimation procedure}
As an initial pre-smoothing step, we construct a highly noisy surface of $(3 \times 3)$-dimensional HPD segmented periodograms based on the multivariate nonstationary EEG time series. The complete EEG time series ($T = 32\, 768$ vector-valued observations) is partitioned into $L_t = 128$ non-overlapping time segments of length $T_t = 256$, and we consider $L_f = 128$ frequency points ranging from 0\,Hz to 50\,Hz, thereby defining a square dyadic ($128 \times 128$)-dimensional time-frequency grid at which the time-varying spectrum will be estimated. For each of the $L_t$ time segments, we compute an initial noisy HPD multitaper spectral estimate based on $L = d = 3$ DPSS tapering functions (time-bandwidth parameter $n_w = 3$) using the function \texttt{pdPgram2D()}. By choosing the number of tapers $L$ equal to the dimension of the time series $d$, we pre-smooth the raw periodograms only by a minimal amount to guarantee positive-definiteness (i.e., full-rank matrices). In this way, the HPD periodograms remain highly noisy objects, and the essential task of smoothing the HPD periodogram surface across time and frequency is performed by the intrinsic nonlinear wavelet estimation procedure. 
\begin{figure}[t]
\centering
\hspace{-3mm}
\includegraphics[scale=0.55]{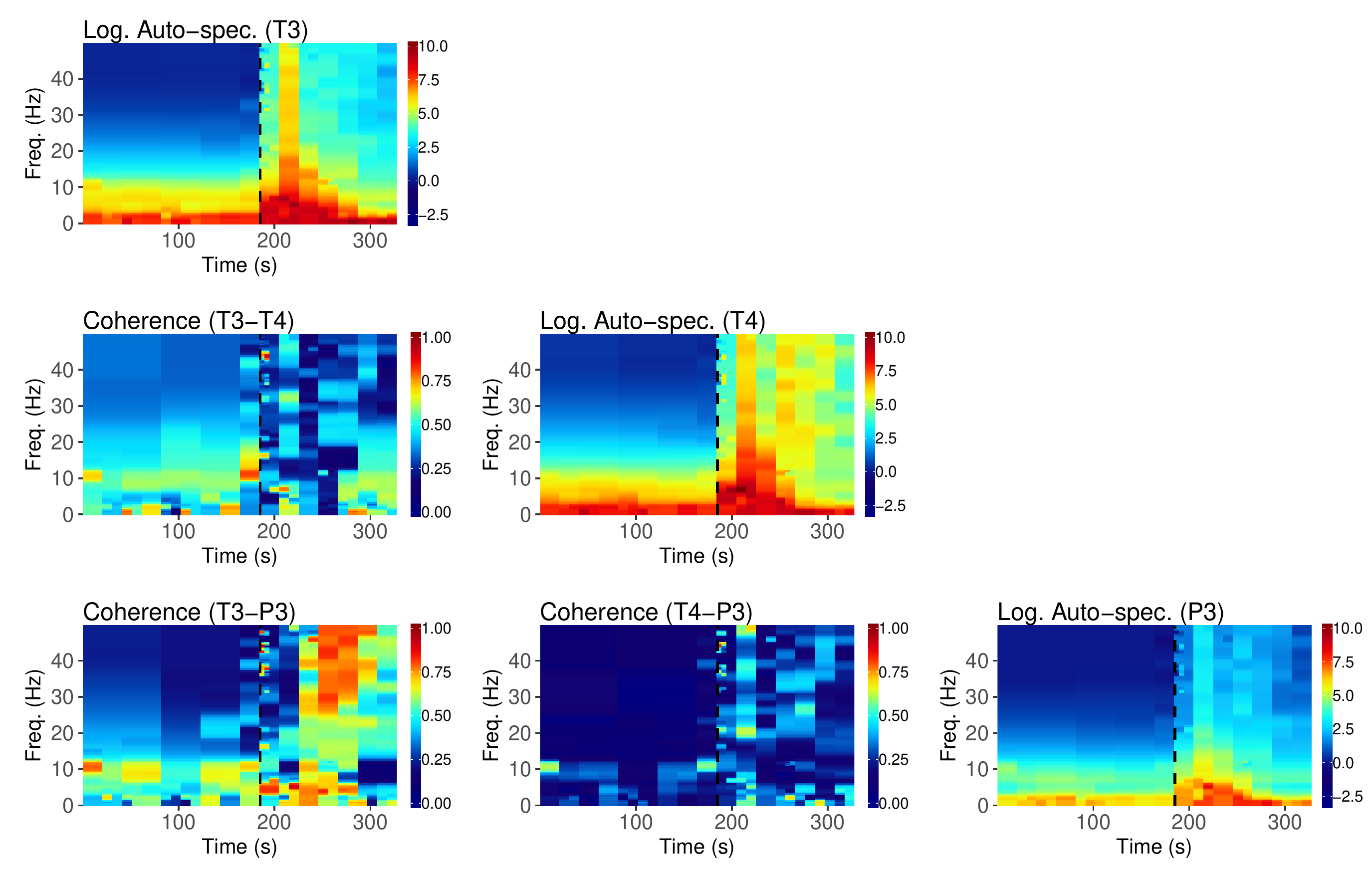}
\caption{Intrinsic wavelet-thresholded HPD spectral estimate of $3$-dimensional EEG time series, with channels \texttt{T3}, \texttt{T4} and \texttt{P3}, at a $(128 \times 128)$ time-frequency grid, obtained via the function \texttt{pdSpecEst2D()}, with affine-invariant metric $g_R$, average-interpolation order $(1,3)$ and universal penalty parameter $\lambda$. The time onset of the epileptic seizure is indicated by the dashed line. \label{fig:12}}
\end{figure}
Figure \ref{fig:12} and \ref{fig:13} display wavelet-denoised HPD time-varying spectral estimates obtained by nonlinear tree-structured wavelet thresholding of the time-varying HPD periodogram surface on a dyadic $(128 \times 128)$ time-frequency grid, with on the x-axes the time parameter (in seconds) and on the y-axes the frequency parameter (in Hertz). The off-diagonal entries display the cross-coherences between the channels across time and frequency, where the (non-squared) coherence at time-frequency $(u,\omega)$ between channels $x$ and $y$ is given by $c_{xy}(u,\omega) = |f_{xy}(u,\omega)|/\sqrt{f_{xx}(u,\omega)f_{yy}(u,\omega)}$. Here, $f_{xy}(u, \omega)$ is the cross-spectrum between $x$ and $y$ at time-frequency $(u, \omega)$, and $f_{xx}(u, \omega)$ and $f_{yy}(u,\omega)$ are the auto-spectra of the components $x$ and $y$. The cross-coherences for the upper-diagonal entries are identical to those in the lower-diagonal entries by symmetry and are therefore omitted. The diagonal entries display the (ordinary) logarithms of the auto-spectra across time and frequency conveying information about the scale of the spectrum. The spectral estimates are computed with the function \texttt{pdSpecEst2D()} based on a natural dyadic refinement pyramid, using the affine-invariant Riemannian metric, maximum non-zero wavelet scale $J = 6$, and penalty parameter $\lambda$ in the CPRESS criterion equal to the universal threshold, (with the noise variance robustly estimated from the finest wavelet scale via the MAD, i.e., median absolute deviation). In Figure \ref{fig:12}, we display the spectral estimate obtained with the average-interpolation order $(N_1,N_2) = (1,3)$, which results in smooth spectral behavior in the frequency direction, but a piecewise constant (Haar wavelet) structure in the time direction. The piecewise constant time structure allows us to capture the abrupt changes over time in the spectrum before and after the onset of the seizure, similar to the piecewise-constant SLEX spectral estimation procedure in \cite{O01} and \cite{OvSG05}. We observe that the estimated time-varying log-auto-spectra and cross-coherences in Figure \ref{fig:12} are highly similar to the estimated time-varying SLEX spectra in \cite{O01} and in \cite{OvSG05}, except that the SLEX spectral estimation procedure by construction can only produce piecewise constant estimates of the spectrum, whereas the wavelet estimation procedure is also able to construct smooth estimates of the spectrum either in the time direction, the frequency direction or both. This is further illustrated in Figure \ref{fig:13}, where we display the spectral estimate obtained with the average-interpolation order $(N_1,N_2) = (3,3)$, which results in smooth spectral behavior in both the time and frequency direction. The wavelet-denoised HPD spectral estimates in Figures \ref{fig:12} and \ref{fig:13} demonstrate the power of the intrinsic nonlinear wavelet estimator. On the one hand, the spectral estimates captures the local power burst after the onset of the epileptic seizure and the change in time-varying spectral behavior before and after the seizure. On the other hand, the spectral estimate is able to capture the smoothly evolving spectral behavior before the seizure across time and frequency. If we compare this to the benchmark intrinsic 2D kernel-based estimators in Section \ref{sec:5.1}, in order to achieve the same level of flexibility, the benchmark estimators ideally require adaptive local bandwidths across time and frequency to capture both smooth and local characteristics in the spectrum. Automatic selection of such local bandwidths by means of e.g., cross-validation quickly becomes computationally expensive, in particular when working with the affine-invariant Riemannian metric. This is in contrast to the intrinsic nonlinear wavelet estimator, which uses a single (primary) tuning parameter $\lambda$ based on a simple universal threshold. 

\section{Concluding remarks}
In this paper, we studied intrinsic average-interpolation wavelet transforms and linear and nonlinear wavelet denoising methods for \emph{surfaces} in the space of HPD matrices, with the primary focus on the space as a Riemannian manifold equipped with the affine-invariant metric $g_R$. The intrinsic wavelet methods for HPD surfaces proposed in this chapter are natural extensions of the intrinsic framework and wavelet methodology for curves of HPD matrices developed in \cite{CvS17}. In the following, we list several unsolved challenges and topics of interest for future research. \\
First, analogous to \cite{CvS17}, in Section \ref{sec:3} we derived the wavelet coefficient decay and convergence rates with respect to the affine-invariant metric in a dyadic framework. Most arguments seem to extend without much effort to other Riemannian metrics as well, such as the Log-Euclidean metric studied in e.g., \cite{A06}, and it is of interest to derive generalized versions of the proofs without restricting necessarily to the affine-invariant Riemannian metric. The generalization of the proofs to \emph{non-dyadic} observation grids seems to be more challenging at this moment, as the proofs currently rely on the fact that the refinement rectangles $I_{j,k_1,k_2}$ in the wavelet transform are of equal size within each resolution scale. This is true for the natural dyadic refinement pyramid, but is no longer the case for general non-dyadic refinement pyramids.\\
Another interesting challenge for future work, in particular in the context of non-dyadic observation grids, is data-driven selection of the refinement pyramid. In the dyadic framework, there is a natural choice for the dyadic refinement pyramid, and this has been the preferred refinement pyramid throughout this paper. In the non-dyadic context, a working solution is to select a refinement pyramid that is \emph{as close as possible} in shape and size to the dyadic refinement pyramid, but ideally we should let the data decide a proper choice of refinement pyramid, as in e.g., \cite{F07} and \cite{FT16} for scalar piecewise constant curves and surfaces. The aim would be to find a data-adaptive refinement pyramid that enforces maximal sparsity of the coefficients in the wavelet domain, and the main challenge is to do this in a computationally efficient way, relying e.g., on greedy top-down or bottom-up decision-tree based approaches to select an optimal refinement pyramid. This is important, as a naive search through the set of all possible refinement pyramids quickly becomes computationally infeasible. \\
Finally, we point out that in this paper we focus only on hard thresholding of entire matrix-valued coefficients in the intrinsic wavelet domain based on their \emph{trace}. This is appropriate from the viewpoint of HPD matrices as single data objects in the Riemannian manifold. However, the performance of nonlinear wavelet estimation through thresholding or (Bayesian) shrinkage of individual components of the wavelet coefficients may be superior in practice due to the additional flexibility. For instance, in the context of (time-varying) spectral estimation, componentwise shrinkage of the matrix-valued wavelet coefficients allows one to capture varying degrees of smoothness across matrix components in the (time-varying) spectrum. At the time of writing, we have experimented with several hard thresholding procedures for the individual components of the matrix-valued wavelet coefficients, which seem to perform well in practice. An important challenge that remains for future work is the development of a proper theoretical background for individual componentwise shrinkage of coefficients and practical selection procedures for the componentwise shrinkage parameters, as the properties derived for nonlinear trace thresholding of the coefficients in Section \ref{sec:4.1} do not have immediate analogs in terms of the components of the matrix-valued wavelet coefficients.

\section*{Acknowledgments}
The authors gratefully acknowledge financial support from the following agencies and projects: the Belgian Fund for Scientific Research FRIA/FRS-FNRS (J. Chau), the contract ``Projet d'Actions de Recherche Concert\'ees'' No. 12/17-045 of the ``Communaut\'e fran\c{c}aise de Belgique'' (J.Chau and R. von Sachs), and the IAP research network P7/06 of the Belgian government (R. von Sachs). We thank Hernando Ombao and the UC Irvine Space-Time Modeling Group for providing access to the EEG seizure data. Computational resources have been provided by the CISM/UCL and the C\'ECI funded by the FRS-FNRS.  


\bibliography{Main}

\begin{thebibliography}{}

\bibitem[\protect\citeauthoryear{Adak}{Adak}{1998}]{A98}
Adak, S. (1998).
\newblock Time-dependent spectral analysis of nonstationary time series.
\newblock {\em Journal of the American Statistical Association\/}~{\em
  93\/}(444), 1488--1501.

\bibitem[\protect\citeauthoryear{Arsigny, Fillard, Pennec, and Ayache}{Arsigny
  et~al.}{2006}]{A06}
Arsigny, V., P.~Fillard, X.~Pennec, and N.~Ayache (2006).
\newblock Log-{E}uclidean metrics for fast and simple calculus on diffusion
  tensors.
\newblock {\em Magnetic Resonance in Medicine\/}~{\em 56\/}(2), 411--421.

\bibitem[\protect\citeauthoryear{Bayram and Baraniuk}{Bayram and
  Baraniuk}{1996}]{BB96}
Bayram, M. and R.~Baraniuk (1996).
\newblock Multiple window time-frequency analysis.
\newblock In {\em Proceedings of the IEEE-SP International Symposium on
  Time-Frequency and Time-Scale Analysis}, pp.\  173--176.

\bibitem[\protect\citeauthoryear{Bhatia}{Bhatia}{2009}]{B09}
Bhatia, R. (2009).
\newblock {\em Positive Definite Matrices}.
\newblock New Jersey: Princeton University Press.

\bibitem[\protect\citeauthoryear{Boothby}{Boothby}{1986}]{B86}
Boothby, W. (1986).
\newblock {\em An Introduction to Differentiable Manifolds and Riemannian
  Geometry}.
\newblock New York: Academic Press.

\bibitem[\protect\citeauthoryear{Boumal and Absil}{Boumal and
  Absil}{2011a}]{BA11b}
Boumal, N. and P.-A. Absil (2011a).
\newblock A discrete regression method on manifolds and its application to data
  on {SO}(n).
\newblock {\em IFAC Proceedings Volumes\/}~{\em 44\/}(1), 2284--2289.

\bibitem[\protect\citeauthoryear{Boumal and Absil}{Boumal and
  Absil}{2011b}]{BA11}
Boumal, N. and P.-A. Absil (2011b).
\newblock Discrete regression methods on the cone of positive-definite
  matrices.
\newblock In {\em IEEE ICASSP, 2011}, pp.\  4232--4235.

\bibitem[\protect\citeauthoryear{Brillinger}{Brillinger}{1981}]{B81}
Brillinger, D. (1981).
\newblock {\em Time Series: Data Analysis and Theory}.
\newblock San Francisco: Holden-Day.

\bibitem[\protect\citeauthoryear{Brockwell and Davis}{Brockwell and
  Davis}{2006}]{BD06}
Brockwell, P. and R.~Davis (2006).
\newblock {\em Time Series: Theory and Methods}.
\newblock New York: Springer.

\bibitem[\protect\citeauthoryear{Chau}{Chau}{2017}]{C17}
Chau, J. (2017).
\newblock {\em pdSpecEst: An Analysis Toolbox for Hermitian Positive Definite
  Matrices}.
\newblock Package version 1.2.2.

\bibitem[\protect\citeauthoryear{Chau}{Chau}{2018}]{C18}
Chau, J. (2018).
\newblock {\em Advances in Spectral Analysis for Multivariate, Nonstationary
  and Replicated Time Series}.
\newblock Ph.\ D. thesis, Universit{\'e} catholique de Louvain.

\bibitem[\protect\citeauthoryear{Chau and von Sachs}{Chau and von
  Sachs}{2017}]{CvS17}
Chau, J. and R.~von Sachs (2017).
\newblock Intrinsic wavelet regression for curves of {H}ermitian positive
  definite matrices.
\newblock {\em ArXiv preprint 1701.03314\/}.

\bibitem[\protect\citeauthoryear{Dahlhaus}{Dahlhaus}{1997}]{Da97}
Dahlhaus, R. (1997).
\newblock Fitting time series models to nonstationary processes.
\newblock {\em The Annals of Statistics\/}~{\em 25\/}(1), 1--37.

\bibitem[\protect\citeauthoryear{Dahlhaus}{Dahlhaus}{2012}]{D12}
Dahlhaus, R. (2012).
\newblock {\em Locally stationary processes}, Chapter in Time Series Analysis:
  Methods and Applications, Vol. 30, pp.\  351--413.
\newblock Amsterdam: Elsevier.

\bibitem[\protect\citeauthoryear{do~Carmo}{do~Carmo}{1992}]{D92a}
do~Carmo, M. (1992).
\newblock {\em Riemannian Geometry}.
\newblock Boston: Birkh\"auser.

\bibitem[\protect\citeauthoryear{Dryden, Koloydenko, and Zhou}{Dryden
  et~al.}{2009}]{D09}
Dryden, I., A.~Koloydenko, and D.~Zhou (2009).
\newblock Non-{E}uclidean statistics for covariance matrices, with applications
  to diffusion tensor imaging.
\newblock {\em The Annals of Applied Statistics\/}~{\em 3\/}(3), 1102--1123.

\bibitem[\protect\citeauthoryear{Fiecas and Ombao}{Fiecas and
  Ombao}{2016}]{FO16}
Fiecas, M. and H.~Ombao (2016).
\newblock Modeling the evolution of dynamic brain processes during an
  associative learning experiment.
\newblock {\em Journal of the American Statistical Association\/}~{\em
  111\/}(516), 1440--1453.

\bibitem[\protect\citeauthoryear{Fryzlewicz}{Fryzlewicz}{2007}]{F07}
Fryzlewicz, P. (2007).
\newblock Unbalanced {H}aar technique for nonparametric function estimation.
\newblock {\em Journal of the American Statistical Association\/}~{\em
  102\/}(480), 1318--1327.

\bibitem[\protect\citeauthoryear{Fryzlewicz and Timmermans}{Fryzlewicz and
  Timmermans}{2016}]{FT16}
Fryzlewicz, P. and C.~Timmermans (2016).
\newblock {SHAH}: {SH}ape-{A}daptive {H}aar wavelets for image processing.
\newblock {\em Journal of Computational and Graphical Statistics\/}~{\em
  25\/}(3), 879--898.

\bibitem[\protect\citeauthoryear{Guo and Dai}{Guo and Dai}{2006}]{GD06}
Guo, W. and M.~Dai (2006).
\newblock Multivariate time-dependent spectral analysis using {C}holesky
  decomposition.
\newblock {\em Statistica Sinica\/}~{\em 16}, 825--845.

\bibitem[\protect\citeauthoryear{Guo, Dai, Ombao, and von Sachs}{Guo
  et~al.}{2003}]{GDOvS03}
Guo, W., M.~Dai, H.~Ombao, and R.~von Sachs (2003).
\newblock Smoothing spline {ANOVA} for time-dependent spectral analysis.
\newblock {\em Journal of the American Statistical Association\/}~{\em
  98\/}(463), 643--652.

\bibitem[\protect\citeauthoryear{Higham}{Higham}{2008}]{H08}
Higham, N.~J. (2008).
\newblock {\em Functions of Matrices: Theory and Computation}.
\newblock Philadelphia: Siam.

\bibitem[\protect\citeauthoryear{Hinkle, Fletcher, and Joshi}{Hinkle
  et~al.}{2014}]{HFJ14}
Hinkle, J., P.~Fletcher, and S.~Joshi (2014).
\newblock Intrinsic polynomials for regression on {R}iemannian manifolds.
\newblock {\em Journal of Mathematical Imaging and Vision\/}~{\em 50\/}(1-2),
  32--52.

\bibitem[\protect\citeauthoryear{Holbrook, Lan, Vandenberg-Rodes, and
  Shahbaba}{Holbrook et~al.}{2018}]{H16}
Holbrook, A., S.~Lan, A.~Vandenberg-Rodes, and B.~Shahbaba (2018).
\newblock Geodesic {L}agrangian {M}onte {C}arlo over the space of positive
  definite matrices: with application to {B}ayesian spectral density
  estimation.
\newblock {\em Journal of Statistical Computation and Simulation\/}~{\em
  88\/}(5), 982--1002.

\bibitem[\protect\citeauthoryear{Le}{Le}{1995}]{Le95}
Le, H. (1995).
\newblock Mean size-and-shapes and mean shapes: a geometric point of view.
\newblock {\em Advances in Applied Probability\/}~{\em 27\/}(1), 44--55.

\bibitem[\protect\citeauthoryear{Lee}{Lee}{2003}]{L03}
Lee, J. (2003).
\newblock {\em Introduction to Smooth Manifolds}.
\newblock New York: Springer-Verlag.

\bibitem[\protect\citeauthoryear{Li and Krafty}{Li and Krafty}{2018}]{LK18}
Li, Z. and R.~Krafty (2018).
\newblock Adaptive {B}ayesian time-frequency analysis of multivariate time
  series.
\newblock {\em Journal of the American Statistical Association\/}, 1--13.

\bibitem[\protect\citeauthoryear{Ombao and Ho}{Ombao and Ho}{2006}]{OH06}
Ombao, H. and M.~Ho (2006).
\newblock Time-dependent frequency domain principal components analysis of
  multi-channel non-stationary signals.
\newblock {\em Computational Statistics and Data Analysis\/}~{\em 50\/}(9),
  2339--2360.

\bibitem[\protect\citeauthoryear{Ombao, Raz, von Sachs, and Malow}{Ombao
  et~al.}{2001}]{O01}
Ombao, H., J.~Raz, R.~von Sachs, and B.~Malow (2001).
\newblock Automatic statistical analysis of bivariate nonstationary time
  series.
\newblock {\em Journal of the American Statistical Association\/}~{\em
  96\/}(454), 543--560.

\bibitem[\protect\citeauthoryear{Ombao, von Sachs, and Guo}{Ombao
  et~al.}{2005}]{OvSG05}
Ombao, H., R.~von Sachs, and W.~Guo (2005).
\newblock {SLEX} analysis of multivariate nonstationary time series.
\newblock {\em Journal of the American Statistical Association\/}~{\em
  100\/}(470), 519--531.

\bibitem[\protect\citeauthoryear{Park, Eckley, and Ombao}{Park
  et~al.}{2014}]{PEO14}
Park, T., I.~Eckley, and H.~Ombao (2014).
\newblock Estimating time-evolving partial coherence between signals via
  multivariate locally stationary wavelet processes.
\newblock {\em IEEE Transactions on Signal Processing\/}~{\em 62\/}(20),
  5240--5250.

\bibitem[\protect\citeauthoryear{Pasternak, Sochen, and Basser}{Pasternak
  et~al.}{2010}]{P10}
Pasternak, O., N.~Sochen, and P.~Basser (2010).
\newblock The effect of metric selection on the analysis of diffusion tensor
  {MRI} data.
\newblock {\em NeuroImage\/}~{\em 49\/}(3), 2190--2204.

\bibitem[\protect\citeauthoryear{Pennec}{Pennec}{2006}]{P06}
Pennec, X. (2006).
\newblock Intrinsic statistics on {R}iemannian manifolds: Basic tools for
  geometric measurements.
\newblock {\em Journal of Mathematical Imaging and Vision\/}~{\em 25\/}(1),
  127--154.

\bibitem[\protect\citeauthoryear{Pennec, Fillard, and Ayache}{Pennec
  et~al.}{2006}]{PFA05}
Pennec, X., P.~Fillard, and N.~Ayache (2006).
\newblock A {R}iemannian framework for tensor computing.
\newblock {\em International Journal of Computer Vision\/}~{\em 66\/}(1),
  41--66.

\bibitem[\protect\citeauthoryear{Said, Bombrun, Berthoumieu, and Manton}{Said
  et~al.}{2017}]{S15}
Said, S., L.~Bombrun, Y.~Berthoumieu, and J.~Manton (2017).
\newblock Riemannian {G}aussian distributions on the space of symmetric
  positive definite matrices.
\newblock {\em IEEE Transactions on Information Theory\/}~{\em 63\/}(4),
  2153--2170.

\bibitem[\protect\citeauthoryear{Skovgaard}{Skovgaard}{1984}]{S84}
Skovgaard, L. (1984).
\newblock A {R}iemannian geometry of the multivariate normal model.
\newblock {\em Scandinavian Journal of Statistics\/}~{\em 11\/}(4), 211--223.

\bibitem[\protect\citeauthoryear{Smith}{Smith}{2000}]{S00}
Smith, S. (2000).
\newblock Intrinsic {C}ram{\'e}r-{R}ao bounds and subspace estimation accuracy.
\newblock In {\em Proceedings of the IEEE Sensor Array and Multichannel Signal
  Processing Workshop}, pp.\  489--493. IEEE.

\bibitem[\protect\citeauthoryear{Walden}{Walden}{2000}]{W00}
Walden, A. (2000).
\newblock A unified view of multitaper multivariate spectral estimation.
\newblock {\em Biometrika\/}~{\em 87\/}(4), 767--788.

\bibitem[\protect\citeauthoryear{Xiao and Flandrin}{Xiao and
  Flandrin}{2007}]{XF07}
Xiao, J. and P.~Flandrin (2007).
\newblock Multitaper time-frequency reassignment for nonstationary spectrum
  estimation and chirp enhancement.
\newblock {\em IEEE Transactions on Signal Processing\/}~{\em 55\/}(6),
  2851--2860.

\bibitem[\protect\citeauthoryear{Yuan, Zhu, Lin, and Marron}{Yuan
  et~al.}{2012}]{Y12}
Yuan, Y., H.~Zhu, W.~Lin, and J.~Marron (2012).
\newblock Local polynomial regression for symmetric positive definite matrices.
\newblock {\em Journal of the Royal Statistical Society: Series B\/}~{\em
  74\/}(4), 697--719.

\bibitem[\protect\citeauthoryear{Zhang}{Zhang}{2016}]{Z16b}
Zhang, S. (2016).
\newblock Adaptive spectral estimation for nonstationary multivariate time
  series.
\newblock {\em Computational Statistics and Data Analysis\/}~{\em 103},
  330--349.

\end{thebibliography}


\section{Supplementary material}

\begin{small}
\subsection{Proof of Proposition \ref{prop:3.1}}
\begin{proof}
By the midpoint relation in eq.(\ref{eq:2.2}) in the main document, $M_{j,\vec{k},n}$ is the intrinsic average over a grid of $n \lambda_2(I_{j,\vec{k}})/ \lambda_2(\mathcal{I})$ observations $(M_{J,\vec{k},n})_{\vec{k}}$ with $\bs{E}[\delta_R(M_{J,\vec{k},n},M_{J,\vec{k}})^2] = O(1)$, where $n = n_1n_2$ denotes the total number of observations, and we write $\vec{k}$ for the bivariate location vector $(k_1,k_2)$. By similar arguments as in the proof of \cite[Proposition 4.1]{CvS17}, it is easily verified that $\bs{E}[\delta_R(M_{j,\vec{k},n}, M_{j,\vec{k}})^2] = O(n^{-1}\lambda_2(\mathcal{I})/\lambda_2(I_{j,\vec{k}}))$ for each scale $j = 0,\ldots,J$ and location $(k_1,k_2)$. For notational convenience, in the remainder of this proof $\epsilon_{j,n}$ denotes a general (not necessarily the same) random error matrix that satisfies $\bs{E} \Vert \epsilon_{j,n} \Vert_F^2 = O(n^{-1}\lambda_2(\mathcal{I})/\lambda_2(I_{j,\vec{k}}))$, where we note that the rate on the right-hand side does not depend on the location $\vec{k}$ as the size $\lambda_2(I_{j,\vec{k}})$ does not depend on $\vec{k}$, see eq.(\ref{eq:3.1}) in the main document. Furthermore, we can appropriately write $M_{j,\vec{k},n} = \Exp_{M_{j,\vec{k}}}(\epsilon_{j,n})$ at the correct rate since, 
\begin{eqnarray*}
\bs{E}[\delta_R(\Exp_{M_{j,\vec{k}}}(\epsilon_{j,n}), M_{j,\vec{k}})^2] &=& \bs{E}\Vert \Log(M^{-1/2}_{j,\vec{k}} \ast \Exp_{M_{j,\vec{k}}}(\epsilon_{j,n})) \Vert_F^2 \nn
&=& \bs{E}\Vert M^{-1/2}_{j,\vec{k}} \ast \epsilon_{j,n} \Vert_F^2 \nn
&=& O(n^{-1}\lambda_2(\mathcal{I})/\lambda_2(I_{j,\vec{k}})),
\end{eqnarray*}
using the definitions of the Riemannian distance function and the logarithmic and exponential maps. In particular, by a first-order Taylor expansion of the matrix exponential, (abusing notation of $\epsilon_{j,n}$), $M_{j,\vec{k},n} = M^{1/2}_{j,\vec{k}} \ast \Exp(\epsilon_{j,n}) = M^{1/2}_{j,\vec{k}} \ast ( \te{Id} + \epsilon_{j,n} + \ldots) = M_{j,\vec{k}} + \epsilon_{j,n}$.\\[3mm]
As in Section \ref{sec:2.2} in the main document, the predicted midpoint $\widetilde{M}_{j,\vec{k},n}$ is a weighted intrinsic mean of $N_1N_2$ coarse-scale midpoints $(M_{j-1,\vec{k}',n})_{\vec{k}'}$ with weights summing up to 1. The rate of $\widetilde{M}_{j,\vec{k},n}$ is therefore upper bounded by the (worst) convergence rate of the individual midpoints $(M_{j-1,\vec{k}',n})_{\vec{k}'}$, i.e., the convergence rate of $\epsilon_{j-1,n}$. Hence, we can also write $\widetilde{M}_{j,\vec{k},n} = \widetilde{M}_{j,\vec{k}} + \epsilon_{j-1,n}$. \\[3mm]
Below, we use several implications verified in the proof of \cite[Proposition 4.3]{CvS17}. Let $M \in \mathcal{M}$ be a deterministic matrix and $\lambda E = O_p(\lambda)$ a random error matrix, such that $\bs{E}\Vert \lambda E \Vert_F = O(\lambda)$, then if $\lambda \to 0$ sufficiently small: (i) $\textnormal{Log}(M + \lambda E) \ =\ \textnormal{Log}(M) + O_p(\lambda)$, (ii) $(M + \lambda E)^{1/2} \ = \ M^{1/2} + O_p(\lambda)$ and (iii) $(M + \lambda E)^{-1} \ = \ M^{-1} + O_p(\lambda)$. Combining (ii) and (iii), it follows in particular also that $(M + \lambda E)^{-1/2} = M^{-1/2} + O_p(\lambda)$.\\
Let $j \ll J$ be sufficiently small, such that we can write $\epsilon_{j,n} = O_p(\lambda)$ with $\lambda \to 0$ sufficiently small and the above identities hold. In this case, we write out for the empirical whitened wavelet coefficient $\widehat{\mathfrak{D}}_{j,\vec{k},n}$, (abusing notation for $\epsilon_{j,n}$),
\begin{eqnarray*}
\widehat{\mathfrak{D}}_{j,\vec{k},n} &=& \sqrt{\frac{\lambda_2(I_{j,\vec{k}})}{\lambda_2(\mathcal{I})}} \, \Log\left( \widetilde{M}_{j,\vec{k},n}^{-1/2} \ast M_{j,\vec{k},n} \right)\nn
&=& \sqrt{\frac{\lambda_2(I_{j,\vec{k}})}{\lambda_2(\mathcal{I})}}\, \Log\left( (\widetilde{M}_{j,\vec{k}} + \epsilon_{j-1,n})^{-1/2} \ast (M_{j,\vec{k}} + \epsilon_{j,n}) \right) \nn
&=& \sqrt{\frac{\lambda_2(I_{j,\vec{k}})}{\lambda_2(\mathcal{I})}}\, \Log\left( (\widetilde{M}_{j,\vec{k}}^{-1/2} + \epsilon_{j-1,n}) \ast (M_{j,\vec{k}} + \epsilon_{j,n}) \right) \nn
&=& \sqrt{\frac{\lambda_2(I_{j,\vec{k}})}{\lambda_2(\mathcal{I})}}\, \Log\left( \widetilde{M}^{-1/2}_{j,\vec{k}} \ast M_{j,\vec{k}} + \epsilon_{j, n} + \ldots \right) \nn
&=& \sqrt{\frac{\lambda_2(I_{j,\vec{k}})}{\lambda_2(\mathcal{I})}}\, \Log\left( \widetilde{M}^{-1/2}_{j,\vec{k}} \ast M_{j,\vec{k}}\right) + \sqrt{\frac{\lambda_2(I_{j,\vec{k}})}{\lambda_2(\mathcal{I})}}\, O_p\Big(\sqrt{n^{-1}\lambda_2(\mathcal{I})/\lambda_2(I_{j,\vec{k}})}\Big) \nn
&=&  \mathfrak{D}_{j,\vec{k}} + O_p(\sqrt{n^{-1}}).
\end{eqnarray*}
Plugging in the above result, it follows that at scales $j \ll J$ sufficiently small,
\begin{eqnarray*} 
\bs{E}\Vert \widehat{\mathfrak{D}}_{j,\vec{k},n} - \mathfrak{D}_{j,\vec{k}} \Vert_F^2\ = \ O(n^{-1})\ = \ O((n_1n_2)^{-1}).
\end{eqnarray*}
\end{proof}

\subsection{Proof of Proposition \ref{prop:3.2}}
\begin{proof}
If $J_1 \neq J_2$ and $j \leq |J_1 - J_2|$ the 2D refinement scheme reduces to the 1D refinement scheme in \cite{CvS17} and the wavelet coefficient decay rates at scales $j$, with $j$ sufficiently large, are equivalent to the rates in \cite[Proposition 4.2]{CvS17}.\\[3mm]
It remains to derive the wavelet coefficient decay for $j > |J_1 - J_2|$. Let $(N_1,N_2) = (2L_1+1,2L_2+2)$, with $(L_1,L_2) \geq (0,0)$, and fix $j$ sufficiently large and a location $(k_1,k_2)$ at scale $j-1$ away from the boundary, such that the neighboring $(j-1)$-midpoints $N_1N_2$ closest neighboring $j-1$-scale midpoints $(M_{j-1,k_1 + \ell_1, k_2+\ell_2})_{\ell_1,\ell_2}$ to $M_{j-1,k_1,k_2}$, with $(\ell_1, \ell_2) \in \{-L_1,\ldots, L_1 \} \times \{-L_2, \ldots, L_2\}$ exist.
\begin{remark} 
For $(k_1,k_2)$ near the boundary, we collect the $N_1N_2$ available closest neighbors of $M_{j-1,k_1,k_2}$ (not necessarily symmetric). The remainder of the proof for the boundary case is exactly analogous to the non-boundary case and follows directly by mimicking the arguments outlined below.
\end{remark}
We predict $(\widetilde{M}_{j,i_1,i_2})_{i_1,i_2}$ at locations $(i_1,i_2)$ corresponding to the rectangles $I_{j,i_1,i_2} \subset I_{j-1,k_1,k_2}$ from $(M_{j-1,k_1 + \ell_1, k_2+\ell_2})_{\ell_1,\ell_2}$ via intrinsic polynomial interpolation of degree $(N_1-1, N_2-1)$ passing through the $N_1N_2$ points $(\widebar{M}_{j-1,r_1,r_2})_{r_1,r_2}$ with $(r_1,r_2) \in \{0,\ldots,N_1-1\} \times \{0,\ldots,N_2-1\}$, where $\widebar{M}_{j-1,r_1,r_2} = M_{t_0,s_0}(\max(I_{j-1,k_1-L_1+r_1,k_2-L_2+r_2}))$ is the cumulative intrinsic average according to eq.(\ref{eq:2.4}) in the main document. For notational simplicity, write $M(t,s) := M_{t_0,s_0}(t,s)$ and $\widehat{M}(t,s) := \widehat{M}_{t_0,s_0}(t,s)$ for the true and estimated intrinsic cumulative mean surfaces respectively, where the latter is an interpolating polynomial surface of order $(N_1-1,N_2-1)$ passing through a rectangular grid of $N_1N_2$ equidistant points $(x_{i_1},y_{i_2})_{i_1,i_2}$ with $i_1=0,\ldots,N_1-1$ and $i_2 = 0,\ldots,N_2-1$ on the surface $M(t,s)$. $M(t,s)$ itself is a smooth surface with existing partial covariant derivatives up to orders $N_1$ and $N_2$ in both marginal directions. The polynomial remainder of the interpolating polynomial surface is obtained by the following two-step argument:
\begin{enumerate}
\item To construct the interpolating polynomial surface, for each $x_{i_1}$, with $i_1 =0,\ldots,N_1-1$, we first fit marginal interpolating polynomial curves $\widehat{f}(s) = \widehat{M}_{x_{i_1}}(x_{i_1},s)$ through the $N_2$ equidistant points $(x_{i_1},y_{i_2})_{i_2=0,\ldots,N_2-1}$ on the smooth curve $f(s) = M(x_{i_1},s)$. By the same arguments as in the proof of \cite[Proposition 4.2]{CvS17}, it follows that for each $x_{i_1}$ and $j$ sufficiently large, the polynomial remainder of $\hat{g}(t)$ with respect to the smooth curve $g(t)$ is upper bounded by:
\begin{eqnarray} \label{C-eq:2.1}
\widehat{M}_{x_{i_1}}(x_{i_1},s) - M(x_{i_1},s) &=& O(|y_0-y_{N_2-1}|^{N_2}), \quad \forall\,s \in [y_0,y_{N_2-1}].
\end{eqnarray}
\item To evaluate $\widehat{M}(t,s)$ at $(t,s) \in [x_0,x_{N_1-1}]\times [y_0,y_{N_2-1}]$, we fit a marginal interpolating polynomial curve through the $N_1$ equidistant points $(x_{i_1},s)_{i_1=0,\ldots,N_1-1}$ on the estimated curves $\widehat{M}_{x_{i_1}}(t, s)$ obtained in step 1. As the interpolated polynomial curve $\widehat{M}(t,s)$ is a weighted intrinsic average of the estimates $\widehat{M}_{x_{i_1}}(x_{i_1},s)$, it follows by eq.(\ref{C-eq:2.1}) that for all $s \in [y_0,y_{N_2-1}]$,
\begin{eqnarray} \label{C-eq:2.2}
\widehat{M}(t,s) &=& \widehat{M}_{s}(t,s) + O(|y_0 - y_{N_2-1}|^{N_2}),
\end{eqnarray}
where $\widehat{g}(t) = \widehat{M}_{s}(t,s)$ is the interpolating polynomial evaluated at $(t,s)$ through the $N_1$ equidistant points $(x_{i_1},s)_{i_1=0,\ldots,N_1-1}$ on the true smooth curve $g(t) = M(t,s)$. Exactly analogous to step 1, for $j$ sufficiently large, the polynomial remainder with respect to the smooth curve is upper bounded by:
\begin{eqnarray*}
\widehat{M}_{s}(t,s) - M(t,s) = O(|x_0 - x_{N_1-1}|^{N_1}), \quad \forall\, t \in [x_0,x_{N_1-1}].
\end{eqnarray*}
Plugging this back into eq.(\ref{C-eq:2.2}) yields $\forall\, (t,s) \in [x_0,x_{N_1-1}] \times [y_0,y_{N_2-1}]$,
\begin{eqnarray*} 
\widehat{M}(t,s) - M(t,s) &=& O(|x_0 - x_{N_1-1}|^{N_1}) + O(|y_0 - y_{N_2-1}|^{N_2}) \nn
&=& O(2^{(-j+J-J_1)N_1} \vee 2^{(-j+J-J_2)N_2}).
\end{eqnarray*}
\end{enumerate}
The final step follows by the fact that, given the dyadic refinement pyramid, by eq.(\ref{eq:3.1}) in the main document, we can bound $|x_0 - x_{N_1-1}| \lesssim \lambda_{2,x}(I_{j,k_1,k_2}) \leq 2^{-j + J - J_1}$ and $|y_0 - y_{N_2-1}| \lesssim \lambda_{2,y}(I_{j,k_1,k_2}) \leq 2^{-j + J - J_2}$, with $J = J_1 \vee J_2$. Here, $\lambda_{2,x}(I_{j,k_1,k_2})$ and $\lambda_{2,y}(I_{j,k_1,k_2})$ respectively denote the width and height of the rectangle $I_{j,k_1,k_2}$, such that $\lambda_2(I_{j,k_1,k_2}) = \lambda_{2,x}(I_{j,k_1,k_2}) \lambda_{2,y}(I_{j,k_1,k_2})$. \\[3mm]
The predicted midpoints $(\widetilde{M}_{j,i_1,i_2})_{i_1,i_2}$ at locations $(i_1,i_2)$ corresponding to the rectangles $I_{j,i_1,i_2} \subset I_{j-1,k_1,k_2}$ are obtained from the known cumulative intrinsic averages $(\widebar{M}_{j-1,r_1,r_2})_{r_1,r_2}$ and the interpolated polynomial $\widehat{M}(t,s)$ evaluated at different locations $(t,s) \in [x_0,x_{N_1-1}] \times [y_0,y_{N_2-1}]$ through several applications of the following two operations (see the prediction equations in Section \ref{C-sec:5} below): 
\begin{enumerate}
\item[(i)] Given $x,y \in \mathcal{M}$, compute $z = \te{Ave}(\{x,y\} ; \{w,1-w\})$ with $w \in [0, 1]$ or $w \in [2, \infty) \cap \mathbb{Z}$.\\[3mm]
The two-point weighted intrinsic mean above can be rewritten as a point on the geodesic segment connecting $x$ and $y$, (see eq.(\ref{eq:2.1}) in the main document):
\begin{eqnarray*}
z \ =\ \eta(x,y,1-w) \ = \ x^{1/2} \ast (x^{-1/2} \ast y)^{1-w}.
\end{eqnarray*}
Suppose that the inputs are given by $x_{\Lambda} = x + \Lambda$ and $y_{\Lambda} = y + \Lambda$, where $\Lambda$ is some arbitrary (not necessarily fixed) error matrix, such that $\Vert \Lambda\Vert_F =  O(\lambda)$, and $\lambda \to 0$ is small. By slight abuse of notation, we also write $\Lambda = O(\lambda)$. Denote $v := 1-w \in [0,1] \cup ([-1, -\infty) \cap \mathbb{Z})$. For fixed $v$, as in the proof of \cite[Proposition 4.2]{CvS17}, it follows that for $x \in \mathcal{M}$,
\begin{eqnarray*}
(x + \Lambda)^v & = & \Exp( v \Log(x + \Lambda) ) \ = \ \sum_{k=0}^\infty \frac{(v\Log(x + \Lambda))^k}{k!} \ =\ \sum_{k=0}^\infty \frac{v^k (\Log(x) + O(\lambda))^k}{k!} \nn
& = & \sum_{k=0}^\infty \frac{v^k (\Log(x))^k}{k!} + O(\lambda)\sum_{k=0}^\infty \frac{v^k}{k!} \ = \ x^v + O(\lambda),
\end{eqnarray*}
where we used that $\Log(x + \Lambda) = \Log(x) + O(\lambda)$ for $\lambda \to 0$ sufficiently small, as verified in the proof of \cite[Proposition 4.3]{CvS17} and the fact that $|\sum_{k=0}^\infty v^k / k!| = |\exp(v)| < \infty$. \\[3mm]
Using the above result and the results in the proof of \cite[Proposition 4.3]{CvS17}, in both cases $w \in [0,1]$ or $w \in [2,\infty) \cap \mathbb{Z}$, the weighted intrinsic mean $\tilde{z}$ obtained from the inputs $x_{\Lambda}$ and $y_{\Lambda}$ satisfies, (abusing notation for $\Lambda$):
\begin{eqnarray} \label{C-eq:2.3}
\tilde{z} &=& x_{\Lambda}^{1/2} \ast(x_{\Lambda}^{-1/2} \ast y_{\Lambda})^{1-w} \nn
&=& (x + \Lambda)^{1/2} \ast ((x + \Lambda)^{-1/2} \ast (y + \Lambda))^{1-w} \nn
&=& (x + \Lambda)^{1/2} \ast ((x^{-1/2} \ast y) + \Lambda)^{1-w} \nn
&=& (x^{1/2} + \Lambda) \ast ((x^{-1/2} \ast y)^{1-w} + \Lambda) \nn
&=& z + O(\lambda),
\end{eqnarray}
where $z$ is the weighted intrinsic mean obtained from the inputs $x$ and $y$.
\item[(ii)] Given $x,y_1,y_2,y_3 \in \mathcal{M}$, compute $z = \Exp_x(-\sum_{i=1}^3 \Log_x(y_i))$.\\[3mm]
Suppose again that the inputs are given by $x_\Lambda = x+\Lambda$ and $y_{i,\Lambda} = y_i + \Lambda$ for $i = 1,2,3$, where $\Lambda$ is some arbitrary, (not necessarily fixed) error matrix, such that $\Vert \Lambda \Vert_F = O(\lambda)$, and $\lambda \to 0$ small. By slight abuse of notation, we also write $\Lambda = O(\lambda)$.
First, observe that:
\begin{eqnarray*}
\Exp(x + \Lambda) \ = \ \sum_{k=0}^\infty \frac{(x + \Lambda)^k}{k!} \ = \ \sum_{k=0}^\infty \frac{x^k + O(\lambda)}{k!} \ = \ \Exp(x) + O(\lambda).
\end{eqnarray*}
By the above result combined with the results in the proof of \cite[Proposition 4.3]{CvS17}, abusing notation for $\Lambda$, the output $\tilde{z}$ obtained from the inputs $x_{\Lambda}$ and $y_{1,\Lambda}$, $y_{2,\Lambda}$, $y_{3,\Lambda}$, satisfies:
\begin{eqnarray} \label{C-eq:2.4}
\tilde{z} &=& \Exp_{x_\Lambda}\bigg( - \sum_{i=1}^3 \Log_{x_{\Lambda}}(y_{i,\Lambda}) \bigg) \nn
&=& (x + \Lambda)^{1/2} \ast \Exp\bigg( - \sum_{i=1}^3 \Log\Big((x + \Lambda)^{-1/2} \ast (y_i + \Lambda)\Big) \bigg) \nn
&=& (x^{1/2} + \Lambda) \ast \Exp\bigg( -\sum_{i=1}^3 \Log\Big( x^{-1/2} \ast y_i + \Lambda \Big) \bigg) \nn
&=& (x^{1/2} + \Lambda) \ast \Exp\bigg( -\sum_{i=1}^3 \Log\Big(x^{-1/2} \ast y_i \Big) + \Lambda \bigg) \nn
&=& (x^{1/2} + \Lambda) \ast \left[\Exp\bigg( -\sum_{i=1}^3 \Log\Big(x^{-1/2} \ast y_i \Big) \bigg) + \Lambda \right] \nn
&=& z + O(\lambda),
\end{eqnarray}
where $z$ is the output obtained from the inputs $x$ and $y_1$, $y_2$ and $y_3$.
\end{enumerate}
To predict the midpoints $(\widetilde{M}_{j,i_1,i_2})_{i_1,i_2}$, we start with as input the cumulative intrinsic averages cumulative intrinsic averages $(\widebar{M}_{j-1,r_1,r_2})_{r_1,r_2}$ and the interpolated polynomial $\widehat{M}(t,s)$ evaluated at $(t,s) \in [x_0,x_{N_1-1}] \times [y_0,y_{N_2-1}]$. The cumulative intrinsic averages are error-free as they are obtained from the true surface $M(t,s)$. The error for the interpolated polynomial $\widehat{M}(t,s)$ for each $(t,s) \in [x_0,x_{N_1-1}] \times [y_0,y_{N_2-1}]$ satisfies $\Lambda = O(\lambda) = O(2^{(-j + J - J_1)N_1} \vee 2^{(-j+J-J_2)N_2})$, such that $\lambda \to 0$ as $j \to \infty$. Hence, for $j$ sufficiently large, after each successive application of one of the above operations, the error rate remains of the same order $O(2^{(-j + J - J_1)N_1} \vee 2^{(-j+J-J_2)N_2})$ by eq.(\ref{C-eq:2.3}) and eq.(\ref{C-eq:2.4}) respectively. As a consequence, it follows that the error in the predicted midpoints is of the same rate as that of the interpolated polynomial, i.e.,
\begin{eqnarray*}
\widetilde{M}_{j,i_1,i_2} &=& M_{j,i_1,i_2} + O(2^{(-j + J - J_1)N_1} \vee 2^{(-j+J-J_2)N_2}), \quad \forall\, (i_1,i_2)\ \te{at scale } j.
\end{eqnarray*}
Substituting the above result in the whitened wavelet coefficient, by similar arguments as used in (i) and (ii) above (and rewriting $J-(J_1+J_2)/2 = |J_1-J_2|/2$), it follows that for $j > |J_1 - J_2|$ sufficiently large,
\begin{eqnarray*}
\Vert \mathfrak{D}_{j,k_1,k_2} \Vert_F &=& \Bigg\Vert \sqrt{\frac{\lambda_2(I_{j,k_1,k_2})}{\lambda_2(\mathcal{I})}}\, \Log\big((M_{j,k_1,k_2} + \Lambda)^{-1/2} \ast M_{j,k_1,k_2} \big) \Bigg\Vert_F\nn 
&=& 2^{-j+J-(J_1+J_2)/2} \Big\Vert \Log\big((M_{j,k_1,k_2}^{-1/2} + \Lambda) \ast M_{j,k_1,k_2}\big) \Big\Vert_F\nn
&=& 2^{-j+|J_1-J_2|/2} \big\Vert \Log\big(\te{Id} + \Lambda \big) \big\Vert_F\nn
& \lesssim &  2^{-j+|J_1-J_2|/2} \big(2^{(-j + J - J_1)N_1} \vee 2^{(-j+J-J_2)N_2}\big),
\end{eqnarray*}
where in the final step we expanded $\Log(\te{Id} + \Lambda) =  O\big(2^{(-j + J - J_1)N_1} \vee 2^{(-j+J-J_2)N_2}\big)$ via its Mercator series (see \cite[Section 11.3]{H08}), using that the spectral radius of $\Lambda$ is smaller than 1 for $j$ sufficiently large. 
\end{proof}

\subsection{Proof of Theorem \ref{thm:3.3}}
\begin{proof} First, we verify that, by construction, $J_0 > |J_1 - J_2|$, (independent of $n = 2^{J_1 + J_2}$). This follows by substituting $J_0 = (\log_2(n) + |J_1-J_2|(1 + 2(N_1 \vee N_2))/(2 + 2(N_1 \wedge N_2))$, and using that $\log_2(n) = J_1 + J_2 > |J_1-J_2|$, (which immediately follows by rewriting $|J_1 - J_2| = 2J - (J_1 + J_2)$ with $J = J_1 \vee J_2$ and using that $J_1 + J_2 > J$), we write out,
\begin{eqnarray*}
J_0 \ > \ & \dfrac{|J_1 - J_2| + |J_1 - J_2|(1+2(N_1 \vee N_2))}{2+2(N_1 \wedge N_2)} \ = \dfrac{2 + 2(N_1 \vee N_2)}{2 + 2(N_1 \wedge N_2)} |J_1 - J_2| \ \geq \ |J_1 - J_2|,
\end{eqnarray*}
where in the last step we used that $(N_1 \vee N_2) \geq (N_1 \wedge N_2)$, since $N_1, N_2 \geq 1$. From the definition of $J_0$, we also immediately observe that $J_0 \to \infty$ as $n \to \infty$.\\[3mm]
For the first part of the theorem, suppose that $n = n_1n_2$ is such that $J_0$ is sufficiently large for the rates in Proposition \ref{prop:3.1} to hold for $j < J_0$ and the rates in Proposition \ref{prop:3.2} to hold for $j \geq J_0 > |J_1 - J_2|$. \\
We write out the total mean squared error in terms of the whitened coefficients, with $\vec{k} = (k_1,k_2)$, as:
\begin{eqnarray} \label{C-eq:3.1}
\sum_{j,\vec{k}} \bs{E}\Vert \widehat{\mathfrak{D}}_{j,\vec{k}} - \mathfrak{D}_{j,\vec{k}} \Vert_F^2 &=& \sum_{j \geq J_0} \sum_{\vec{k}} \Vert \mathfrak{D}_{j,\vec{k}}\Vert_F^2 + \sum_{j < J_0} \sum_{\vec{k}} \bs{E}\Vert \widehat{\mathfrak{D}}_{j,\vec{k}} - \mathfrak{D}_{j,\vec{k}} \Vert_F^2\nn
&\lesssim & \sum_{j \geq J_0} 2^{j - J + J_1}2^{j-J+J_2} \big(2^{-2j+|J_1-J_2|} \big(2^{2(-j + J - J_1)N_1} \vee 2^{2(-j+J-J_2)N_2}\big)\big) \nn
&& + \sum_{j < J_0} (2^{j-J+J_1} \vee 1)(2^{j-J+J_2} \vee 1)\, n^{-1} \nn
&=& \sum_{j \geq J_0} 2^{2(-j + J - J_1)N_1} \vee 2^{2(-j+J-J_2)N_2} \nn
&& + n^{-1}\sum_{j < J_0} (2^{j-J+J_1} \vee 1)(2^{j-J+J_2} \vee 1), 
\end{eqnarray}
where we used that $2J - (J_1 + J_2) = |J_1 - J_2|$, since $J = J_1 \vee J_2$. For the first sum on the right-hand side in eq.(\ref{C-eq:3.1}) above, using again that $2J - (J_1 + J_2) = |J_1 - J_2|$ and the fact that $(N_1 \wedge N_2) \geq 1$, we upper bound: 
\begin{footnotesize}
\begin{eqnarray} \label{C-eq:3.2}
\sum_{j \geq J_0} 2^{2(-j + J - J_1)N_1} \vee 2^{2(-j+J-J_2)N_2} &\lesssim & 2^{2|J_1 - J_2| (N_1 \vee N_2)} \sum_{J_0 \leq j \leq J} 2^{-2j(N_1 \wedge N_2)} \nn
& \lesssim & 2^{2|J_1 - J_2| (N_1 \vee N_2)}2^{-2(N_1 \wedge N_2)J_0}.
\end{eqnarray}
\end{footnotesize}
For the second sum on the right-hand side in eq.(\ref{C-eq:3.1}) above, we upper bound: 
\begin{footnotesize}
\begin{eqnarray} \label{C-eq:3.3}
n^{-1}\sum_{j < J_0} (2^{j-J+J_1} \vee 1)(2^{j-J+J_2} \vee 1) &=& n^{-1} \Big( \sum_{j < |J_1 - J_2|} 2^j \Big) - n^{-1} 2^{-|J_1-J_2|} \Big(\sum_{|J_1 - J_2| \leq j < J_0} 2^{2j} \Big) \nn
&=& n^{-1} \Bigg( \frac{1-2^{|J_1-J_2|}}{1-2} - 1\Bigg) - n^{-1}2^{-|J_1-J_2|}\Bigg(\frac{2^{2|J_1-J_2|} - 2^{2J_0}}{1-2^2} \Bigg) \nn
&\lesssim & n^{-1}2^{-|J_1-J_2|}2^{2J_0},
\end{eqnarray}
\end{footnotesize}
where in the second step, the first term on the right-hand side is dominated by the second term since $J_0 > |J_1-J_2|$. Plugging the bounds in eq.(\ref{C-eq:3.2}) and eq.(\ref{C-eq:3.3}) back into the right-hand side of eq.(\ref{C-eq:3.1}), completes the first part of the theorem:
\begin{eqnarray} \label{C-eq:3.4}
\sum_{j,\vec{k}} \bs{E}\Vert \widehat{\mathfrak{D}}_{j,\vec{k}} - \mathfrak{D}_{j,\vec{k}} \Vert_F^2 &\lesssim & 2^{2|J_1 - J_2| (N_1 \vee N_2)}2^{-2(N_1 \wedge N_2)J_0} + n^{-1}2^{-|J_1-J_2|}2^{2J_0} \nn
& \lesssim & 2^{\frac{2(N_1\vee N_2)}{1+(N_1\wedge N_2)}|J_1-J_2|}\,(n_1 \vee n_2)^{\frac{-2(N_1 \wedge N_2)}{1 + (N_1 \wedge N_2)}}.
\end{eqnarray}
The second step in the above equation follows by substituting $J_0 = (\log_2(n)+|J_1-J_2|(1+2(N_1\vee N_2))/(2 + 2(N_1 \wedge N_2))$, since for the first term on the right-hand side in eq.(\ref{C-eq:3.4}),
\begin{footnotesize}
\begin{eqnarray*}
2^{2|J_1-J_2|(N_1 \vee N_2)}2^{-2(N_1\wedge N_2)J_0} & = & \exp\bigg[ 2|J_1-J_2|(N_1\vee N_2)\log(2) - \ 2(N_1\wedge N_2)\log(2) J_0 \bigg] \nn
&=& \exp\Bigg[ 2\log(2)|J_1-J_2|(N_1 \vee N_2)\left(1 - \frac{(N_1 \wedge N_2)}{1 + (N_1 \wedge N_2)} \right) \nn
&& -\ (J_1+J_2)\log(2)\frac{(N_1 \wedge N_2)}{1 + (N_1 \wedge N_2)} - \log(2)\frac{|J_1-J_2|(N_1 \wedge N_2)}{1 + (N_1 \wedge N_2)} \Bigg] \nn
&=& 2^{\frac{2(N_1\vee N_2)}{1+(N_1\wedge N_2)}|J_1-J_2|}\, \exp\left[ -\log(2)\frac{(N_1\wedge N_2)}{1+(N_1 \wedge N_2)} (J_1 + J_2 + |J_1-J_2|) \right] \nn
&=& 2^{\frac{2(N_1\vee N_2)}{1+(N_1\wedge N_2)}|J_1-J_2|}\,(n_1 \vee n_2)^{\frac{-2(N_1 \wedge N_2)}{1 + (N_1 \wedge N_2)}}, 
\end{eqnarray*}
\end{footnotesize}
where in the final step we used that $J_1 + J_2 + |J_1 - J_2| = 2J = 2\log_2(n_1 \vee n_2)$. Also, for the second term on the right-hand side in eq.(\ref{C-eq:3.4}), plugging in $J_0$ yields,
\begin{footnotesize}
\begin{eqnarray*}
n^{-1}2^{-|J_1-J_2|}2^{2J_0} &\lesssim & \exp\Bigg[ -(J_1+J_2)\log(2) - |J_1-J_2|\log(2) \nn
&& +\ 2\log(2)\Bigg(\frac{\log(n)}{2\log(2)(1+N_1\wedge N_2)} + \frac{|J_1-J_2|(1+2(N_1\vee N_2))}{2(1+(N_1\wedge N_2))}\Bigg)\Bigg] \nn
&=&  \exp\Bigg[-(J_1+J_2)\log(2)\frac{(N_1 \wedge N_2)}{1+(N_1\wedge N_2)} + 2\log(2)\frac{|J_1-J_2|(N_1\vee N_2)}{1+(N_1 \wedge N_2)} \nn
&& -\ \log(2)\frac{|J_1-J_2|(N_1 \wedge N_2)}{1 + (N_1 \wedge N_2)}\Bigg] \nn
&=& 2^{\frac{2(N_1\vee N_2)}{1+(N_1\wedge N_2)}|J_1-J_2|}\, \exp\left[ -\log(2)\frac{(N_1\wedge N_2)}{1+(N_1 \wedge N_2)} (J_1 + J_2 + |J_1-J_2|) \right] \nn
&=& 2^{\frac{2(N_1\vee N_2)}{1+(N_1\wedge N_2)}|J_1-J_2|}\,(n_1 \vee n_2)^{\frac{-2(N_1 \wedge N_2)}{1 + (N_1 \wedge N_2)}}.
\end{eqnarray*}
\end{footnotesize}
For the second part of the theorem, suppose again that $n$ is such that $J_0$ is sufficiently large for the rates in Proposition \ref{prop:3.1} to hold for $j < J_0$ and the rates in Proposition \ref{prop:3.2} to hold for $j \geq J_0 > |J_1 - J_2|$.\\
If for each $\vec{k} = (k_1,k_2) \in \{0,\ldots,n_1-1 \} \times \{0,\ldots,n_2-1 \}$, we can verify that $\bs{E}[\delta_R(M_{J,\vec{k}}, \widehat{M}_{J,\vec{k},n})^2] \lesssim 2^{\frac{2(N_1\vee N_2)}{1+(N_1\wedge N_2)}|J_1-J_2|}\,(n_1 \vee n_2)^{\frac{-2(N_1 \wedge N_2)}{1 + (N_1 \wedge N_2)}}$ the proof is finished.\\[3mm]
At scales $j=1,\ldots,J$, based on the estimated midpoints $(\widehat{M}_{j-1,\ell_1,\ell_2,n})_{\ell_1,\ell_2}$ and the estimated wavelet coefficient $\widehat{D}_{j,k_1,k_2,n}$, in the inverse wavelet transform, the finer-scale midpoint $\widehat{M}_{j,k_1,k_2,n}$ is estimated through,
\begin{eqnarray*}
\widehat{M}_{j,k_1,k_2,n} &=& \Exp_{\widehat{\widetilde{M}}_{j,k_1,k_2,n}}\left( \lambda_2(I_{j,k_1,k_2})^{-1/2} \widehat{D}_{j,k_1,k_2,n} \right),
\end{eqnarray*}
where $\widehat{\widetilde{M}}_{j,k_1,k_2,n}$ is the predicted midpoint at scale-location $(j,k_1,k_2)$ based on $(\widehat{M}_{j-1,\ell_1,\ell_2,n})_{\ell_1,\ell_2}$. In particular, at scale $j = 1$, we have that $\widehat{\widetilde{M}}_{1,\vec{k},n} = \widetilde{M}_{1,\vec{k},n}$, as the estimated coarsest midpoint $\widehat{M}_{0,\vec{0},n}$ corresponds to the empirical coarsest midpoint $M_{0,\vec{0},n}$.\\[3mm]
At scales $j = 1,\ldots, J_0 - 1$, we do not alter the wavelet coefficients, and since $n$ is assumed to be sufficiently large such that the rate in Proposition \ref{prop:3.1} holds, we can write $\widehat{\mathfrak{D}}_{j,\vec{k},n} = \mathfrak{D}_{j,\vec{k}} + \eta_n$, where $\eta_n$ denotes a general (not always the same) random errror matrix satisfying $\bs{E}\Vert \eta_n \Vert_F = O(n^{-1/2})$. Also, as in the proof of Proposition \ref{prop:3.1}, we can write $\widetilde{M}_{j,\vec{k},n} = \widetilde{M}_{j,\vec{k}} + \epsilon_{j,n}$, where $\epsilon_{j,n}$ is a general (not always the same) random error matrix satisfying $\bs{E}\Vert \epsilon_{j,n} \Vert_F = O(n^{-1/2}\lambda_2(I_{j,\vec{k}})^{-1/2})$. \\[3mm]
In particular, at scale $j = 1$,
\begin{eqnarray} \label{C-eq:3.5}
\widehat{M}_{1,\vec{k},n} &=& \Exp_{\widehat{\widetilde{M}}_{1,\vec{k},n}}\big( \lambda_2(I_{1,\vec{k}})^{-1/2} \widehat{D}_{1,\vec{k},n} \big) \nn
&=& \widetilde{M}_{1,\vec{k},n}^{1/2} \ast \Exp\big(\lambda_2(I_{1,\vec{k}})^{-1/2} \widetilde{M}_{1,\vec{k},n}^{-1/2} \ast \widehat{D}_{1,\vec{k},n}\big) \nn
&=& \widetilde{M}_{1,\vec{k},n}^{1/2} \ast \Exp\big( \lambda_2(I_{1,\vec{k}})^{-1/2} \widehat{\mathfrak{D}}_{1,\vec{k},n} \big) \nn
&=&\left(\widetilde{M}_{1,\vec{k}} + \epsilon_{1,n} \right)^{1/2} \ast \Exp\left( \lambda_2(I_{1,\vec{k}})^{-1/2} \big( \mathfrak{D}_{1,\vec{k}} + \eta_n \big) \right) \nn
&=& \left(\widetilde{M}_{1,\vec{k}}^{1/2} + \epsilon_{1,n} \right) \ast \left(\Exp(\lambda_2(I_{1,\vec{k}})^{-1/2} \mathfrak{D}_{1,\vec{k}}) + \lambda_2(I_{1,\vec{k}})^{-1/2} \eta_n \right) \nn
&=& M_{1,\vec{k}} + O_p\big(n^{-1/2}\lambda_2(I_{1,\vec{k}})^{-1/2}\big).
\end{eqnarray}
Here, we used that $(M + \lambda E)^{1/2} = M^{1/2} + O_p(\lambda)$ for $\lambda \to 0$ sufficiently small as in the proof of \cite[Proposition 4.3]{CvS17}, and $\Exp(D + \Lambda) = \Exp(D) + O_p(\lambda)$ for a random error matrix $\Lambda = O_p(\lambda)$, with $\lambda \to 0$ sufficiently small, as in the proof of Proposition \ref{prop:3.2}. As $\widehat{\widetilde{M}}_{2,\vec{k},n}$ is a weighted intrinsic average of $(\widehat{M}_{1, \vec{k}',n})_{\vec{k}'}$, the error rate of $\widehat{\widetilde{M}}_{2,\vec{k},n}$ is upper bounded by the (worst) convergence rate of the $(\widehat{M}_{1,\vec{k}',n})_{\vec{k}'}$, and we can surely write $\widehat{\widetilde{M}}_{2,\vec{k},n} = \widetilde{M}_{2,\vec{k}} + \epsilon_{2,n}$.\\[3mm]
Iterating this same argument for each scale $j = 2,\ldots,J_0-1$, using that by construction $\lambda_2(I_{j,\vec{k}})^{-1} = (2^{j-J+J_1} \vee 1)(2^{j - J + J_2} \vee 1)$, we find that:
\begin{eqnarray} \label{C-eq:3.6}
\widehat{M}_{J_0-1,\vec{k},n} &=& M_{J_0-1,\vec{k}} + \sum_{j<J_0} O_p\big(n^{-1/2} \lambda_2(I_{j,\vec{k}})^{-1/2}\big) \nn
&=& M_{J_0-1,\vec{k}} + \sum_{j<J_0} O_p\big( \big( n^{-1} (2^{j-J+J_1} \vee 1)(2^{j-J+J_2} \vee 1)\big)^{1/2} \big) \nn
&=& M_{J_0-1,\vec{k}} + O_p\big( n^{-1/2} 2^{-|J_1-J_2|/2}2^{J_0} \big),
\end{eqnarray}
where the final step follows in the same fashion as in eq.(\ref{C-eq:3.3}) above. In particular, by the same argument as above, we can write, 
\begin{eqnarray*}
\widehat{\widetilde{M}}_{J_0, \vec{k}, n} \ = \ \widetilde{M}_{J_0, \vec{k}} + \epsilon_{J_0,n} \ =\ \widetilde{M}_{J_0,\vec{k}} + O_p\big( n^{-1/2} 2^{-|J_1-J_2|/2}2^{J_0} \big),
\end{eqnarray*}
by definition of the random error matrix $\epsilon_{J_0,n}$, using that $\lambda_2(I_{J_0,\vec{k}})^{-1} = 2^{2J_0-|J_1-J_2|}$.
\\[3mm]
At scales $j = J_0,\ldots,J$, we set $\widehat{D}_{j,\vec{k},n} = \bs{0}$ at each location $\vec{k}=(k_1,k_2)$. Since $n$ is assumed to be sufficiently large such that the rate in Proposition \ref{prop:3.2} holds for $j \geq J_0 > |J_1 - J_2|$, we can write $\widehat{D}_{j,\vec{k},n} = \bs{0} = \mathfrak{D}_{j,\vec{k}} + \zeta_{j,N_1,N_2}$, with $\zeta_{j,N_1,N_2}$ a general (not always the same) deterministic error matrix satisfying $\Vert \zeta_{j,N_1,N_2} \Vert_F = O\left( 2^{-j+|J_1-J_2|/2} \big(2^{(-j + J - J_1)N_1} \vee 2^{(-j+J-J_2)N_2}\big)\right)$. \\[3mm]
In particular, at scale $j = J_0$, using the result in eq.(\ref{C-eq:3.6}) above and the fact that $\lambda_2(I_{J_0,\vec{k}})^{-1} = 2^{2J_0-|J_1-J_2|}$,
\begin{footnotesize}
\begin{eqnarray*}
\widehat{M}_{J_0,\vec{k},n} &=& \Exp_{\widehat{\widetilde{M}}_{J_0,\vec{k},n}}\big(\lambda_2(I_{J_0,\vec{k}})^{-1/2} \widehat{D}_{J_0,\vec{k},n} \big) \nn
&=& \big( \widetilde{M}_{J_0,\vec{k}} + \epsilon_{J_0, n} \big)^{1/2} \ast \Exp\left( \big(\widetilde{M}_{J_0,\vec{k}} + \epsilon_{J_0,n}\big)^{-1/2} \ast \lambda_2(I_{J_0,\vec{k}})^{-1/2} \big(\mathfrak{D}_{J_0,\vec{k}} + \zeta_{J_0, N_1,N_2} \big)\right) \nn
&=& \big( \widetilde{M}_{J_0,\vec{k}}^{1/2} + \epsilon_{J_0,n} \big) \ast \Exp\left( \big(\widetilde{M}_{J_0,\vec{k}}^{-1/2} + \epsilon_{J_0,n}\big) \ast \big( \lambda_2(I_{j,\vec{k}})^{-1/2} \mathfrak{D}_{J_0,\vec{k}} + \lambda_2(I_{J_0,\vec{k}})^{-1/2}\zeta_{J_0, N_1,N_2} \big)\right) \nn
&\lesssim & \left( \widetilde{M}_{J_0,\vec{k}}^{1/2} + \epsilon_{J_0,n} \right) \ast \left( \Exp(\lambda_2(I_{J_0,\vec{k}})^{-1/2}D_{J_0,\vec{k}}) + \lambda_2(I_{J_0,\vec{k}})^{-1/2} \epsilon_{J_0,n} \mathfrak{D}_{J_0,\vec{k}} + \lambda_2(I_{J_0,\vec{k}})^{-1/2} \zeta_{J_0,N_1,N_2} \right) \nn
&=& \left( \widetilde{M}_{J_0,\vec{k}}^{1/2} + \epsilon_{J_0,n} \right) \ast \left( \Exp(\lambda_2(I_{J_0,\vec{k}})^{-1/2}D_{J_0,\vec{k}}) + O_p\big(2^{(-j + J - J_1)N_1} \vee 2^{(-j+J-J_2)N_2}\big) \right) \nn
&=& M_{J_0,\vec{k}} +  O_p\big( n^{-1/2} 2^{-|J_1-J_2|/2}2^{J_0} \big) + O_p\big(2^{(-j + J - J_1)N_1} \vee 2^{(-j+J-J_2)N_2}\big),
\end{eqnarray*}
\end{footnotesize}
which follows in the same way as in eq.(\ref{C-eq:3.5}) above, combined with the bound:
\begin{eqnarray*}
\lambda_2(I_{J_0,\vec{k}})^{-1/2} \epsilon_{J_0,n} \mathfrak{D}_{J_0,\vec{k}} \ \lesssim \ \lambda_2(I_{J_0,\vec{k}})^{-1/2} \mathfrak{D}_{J_0,\vec{k}} \ = \ O_p\big(2^{(-j + J - J_1)N_1} \vee 2^{(-j+J-J_2)N_2}\big),
\end{eqnarray*}
since $\Vert \lambda_2(I_{j,\vec{k}})^{-1/2} \mathfrak{D}_{J_0,\vec{k}}\Vert_F \sim \Vert \lambda_2(I_{j,\vec{k}})^{-1/2} \zeta_{J_0,N_1,N_2} \Vert_F = O\big(2^{(-j + J - J_1)N_1} \vee 2^{(-j+J-J_2)N_2}\big)$ by Proposition \ref{prop:3.2}. Iterating this same argument for each scale $j = J_0+1,\ldots,J$ yields, 
\begin{eqnarray*}
\widehat{M}_{J,\vec{k},n} &=& M_{J,\vec{k}} + O_p\big( n^{-1/2} 2^{-|J_1-J_2|/2}2^{J_0} \big) + \sum_{j=J_0}^{J} O_p\big(2^{(-j + J - J_1)N_1} \vee 2^{(-j+J-J_2)N_2}\big) \nn
&=& M_{J,\vec{k}} + O_p\big( n^{-1/2} 2^{-|J_1-J_2|/2}2^{J_0} \big) + O_p\big(2^{|J_1 - J_2| (N_1 \vee N_2)}2^{-(N_1 \wedge N_2)J_0}\big).
\end{eqnarray*}
The second step follows by the same arguments as in as in eq.(\ref{C-eq:3.2}) above. Plugging in $J_0 = (\log_2(n)+|J_1-J_2|(1+2(N_1\vee N_2))/(2 + 2(N_1 \wedge N_2))$, as previously demonstrated (following eq.(\ref{C-eq:3.4})), the above expression reduces to: 
\begin{eqnarray*}
\widehat{M}_{J,\vec{k},n} &=& M_{J,\vec{k}} + O_p\bigg(2^{\frac{(N_1\vee N_2)}{1+(N_1\wedge N_2)}|J_1-J_2|}\,(n_1 \vee n_2)^{\frac{-(N_1 \wedge N_2)}{1 + (N_1 \wedge N_2)}} \bigg), \quad \quad \te{for each } \vec{k} = (k_1,k_2).
\end{eqnarray*}
For notational convenience, denote by $\xi_{n,N_1,N_2}$ a general (not always the same)  random error matrix such that $\bs{E}\Vert \xi_{n,N_1,N_2} \Vert_F \lesssim 2^{\frac{(N_1\vee N_2)}{1+(N_1\wedge N_2)}|J_1-J_2|}\,(n_1 \vee n_2)^{\frac{-(N_1 \wedge N_2)}{1 + (N_1 \wedge N_2)}}$.  For each $\vec{k} \in \{0,\ldots,n_1-1 \} \times \{0, \ldots, n_2-1 \}$, by the above result:
\begin{eqnarray*}
\bs{E}\left[\delta_R(M_{J,\vec{k}}, \widehat{M}_{J,\vec{k},n})^2\right] &=& \bs{E}\left[\delta_R\big(M_{J,\vec{k}}, M_{J,\vec{k}} + \xi_{n,N_1,N_2} \big)^2\right] \nn
&=& \bs{E}\left\Vert \Log \left( M_{J,\vec{k}}^{1/2} \ast \big(M_{J,\vec{k}} + \xi_{n,N_1,N_2} \big)\right) \right\Vert_F^2 \nn
&=& \bs{E}\left\Vert \Log \big( \te{Id} + \xi_{n,N_1,N_2} \big) \right\Vert_F^2 \nn
& \lesssim & 2^{\frac{2(N_1\vee N_2)}{1+(N_1\wedge N_2)}|J_1-J_2|}\,(n_1 \vee n_2)^{\frac{-2(N_1 \wedge N_2)}{1 + (N_1 \wedge N_2)}},
\end{eqnarray*}
where in the final step we expanded $\Log(\te{Id} + \xi_{n,N_1,N_2})$ via its Mercator series, using that the spectral radius of $\xi_{n,N_1,N_2}$ is smaller than 1 almost surely for $n$ sufficiently large.
\end{proof}

\subsection{Proof of Proposition \ref{prop:4.1}}
\begin{proof}
Set $M_{J,\vec{k}}^X = X_{\vec{k}}$, $M_{J,\vec{k}}^{\gamma} = \gamma_{\vec{k}}$ and $M_{J,\vec{k}}^{\epsilon} = \epsilon_{\vec{k}}$. Then, for each location $\vec{k}$ at scale $J$, 
\begin{eqnarray*}
\tr(\Log(M_{J,\vec{k}}^X)) \ = \ \tr(\Log((M_{J,\vec{k}}^{\gamma})^{1/2} \ast M_{J,\vec{k}}^{\epsilon})) \ = \ \tr(\Log(M_{J,\vec{k}}^{\gamma})) + \tr(\Log(M_{J,\vec{k}}^{\epsilon})), 
\end{eqnarray*}
using that $\tr(\Log(xy)) = \tr(\Log(x)) + \tr(\Log(y))$ and $\Log(x^t) = t \Log(x)$ for any $x,y \in \mathcal{M}$, $t \in \mathbb{R}$. Next, we verify that the same is true for \emph{any} scale $0 \leq j \leq J$, i.e.,
\begin{eqnarray} \label{C-eq:4.1}
\tr(\Log(M^X_{j,\vec{k}})) &=& \tr(\Log(M^\gamma_{j,\vec{k}})) + \tr(\Log(M^\epsilon_{j,\vec{k}})) \quad \quad \te{for all } j,\vec{k},
\end{eqnarray}
where $M^X_{j,\vec{k}}$, $M^\gamma_{j,\vec{k}}$, and $M^\epsilon_{j,\vec{k}}$ are the midpoints at scale-location $(j,\vec{k})$ based on the sequences $(X_{\vec{\ell}})_{\vec{\ell}}$, $(\gamma_{\vec{\ell}})_{\vec{\ell}}$, and $(\epsilon_{\vec{\ell}})_{\vec{\ell}}$ respectively. Fix a location $\vec{k} = (k_1,k_2)$ at scale $j = 0,\ldots,J$, and consider the subsets of locations $\vec{\ell} = (\ell_1,\ell_2)$ at scale $j + r$ given by $\mathcal{L}_{r} = \{ (\ell_1,\ell_2)\, :\, I_{j+r,\ell_1,\ell_2} \subset I_{j,k_1,k_2} \}$ for $r = 1,\ldots,J-j$. Using that $\tr(\Log(xy)) = \tr(\Log(x)) + \tr(\Log(y))$, $\Log(x^t) = t \Log(x)$, $a \ast \Log_{x}(y) = \Log_{a \ast x}(a \ast y)$ and $a \ast \Exp_{x}(y) = \Exp_{a \ast x}(g \ast y)$ for any $x,y \in \mathcal{M}$, $t \in \mathbb{R}$ and $a \in \te{GL}(d,\mathbb{C})$, we decompose:\\[0mm]
\begin{footnotesize}
\begin{eqnarray*}
\tr(\Log(M^X_{j,\vec{k}})) &=& \tr\Bigg[\Log\Bigg[\Exp_{M^X_{j,\vec{k}}}\Bigg[ \sum_{\vec{\ell} \in \mathcal{L}_{1}} \frac{\lambda_2(I_{j+1,\vec{\ell}})}{\lambda_2(I_{j,\vec{k}})}\, \Log_{M^X_{j,\vec{k}}}(M^X_{j+1, \vec{\ell}}) \Bigg] \Bigg] \Bigg]\nn
&=& \tr(\Log(M_{j,\vec{k}}^X)) - \sum_{\vec{\ell} \in \mathcal{L}_{1}} \frac{\lambda_2(I_{j+1,\vec{\ell}})}{\lambda_2(I_{j,\vec{k}})} \tr(\Log(M_{j,\vec{k}}^X)) + \sum_{\vec{\ell} \in \mathcal{L}_{1}} \frac{\lambda_2(I_{j+1,\vec{\ell}})}{\lambda_2(I_{j,\vec{k}})} \tr(\Log(M_{j+1,\vec{\ell}}^X)) \nn
&=& \sum_{\vec{\ell} \in \mathcal{L}_{1}} \frac{\lambda_2(I_{j+1,\vec{\ell}})}{\lambda_2(I_{j,\vec{k}})} \tr(\Log(M^X_{j+1,\vec{\ell}})) \nn
& \vdots & \nn
&=& \sum_{\vec{\ell} \in \mathcal{L}_{1}} \frac{\lambda_2(I_{j+1,\vec{\ell}})}{\lambda_2(I_{j,\vec{k}})} \left( \sum_{\{\vec{m}\, :\, I_{j+2,\vec{m}} \subset I_{j+1,\vec{\ell}}\}} \frac{\lambda_2(I_{j+2,\vec{m}})}{\lambda_2(I_{j+1,\vec{\ell}})} \tr(\Log(M^X_{j+2,\vec{m}})) \right) \nn
&=& \sum_{\vec{\ell} \in \mathcal{L}_{2}} \frac{\lambda_2(I_{j+2,\vec{\ell}})}{\lambda_2(I_{j,\vec{k}})} \tr(\Log(M^X_{j+2,\vec{\ell}})) \nn
& \vdots & \nn
&=& \sum_{\vec{\ell} \in \mathcal{L}_{J-j}} \frac{\lambda_2(I_{J,\ell})}{\lambda_2(I_{j,\vec{k}})} \tr(\Log(M^X_{J,\vec{\ell}})) \nn
&=& \sum_{\vec{\ell} \in \mathcal{L}_{J-j}} \frac{\lambda_2(I_{J,\ell})}{\lambda_2(I_{j,\vec{k}})} \tr(\Log(M^\gamma_{J,\vec{\ell}})) + \sum_{\vec{\ell} \in \mathcal{L}_{J-j}} \frac{\lambda_2(I_{J,\ell})}{\lambda_2(I_{j,\vec{k}})} \tr(\Log(M^\epsilon_{J,\vec{\ell}})) \nn
&=& \vdots \nn
&=& \sum_{\vec{\ell} \in \mathcal{L}_{1}} \frac{\lambda_2(I_{j+1,\ell})}{\lambda_2(I_{j,\vec{k}})} \tr(\Log(M^\gamma_{j+1,\vec{\ell}})) + \sum_{\vec{\ell} \in \mathcal{L}_{1}} \frac{\lambda_2(I_{j+1,\ell})}{\lambda_2(I_{j,\vec{k}})} \tr(\Log(M^\epsilon_{j+1,\vec{\ell}})) \nn
&=& \tr(\Log(M^\gamma_{j,\vec{k}})) + \tr(\Log(M^\epsilon_{j,\vec{k}})).
\end{eqnarray*}
\end{footnotesize}
\vspace{2mm}
Second, we also verify that: 
\begin{eqnarray} \label{C-eq:4.2}
\tr(\Log(\widetilde{M}^X_{j,\vec{k}})) &=& \tr(\Log(\widetilde{M}^\gamma_{j,\vec{k}})) + \tr(\Log(\widetilde{M}^\epsilon_{j,\vec{k}})) \quad \te{for all } j, \vec{k},
\end{eqnarray}
where $\widetilde{M}^X_{j,\vec{k}}, \widetilde{M}^\gamma_{j,\vec{k}}$, and $\widetilde{M}^W_{j,\vec{k}}$ are the predicted midpoints at scale-location $(j,\vec{k})$ based on the sequences $(X_{\vec{\ell}})_{\vec{\ell}}$, $(\gamma_{\vec{\ell}})_{\vec{\ell}}$, and $(\epsilon_{\vec{\ell}})_{\vec{\ell}}$ respectively. By Section \ref{sec:2.2} in the main document, the predicted midpoints can be written as weighted intrinsic averages of the coarse-scale midpoints in the general form:
\begin{eqnarray*}
\widetilde{M}^X_{j,\vec{k}} &=& \Exp_{\widetilde{M}^X_{j,\vec{k}}}\Bigg( \sum_{\ell \in \Lambda_{\vec{k}}} w_{\vec{\ell}}\ \Log_{\widetilde{M}^X_{j,\vec{k}}}(M^X_{j-1,\vec{\ell}}) \Bigg),
\end{eqnarray*}
where $\Lambda_{\vec{k}}$ denotes the set of locations $\vec{\ell} = (\ell_1,\ell_2)$ corresponding to the collection of neighboring $(j-1)$-scale midpoints to predict the $j$-scale midpoint at location $\vec{k} = (k_1,k_2)$. The filter weights $(w_{\vec{\ell}})_{\ell}$ depend on the average-interpolation order and sum up to 1, i.e., $\sum_{\vec{\ell} \in \Lambda_{\vec{k}}} w_{\vec{\ell}} = 1$. Using eq.(\ref{C-eq:4.1}) and the same identities as above, we decompose, 
\begin{eqnarray*}
\tr(\Log(\widetilde{M}^{X}_{j,\vec{k}})) &=& \tr\Big(\Log\Big(\Exp_{\widetilde{M}^X_{j,\vec{k}}}\Big( \sum_{\vec{\ell} \in \Lambda_{\vec{k}}} w_{\vec{\ell}}\ \Log_{\widetilde{M}^X_{j,\vec{k}}}(M^X_{j-1,\vec{\ell}}) \Big)\Big)\Big)\nn
&=& \tr(\Log(\widetilde{M}^{X}_{j,\vec{k}})) + \tr\Big( \sum_{\vec{\ell}\in \Lambda_{\vec{k}}} w_{\vec{\ell}}\ \Log\Big( (\widetilde{M}^{X}_{j,\vec{k}})^{-1/2} \ast M^{X}_{j-1,\vec{\ell}} \Big) \Big) \nn
&=& \tr(\Log(\widetilde{M}^{X}_{j,\vec{k}})) + \sum_{\vec{\ell} \in \Lambda_{\vec{k}}} w_{\vec{\ell}}\ \Big( \tr(\Log(M^{X}_{j-1,\vec{\ell}})) - \tr(\Log(\widetilde{M}^{X}_{j,\vec{k}})) \Big) \nn
&=& \sum_{\vec{\ell} \in \Lambda_{\vec{k}}} w_{\vec{\ell}}\ \tr(\Log(M^{X}_{j-1,\vec{\ell}})) \nn
&=& \sum_{\vec{\ell} \in \Lambda_{\vec{k}}} w_{\vec{\ell}}\ \tr(\Log(M^{\gamma}_{j-1,\vec{\ell}})) + \sum_{\ell} w_{\vec{\ell}}\ \tr(\Log(M^{\epsilon}_{j-1,\vec{\ell}})) \nn
&\vdots & \nn
&=& \tr(\Log(\widetilde{M}^\gamma_{j,\vec{k}})) + \tr(\Log(\widetilde{M}^\epsilon_{j,\vec{k}}))
\end{eqnarray*}
The first claim in the Proposition now follows from eq.(\ref{C-eq:4.1}) and eq.(\ref{C-eq:4.2}) through: 
\begin{eqnarray}
\tr(\mathfrak{D}^{X}_{j,\vec{k}}) &=& \sqrt{\frac{\lambda_2(I_{j,\vec{k}})}{\lambda_2(\mathcal{I})}}\, \tr\Big(\Log\Big((\widetilde{M}^{X}_{j,\vec{k}})^{-1/2} \ast M^{X}_{j,\vec{k}} \Big)\Big) \nn
&=& \sqrt{\frac{\lambda_2(I_{j,\vec{k}})}{\lambda_2(\mathcal{I})}}\, \Big(\tr(\Log( M^{X}_{j,\vec{k}})) - \tr(\Log( \widetilde{M}^{X}_{j,\vec{k}} ))\Big) \nn
&=& \sqrt{\frac{\lambda_2(I_{j,\vec{k}})}{\lambda_2(\mathcal{I})}}\, \Bigg[\tr(\Log( M^\gamma_{j,\vec{k}})) + \tr(\Log(M^\epsilon_{j,\vec{k}})) - \tr(\Log( \widetilde{M}^{\gamma}_{j,\vec{k}} )) - \tr(\Log( \widetilde{M}^{\epsilon}_{j,\vec{k}} )) \Bigg]  \nn
&=& \tr(\mathfrak{D}^\gamma_{j,\vec{k}}) + \tr(\mathfrak{D}^\epsilon_{j,\vec{k}}). \label{C-eq:4.3}
\end{eqnarray}
For the second claim in the proposition, first observe that:
\begin{eqnarray*}
\bs{E}[\tr(\Log(M^\epsilon_{j,\vec{k}}))] \ =\ \sum_{\vec{\ell} \in \mathcal{L}_{J-j}} \frac{\lambda_2(I_{J,\vec{\ell}})}{\lambda_2(I_{j,\vec{k}})}\bs{E}[\tr(\Log(M^\epsilon_{J,\vec{\ell}}))] \ = \ 0, \quad \quad \te{for each } j,\vec{k},
\end{eqnarray*}
using that $\bs{E}[\tr(\Log(M^\epsilon_{J,\vec{\ell}})] = 0$ for each $\vec{\ell} \in \{0,\ldots,n_1-1\} \times \{0,\ldots, n_2-1\}$, which is implied by $\bs{E}[\Log_{\te{Id}}(M^\epsilon_{J,\vec{\ell}})] = \bs{0}$ as the intrinsic mean of $M^{\epsilon}_{J,\vec{\ell}}$ equals the identity matrix. As a consequence, also,
\begin{eqnarray*}
\bs{E}[\tr(\Log(\widetilde{M}^\epsilon_{j,\vec{k}}))] \ = \ \sum_{\vec{\ell} \in \Lambda_{\vec{k}}} w_{\vec{\ell}}\ \bs{E}[\tr(\Log(M^{\epsilon}_{j-1,\vec{\ell}}))] \ = \ 0 \quad \quad \te{for each } j,\vec{k},
\end{eqnarray*}
and therefore, 
\begin{eqnarray*}
\bs{E}[\tr(\mathfrak{D}^X_{j,\vec{k}})] &=& \tr(\mathfrak{D}^\gamma_{j,\vec{k}}) + \bs{E}[\tr(\mathfrak{D}^\epsilon_{j,\vec{k}})] \nn
&=& \tr(\mathfrak{D}^\gamma_{j,\vec{k}}) + \bs{E}\left[ \tr(\Log(M^\epsilon_{j,\vec{k}})) - \tr(\Log(\widetilde{M}^\epsilon_{j,\vec{k}})) \right] \nn
&=& \tr(\mathfrak{D}^\gamma_{j,\vec{k}}).
\end{eqnarray*}
For the variance of $\tr(\mathfrak{D}_{j,\vec{k}}^X)$, we first note that the random variables $(M^{\epsilon}_{J,\vec{k}})_{\vec{k}}$ are independent across locations, implying that the random variables $(\tr(\Log(M^\epsilon_{j,\vec{k}}))_{\vec{k}}$ are independent within every scale $j=0,\ldots,J$. Denote $\vec{m} = (m_1,m_2)$ for the location at scale $j-1$, such that $I_{j,\vec{k}} \subset I_{j-1,\vec{m}}$ and let $\mathcal{L}_{\vec{m}} := \{ (\ell_1,\ell_2)\, :\, I_{j,\ell_1,\ell_2} \subset I_{j-1,m_1,m_2} \}$. We write out:
\begin{footnotesize}
\begin{eqnarray} \label{C-eq:4.4}
\var(\tr(\mathfrak{D}_{j,\vec{k}}^{X})) &=& \frac{\lambda_2(I_{j,\vec{k}})}{\lambda_2(\mathcal{I})}\ \var\left(\tr(\Log(M^\epsilon_{j,\vec{k}})) - \tr(\Log(\widetilde{M}^\epsilon_{j,\vec{k}}))\right) \nn
&=& \frac{\lambda_2(I_{j,\vec{k}})}{\lambda_2(\mathcal{I})}\ \var\Big(\tr(\Log(M^\epsilon_{j,\vec{k}})) -  \sum_{\vec{\ell} \in \Lambda_{\vec{k}}} w_{\vec{\ell}}\ \tr(\Log(M^{\epsilon}_{j-1,\vec{\ell}}))\Big) \nn
&=& \frac{\lambda_2(I_{j,\vec{k}})}{\lambda_2(\mathcal{I})}\ \Bigg[ \var\Big(\tr(\Log(M^\epsilon_{j,\vec{k}})) - \tr(\Log(M^\epsilon_{j-1,\vec{m}}))\Big) \nn
&& + \sum_{\vec{\ell} \in \Lambda_{\vec{k}}\, :\, \vec{\ell} \neq \vec{m}} w_{\vec{\ell}}^2\ \var(\tr(\Log(M^\epsilon_{j-1,\vec{\ell}}))) \Bigg] \nn
&=& \frac{\lambda_2(I_{j,\vec{k}})}{\lambda_2(\mathcal{I})}\ \Bigg[ \var\Bigg(\tr(\Log(M^\epsilon_{j,\vec{k}})) - \sum_{\vec{\ell} \in \mathcal{L}_{\vec{m}}} \frac{\lambda_2(I_{j,\vec{\ell}})}{\lambda_2(I_{j-1,\vec{m}})}\, \tr(\Log(M^\epsilon_{j,\vec{\ell}}))\Bigg) \nn
&& + \sum_{\vec{\ell} \in \Lambda_{\vec{k}}\, :\, \vec{\ell} \neq \vec{m}} w_{\vec{\ell}}^2\ \var(\tr(\Log(M^\epsilon_{j-1,\vec{\ell}}))) \Bigg] \nn
&=& \frac{\lambda_2(I_{j,\vec{k}})}{\lambda_2(\mathcal{I})}\ \Bigg[ \sum_{\vec{\ell} \in \mathcal{L}_{\vec{m}}\, :\, \vec{\ell} \neq \vec{k}} \frac{\lambda_2(I_{j,\vec{\ell}})^2}{\lambda_2(I_{j-1,\vec{m}})^2}\ \var(\tr(\Log(M^\epsilon_{j,\vec{\ell}}))) \nn
&& +\ \Bigg(1 - \frac{\lambda_2(I_{j,\vec{k}})}{\lambda_2(I_{j-1,\vec{m}})}\Bigg)^2 \var(\tr(\Log(M^\epsilon_{j,\vec{k}}))) + \sum_{\vec{\ell} \in \Lambda_{\vec{k}} : \vec{\ell} \neq \vec{m}} w_{\vec{\ell}}^2\ \var(\tr(\Log(M^\epsilon_{j-1,\vec{\ell}}))) \Bigg] \nn
&=& \frac{\lambda_2(I_{j,\vec{k}})}{\lambda_2(\mathcal{I})}\ \Bigg[ \sum_{\vec{\ell} \in \Lambda_{\vec{k}}} w_{\vec{\ell}}^2\ \var(\tr(\Log(M^\epsilon_{j-1,\vec{\ell}}))) + \Bigg(1 - \frac{2\lambda_2(I_{j,\vec{k}})}{\lambda_2(I_{j-1,\vec{m}})}\Bigg) \var(\tr(\Log(M^\epsilon_{j,\vec{k}}))) \Bigg], \nn
\end{eqnarray}
\end{footnotesize}
where $\Lambda_{\vec{k}}$ denotes again the set of locations corresponding to the collected $(j-1)$-scale midpoints to predict the $j$-scale midpoint at location $\vec{k}$, and we used that $w_{\vec{m}} = 1$, see eq.(\ref{eq:2.5}) in the main document. For each scale-location $(j,\vec{k})$, analogous to the decomposition below eq.(\ref{C-eq:4.1}) and with the same notation $\mathcal{L}_r$, using the independence of $(M^\epsilon_{j,\vec{\ell}})_{\ell}$, $\var(\tr(\Log(M^{\epsilon}_{j,\vec{k}}))$ can be decomposed as:
\begin{eqnarray} \label{C-eq:4.5}
\var(\tr(\Log(M^{\epsilon}_{j,\vec{k}})) &=& \var\Bigg( \sum_{\vec{\ell} \in \mathcal{L}_{J-j}} \frac{\lambda_2(I_{J,\vec{\ell}})}{\lambda_2(I_{j,\vec{k}})} \tr(\Log(M_{J,\vec{\ell}}^\epsilon)) \Bigg) \nn
&=& \sum_{\vec{\ell} \in \mathcal{L}_{J-j}} \frac{\lambda_2(I_{J,\vec{\ell}})^2}{\lambda_2(I_{j,\vec{k}})^2} \var(\tr(\zeta)) \nn
&=& | \mathcal{L}_{J-j} | \frac{\lambda_2(I_{J,0,0})^2}{\lambda_2(I_{j,\vec{k}})^2} \var(\tr(\zeta)) \nn
&=& \frac{\lambda_2(I_{j,\vec{k}})}{\lambda_2(I_{J,0,0})} \frac{\lambda_2(I_{J,0,0})^2}{\lambda_2(I_{j,\vec{k}})^2} \var(\tr(\zeta)) \ =\ \frac{\lambda_2(I_{J,0,0})}{\lambda_2(I_{j,\vec{k}})} \var(\tr(\zeta)),
\end{eqnarray}
using that $\Log(M_{J,\vec{\ell}}^\epsilon) \overset{d}{=} \zeta$ and the fact that the size $\lambda_2(I_{J,\vec{\ell}})$ is the same for each location $\vec{\ell} = (\ell_1,\ell_2)$ at scale $J$ in the natural dyadic refinement pyramid. Given the natural dyadic refinement pyramid, by eq.(\ref{eq:3.1}) in the main document:
\begin{eqnarray*}
\lambda_2(I_{j,\vec{k}}) &=& \left\{
\begin{array}{ll}
2^{-j} & \te{if } j \leq |J_1 - J_2| \\
2^{-2j + |J_1 - J_2|} & \te{if } j > |J_1 - J_2|
\end{array}
\right.
\end{eqnarray*}
For the first case, $j \leq |J_1-J_2|$, combining eq.(\ref{C-eq:4.4}) and eq.(\ref{C-eq:4.5}) above and using that $\lambda_2(\mathcal{I}) = 1$,
\begin{eqnarray*}
\var(\tr(\mathfrak{D}_{j,\vec{k}}^X)) &=&  \frac{\lambda_2(I_{j,\vec{k}})}{\lambda_2(\mathcal{I})}\ \Bigg[ \sum_{\vec{\ell} \in \Lambda_{\vec{k}}} w_{\vec{\ell}}^2\ \frac{\lambda_2(I_{J,0,0})}{\lambda_2(I_{j-1,\vec{\ell}})} \var(\tr(\zeta)) + \Bigg(1 - \frac{2\lambda_2(I_{j,\vec{k}})}{\lambda_2(I_{j-1,\vec{m}})}\Bigg) \frac{\lambda_2(I_{J,0,0})}{\lambda_2(I_{j,\vec{k}})} \var(\tr(\zeta)) \Bigg] \nn
&=& 2^{-j}\Bigg[ \sum_{\vec{\ell} \in \Lambda_{\vec{k}}} w_{\vec{\ell}}^2\ \frac{2^{j-1}}{n}\var(\tr(\zeta)) + \Big(1 - 2\,\frac{2^{-j}}{2^{-(j-1)}}\Big) \frac{2^j}{n} \var(\tr(\zeta)) \Bigg] \nn
&=& \frac{1}{n}\Bigg( \frac{1}{2} \sum_{\vec{\ell} \in \Lambda_{\vec{k}}} w_{\vec{\ell}}^2 \Bigg) \var(\tr(\zeta)).
\end{eqnarray*}
For the second case, $j > |J_1 - J_2|$, in the same way, using in particular that $2J = J_1 + J_2 + |J_1 - J_2|$, (since $J = J_1 \vee J_2$), we find:
\begin{eqnarray*}
\var(\tr(\mathfrak{D}_{j,\vec{k}}^X)) &=& 2^{-2j + |J_1-J_2|} \Bigg[ \sum_{\vec{\ell} \in \Lambda_{\vec{k}}} w_{\vec{\ell}}^2\ 2^{-2(J-(j-1))} \var(\tr(\zeta)) + (1-2^{-1})2^{-2(J-j)}\var(\tr(\zeta)) \Bigg] \nn
&=& \frac{1}{n}\Bigg( \frac{1}{2} + \frac{1}{4} \sum_{\vec{\ell} \in \Lambda_{\vec{k}}} w_{\vec{\ell}}^2 \Bigg) \var(\tr(\zeta)),
\end{eqnarray*}
which finishes the proof.
\end{proof}

\subsection{Dyadic midpoint prediction} \label{C-sec:5}

Without loss of generality, assume that $\mathcal{I} = [0,1] \times [0,1]$ as in Section \ref{sec:2.2} in the main document, and the refinement pyramid $(I_{j,\vec{k}})_{j,\vec{k}}$ corresponds to the natural dyadic refinement pyramid. With the same notation as in Section \ref{sec:2} in the main document, the predicted midpoints $(\widetilde{M}_{j,i_1,i_2})_{i_1,i_2}$ at locations $(i_1,i_2)$ corresponding to the equally-sized square refinement rectangles $I_{j,i_1,i_2} \subset I_{j-1,k_1,k_2}$ are obtained from the cumulative intrinsic averages $(\widebar{M}_{j-1,r_1,r_2})_{r_1,r_2}$ and the interpolating polynomial surface $\widehat{M}(t,s)$ evaluated at different locations $(\tilde{t},\tilde{s})$.\\[3mm] 
Fix a location $(k_1,k_2)$ at resolution scale $j-1$ and an average-interpolation order $(N_1,N_2) = (2L_1+1,2L_2+1) \geq (1,1)$. For convenience suppose that the midpoints $(M_{j-1, k_1+\ell_1, k_2+\ell_2})_{\ell_1,\ell_2}$ with symmetric neighboring locations $(\ell_1,\ell_2) \in \{-L_1,\ldots,L_1\} \times \{ -L_2, \ldots, L_2\}$ exist. Below, we describe the exact expressions to predict the finer-scale midpoints $M_{j,2k_1,2k_2}$, $M_{j, 2k_1+1, 2k_2}$, $M_{j,2k_1,2k_2+1}$ and $M_{j,2k_1+1,2k_2+1}$. The interpolating surface $\widehat{M}(t,s)$ needs to be evaluated at the following locations:
\begin{eqnarray*}
\widehat{M}_{\ell_1, \ell_2} & := & \widehat{M}((2k_1 + \ell_1) 2^{-(j+1)}, (2k_2 + \ell_2) 2^{-(j+1)}),
\end{eqnarray*}
with $(\ell_1,\ell_2) \in \{ (0,1), (1,0), (1,1), (0,2), (2,0), (2,1), (1,2) \}$. This can be done efficiently via Neville's algorithm. Note that $\widehat{M}_{2,2} = \widebar{M}_{L_1, L_2}$ and $\widehat{M}_{0,0} = \widebar{M}_{L_1 - 1, L_2 - 1}$ by construction of the interpolating surface.
\paragraph{Prediction of $M_{j,2k_1+1,2k_2+1}$} Predicting the midpoint $M_{j,2k_1+1,2k_2+1}$ requires a few extra --but straightforward-- steps in comparison to the intrinsic 1D AI refinement scheme, which are easily verified by drawing an image of the different areas. 
\begin{enumerate}
\item Compute $A_1 := \te{Ave}(\{ \widehat{M}_{2,1}, \widehat{M}_{1,1} \} ; \{ N_1 + 1, -N_1 \})$.
\item Compute $A_2 := \te{Ave}\left(\{ \widehat{M}_{1,2}, A_1 \} ; \left\{ \frac{N_1(N_2+1)}{(N_1+1)(N_2+1)-1}, \frac{N_2}{(N_1+1)(N_2+1)-1} \right\} \right)$.
\item Compute the predicted midpoint $\widetilde{M}_{j,2k_1+1, 2k_2+1}$ as the following intrinsic weighted average:
\begin{eqnarray*}
\widetilde{M}_{j,2k_1+1, 2k_2+1} &=& \te{Ave}\left( \{ \widebar{M}_{j-1,L_1,L_2}, A_2 \} ; \{ (N_1+1)N_2+N_1+1, -((N_1+1)N_2)+N_1) \} \right).
\end{eqnarray*}
\end{enumerate}

\paragraph{Prediction of $M_{j+1,2k_1,2k_2+1}$ and $M_{j+1,2k_1+1,2k_2}$} 
First, in order to predict $M_{j+1,2k_1, 2k_2+1}$, proceed as follows:
\begin{enumerate}
\item Compute $B_1 := \te{Ave}( \{ \widehat{M}_{0,2}, \widehat{M}_{0,1} \}; \{ N_2 + 1, -N_2 \} )$ 
\item Compute $B_2 := \te{Ave}\left( \{ \widehat{M}_{2,1}, B_1 \} ; \left\{ \frac{(N_1+1)N_2}{(N_1+1)(N_2+1)-2}, \frac{N_1-1}{(N_1+1)(N_2+1)-2} \right\} \right)$.
\item Compute $B_3 := \te{Ave}\left( \{ \widebar{M}_{j-1,L_1,L_2}, B_2 \} ; \left\{ \frac{(N_1+1)(N_2+1)-2}{2}, -\frac{(N_1+1)(N_2+1)-2}{2} \right\} \right)$.
\item Compute the predicted midpoint $\widetilde{M}_{j,2k_1,2k_2+1}$ as the intrinsic weighted average: 
\begin{eqnarray*}
\widetilde{M}_{j,2k_1,2k_2+1} &=& \te{Ave}(\{ B_3, \widetilde{M}_{j,2k_1+1,2k_2+1} \} ; \{ 2, -1 \}).
\end{eqnarray*}
\end{enumerate}
Second, in order to predict $\widetilde{M}_{j+1,2k_1+1,2k_2}$, proceed in an analogous fashion as for the prediction of $\widetilde{M}_{j+1,2k_1,2k_2+1}$: 
\begin{enumerate}
\item Compute $C_1 := \te{Ave}( \{ \widehat{M}_{2,0}, \widehat{M}_{1,0} \}; \{ N_1 + 1, -N_1 \} )$ 
\item Compute $C_2 := \te{Ave}\left( \{ \widehat{M}_{1,2}, C_1 \} ; \left\{ \frac{(N_2+1)N_1}{(N_2+1)(N_1+1)-2}, \frac{N_2-1}{(N_2+1)(N_1+1)-2} \right\} \right)$.
\item Compute $C_3 := \te{Ave}\left( \{ \widebar{M}_{j-1,L_1,L_2}, C_2 \} ; \left\{ \frac{(N_1+1)(N_2+1)-2}{2}, -\frac{(N_1+1)(N_2+1)-2}{2} \right\} \right)$.
\item Compute the predicted midpoint $\widetilde{M}_{j,2k_1+1,2k_2}$ as the intrinsic weighted average: 
\begin{eqnarray*}
\widetilde{M}_{j,2k_1+1,2k_2} &=& \te{Ave}(\{ C_3, \widetilde{M}_{j,2k_1+1,2k_2+1} \} ; \{ 2, -1 \}).
\end{eqnarray*}
\end{enumerate}
\paragraph{Prediction of $M_{j+1,2k_1,2k_2}$} Given the three predicted midpoints calculated above and the coarse-scale midpoint $M_{j,k_1,k_2}$, the predicted midpoint $\widetilde{M}_{j+1,2k_1,2k_2}$ follows from the midpoint relation in eq.(\ref{eq:3.1}) in the main document according to:
\begin{eqnarray} 
\widetilde{M}_{j+1,2k_1,2k_2} &=& \Exp_{M_{j,k_1,k_2}}\Big( - \sum_{\substack{(\ell_1,\ell_2) \in \\ \{(1,0),(0,1),(1,1)\}}} \Log_{M_{j,k_1,k_2}}(\widetilde{M}_{j+1,2k_1+\ell_1, 2k_2+\ell_2}) \Big). \quad \quad
\end{eqnarray}

\end{small}

\end{document}